\documentclass[12pt]{article}
\usepackage[utf8]{inputenc}
\usepackage[T1]{fontenc}
\usepackage{graphicx} % Required for inserting images
\usepackage[left=2cm,right=2cm,top=2cm,bottom=2cm]{geometry}
\usepackage{authblk}
\usepackage{siunitx}
\usepackage{hyperref} 
\usepackage[doi=true, articletitle=true]{achemso} % Loads/configures natbib
\usepackage{doi}
\usepackage{booktabs}
\usepackage{braket}
\usepackage{orcidlink}
\setcitestyle{numbers}
\DeclareSIUnit\angstrom{\text {Å}}
\DeclareSIUnit\celcius{C}
\DeclareSIUnit\percubiccm{\per\cubic\centi\metre}
\DeclareSIUnit\ps{\pico\second}

\title{Light-Stabilized Metastable Electronic State in NiO with Enhanced Orbital Hybridization}

\author[1,2]{Thomas C. Rossi\orcidlink{0000-0002-7448-8948}}
\author[1]{Fabio G. Santomauro}
\author[1]{Lars Mewes\orcidlink{0000-0001-9269-1475}}
\author[1]{Janina L\"offler\orcidlink{0000-0003-3147-8640}\thanks{Current affiliation: HES-SO Valais-Wallis, Institute of Energy and Environment, Rue de l'industrie 23, 1950 Sion, Switzerland}}
\author[1]{Dominik Kinschel\orcidlink{0000-0002-0269-8567}}
\author[1]{Giacomo Rossi\orcidlink{0000-0001-7122-9484}}
\author[1,7]{Mario Guti\'errez\orcidlink{0000-0001-9222-3647}}
\author[1]{Oliviero Cannelli\orcidlink{0000-0002-1844-4799}\thanks{Current affiliation: Centre for Free-electron Laser Science, Deutsches Elektronen-Synchrotron, Notkestr.\ 85, 22607 Hamburg, Germany}}
\author[1]{Boris V. Sorokin\orcidlink{0000-0002-4950-5654}}
\author[1]{Jochen Rittmann}
\author[1]{Jakob Kel\thanks{Current affiliation: Department of Neurology, University Hospital Basel, Basel, Switzerland \& Research Center for Clinical Neuroimmunology and Neuroscience Basel (RC2NB), University Hospital Basel and University of Basel, Switzerland}}
\author[1]{James Budarz\orcidlink{0000-0002-7673-6626}}
\author[3]{Anna Wach\orcidlink{0000-0003-3112-2759}\thanks{Current affiliation: SOLARIS National Synchrotron Radiation Centre, Jagiellonian University, Kraków 30-392, Poland}}
\author[3]{Adam H. Clark\orcidlink{0000-0002-5478-9639}}
\author[4]{Emiliano Dal Molin\orcidlink{0000-0002-1377-3999}}
\author[4]{Maged F. Bekheet\orcidlink{0000-0003-1778-0288}}
\author[2]{Albert Gili\orcidlink{0000-0001-7944-7881}}
\author[5]{Sebastian Praetz\orcidlink{0000-0001-6852-7616}}
\author[5]{Daniel Gr\"otzsch}
\author[6]{Delphine Cabaret\orcidlink{0000-0001-5875-9667}}
\author[2,5]{Renske M. van der Veen\orcidlink{0000-0003-0584-4045}}
\author[1,8]{Majed Chergui\orcidlink{0000-0002-4856-226X}}
\affil[1]{Laboratory of Ultrafast Spectroscopy, Lausanne Centre for Ultrafast Science (LACUS), \'Ecole Polytechnique F\'ed\'erale de Lausanne SB-ISIC, Station 6, CH-1015 Lausanne, Switzerland}
\affil[2]{Department of Atomic-Scale Dynamics in Light-Energy Conversion (PS-ADLU), Helmholtz-Zentrum Berlin f\"ur Materialien und Energie GmbH, Magnusstrasse 2, D-12489 Berlin, Germany}
\affil[3]{Center for Photon Science, Paul Scherrer Institute, CH-5232 Villigen, Switzerland}
\affil[4]{Technische Universit\"at Berlin, Faculty III Process Sciences, Institute of Materials Science and Technology, Chair of Advanced Ceramic Materials, Straße des 17. Juni 135, 10623 Berlin, Germany}
\affil[5]{Institute for Physics and Astronomy, Technische Universit\"at Berlin, Hardenbergstraße 36, D-10623 Berlin, Germany}
\affil[6]{Institut de Min\'eralogie, de Physique des Mat\'eriaux et de Cosmochimie, Universit\'e Pierre et Marie Curie, 75005 Paris, France}
\affil[7]{Departamento de Qu\'imica F\'isica, Faculdad de Ciencias Ambientales y Bioqu\'imica, INAMOL, Universidad de Castilla-La Mancha, Toledo 45071, Spain}
\affil[8]{Elettra-Sincrotrone Trieste, SS 14, km 163.5, 34149 Basovizza, Trieste, Italy}

\date{\today}

\usepackage{amsmath}
\usepackage{pifont}
\usepackage{array}
\usepackage[section]{placeins}
\DeclareSIUnit{\eV}{\text{eV}}
\DeclareSIUnit\bar{bar}

\begin{document}

\maketitle

\begin{center}
\small
\textit{Corresponding authors:}\\
Thomas C. Rossi (\texttt{thomas.rossi@helmholtz-berlin.de})\\
Majed Chergui (\texttt{majed.chergui@elettra.eu})
\end{center}

\cleardoublepage

\begin{abstract}
Light-driven control of electronic structure in correlated metal oxides offers new opportunities for optimizing materials used in photovoltaic and photoelectrochemical technologies. Here, we show that photoexcitation of NiO, a prototypical transparent semiconductor and hole-transport material, across its charge-transfer gap, produces a long-lived metastable state with enhanced Ni \(3d\)–O \(2p\) orbital hybridization. We characterize this state using Ni K-edge X-ray absorption spectroscopy (XAS), which probes how structural and electronic changes affect the unoccupied \(p\) density of states during both continuous (\emph{in situ}) and pulsed UV excitation. Under pulsed excitation, high carrier densities (\(\sim\SI{1e20}{\percubiccm}\)) generate a state with a lifetime of \(\sim\SI{600}{\ps}\), in which enhanced hybridization coexists with lattice heating. By contrast, continuous UV irradiation at much lower carrier densities (\(\sim\SI{1e13}{\percubiccm}\)) stabilizes a similar electronic state with negligible lattice heating, showing that its formation is not solely thermally driven. First-principles DFT+\(U\)+\(V\) calculations attribute the spectral changes to stronger Ni \(3d\)–O \(2p\) hybridization, which alters the unoccupied Ni \(4p\) states probed by dipole-allowed K-edge transitions. We attribute this change to the dynamic screening of on-site electronic correlations following photoexcitation, which redistributes the charge density. Because orbital hybridization governs carrier transport and charge-transfer energetics, our results identify photoinduced screening as a mechanism for dynamically tuning correlated oxides and suggest new design principles for optoelectronic materials.
\end{abstract}

\cleardoublepage

\section*{Introduction}

% Motivation: why controlling electronic correlations matters
Controlling electron--electron interactions and metal-ligand hybridization in transition metal oxides offers a powerful route to tune charge transport, catalytic activity, and energy conversion processes in photovoltaic and photoelectrochemical devices. In materials where valence electrons occupy spatially localized orbitals such as $3d$, strong on-site Coulomb repulsion makes them highly sensitive to screening, which can alter charge-transfer energetics, orbital hybridization, and carrier mobility. However, current strategies to tailor these properties—such as chemical doping or defect engineering—are largely static and often introduce structural disorder~\cite{Li2025:123292,Wrobel2020}. This raises a key question: can light be used to dynamically and reversibly control electron--electron interactions in correlated oxides under operating conditions?

% Background: strongly correlated oxides
Transition metal oxides with partially filled $3d$ shells exhibit rich electronic behavior governed by a competition between strong on-site Coulomb repulsion (correlation) in metal $3d$ orbitals and covalent hybridization with oxygen $2p$ states~\cite{Cox2010}. This interplay dictates metal--ligand covalency, charge-transfer energetics, and redox flexibility, underpinning phenomena such as metal--insulator transitions~\cite{Imada:1998er}, catalytic activity~\cite{Grimaud2017:223225}, and interfacial charge transport~\cite{Hu:2014gw,Wang:2014kj}.

\begin{figure}
	\centering
	\includegraphics[width=0.7\linewidth]{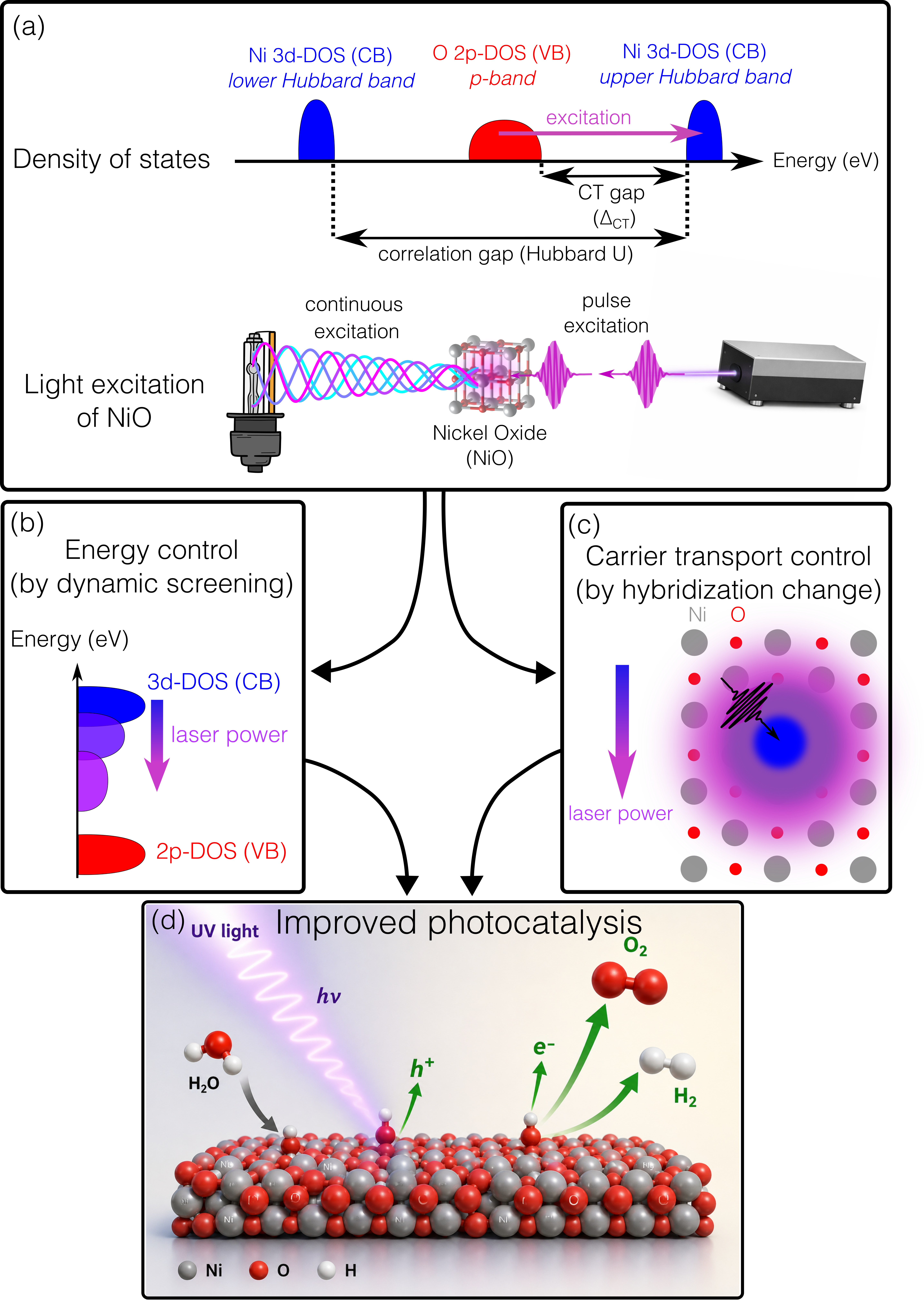}
	\caption{\textbf{Photoinduced changes in the electronic structure of correlated oxides can lead to improved photocatalytic performances.} (a) Schematic density of states near the charge-transfer gap ($\Delta_{\text{CT}}$) of NiO. Different UV light excitations are applied to NiO in this work. (b) Dynamic screening of the $3d$ conduction band leading to a renormalization of the charge-transfer gap and a broadening of the $d$ states, which enables energy control with laser power. (c) Hybridization change due to dynamic screening leading to tunable carrier transport with laser power. (d) Photoinduced tunability of energy and carrier transport is a candidate to transfer more efficiently charge carriers to catalytic sites and increase photocatalytic efficiency of strongly correlated materials.}
	\label{fig:experiment_illustration}
\end{figure}

% Case study: NiO
Nickel oxide (NiO) provides a prototypical platform to investigate these effects. NiO is a charge-transfer insulator with a wide band gap ($\Delta_{\text{CT}}\sim\SI{3.7}{\electronvolt}$). Optical excitation across this gap transfers electrons from oxygen $2p$ states into unoccupied nickel $3d$ Hubbard states (Figure~\ref{fig:experiment_illustration}a). The energy and occupation of the Hubbard bands are highly sensitive to electron--electron repulsion~\cite{Sawatzky:1984kh,Kunes:2007dv}, and can be modified by chemical doping~\cite{Ostrom2025:101832} or lattice distortions~\cite{Anonymous:2013jw}. Owing to its pronounced ligand-to-metal charge-transfer character, NiO is widely used as a hole-transport layer in dye-sensitized~\cite{Mori:2008cw} and perovskite solar cells~\cite{Hu:2014gw,Wang:2014kj}, where its electronic structure directly impacts device performance~\cite{Bai:2013cn,Manders:2013et,Kim:2014bw}.

% Light as a control knob
Ultrafast optical excitation offers a promising alternative to transiently alter correlations through photoinduced Coulomb screening~\cite{Tancogne-Dejean2020}. In strongly correlated oxides, such screening can renormalize band gaps and even drive transitions toward more conductive states~\cite{Cavalleri:2001be,Basov2017:238941}. Yet, the microscopic nature, lifetime, and structural consequences of these photoinduced states remain poorly understood, limiting the practical use of this class of materials. A central challenge is to disentangle intrinsic electronic effects from lattice heating and structural distortions, since electronic and lattice degrees of freedom are strongly coupled~\cite{Ahn2021:291841}.

% State of the art in NiO
Recent ultrafast studies on NiO have shown that photoexcitation across the charge-transfer gap modifies charge-transfer energetics~\cite{Lojewski2023,Cazali2025:166775,Rossi2025:52718} and generates bound electron--hole pairs with significant metal--ligand character~\cite{Merzoni2025:187430}. Consequently, these excitations are expected to influence Ni--O bonding~\cite{Windsor2021} and magnetic interactions~\cite{Wang2022a}. However, direct, element-specific measurements that correlate changes in local electronic structure to orbital hybridization and lattice response are still lacking.

% This work
Here, we address this challenge using Ni K-edge X-ray absorption spectroscopy (XAS) under both \emph{in situ} continuous and pulsed UV excitation (Figure~\ref{fig:experiment_illustration}a) to directly probe the coupled electronic and structural response in monocrystalline NiO. The use of nearly stoichiometric NiO minimizes the influence of intrinsic nickel vacancies~\cite{Mrowec2004:163117}. We show that high-fluence pulsed excitation generates a $\sim\SI{600}{\pico\second}$-lived electronically excited state that coexists with thermal lattice expansion and disorder, whereas continuous low-intensity irradiation stabilizes a similar excited state without significant heating. The electronically excited state is characterized by a red shift of the Ni K-edge interpreted as an enhancement of Ni--O orbital hybridization, which modifies the unoccupied $p$-projected density of states probed by dipole allowed transitions. These results demonstrate that light can selectively alter metal-centered and intersite Coulomb interactions in NiO, independent of bulk thermal effects.

% Implications
Our findings establish light-driven electronic screening as a strategy to reversibly tune the electronic structure of correlated oxides and their functional properties~\cite{Yang2021:207724,Woo2024:107143}. More broadly, they provide a pathway toward the dynamic control of energy states (Figure~\ref{fig:experiment_illustration}b) and charge transport (Figure~\ref{fig:experiment_illustration}c) in correlated oxides, providing new design principles for photoelectrochemical and catalytic materials with improved performance (Figure~\ref{fig:experiment_illustration}d).

\cleardoublepage

%%%%%%%%%%%%%%%%%%%%%%%%%%%%%%%%%%%%%%%%%%%%%%%%%%%%%%%%%%%%%%%%%%%%%%%%%%%%%%%%%%%%%%%%%
%%%%%%%%%%%%%%%%%%%%%%%%%%%%%%%%%%%%%%%%%%%%%%%%%%%%%%%%%%%%%%%%%%%%%%%%%%%%%%%%%%%%%%%%%
%%%%%%%%%%%%%%%%%%%%%%%%%%%%%%%%%%%%%%%%%%%%%%%%%%%%%%%%%%%%%%%%%%%%%%%%%%%%%%%%%%%%%%%%%
%%%%%%%%%%%%%%%%%%%%%%%%%%%%%%%%%%%%%%%%%%%%%%%%%%%%%%%%%%%%%%%%%%%%%%%%%%%%%%%%%%%%%%%%%

\section{Effect of continuous light irradiation on the electronic and lattice structure of NiO: \emph{in situ} XAS}

\begin{figure}
    \centering
    \includegraphics[width=0.6\linewidth]{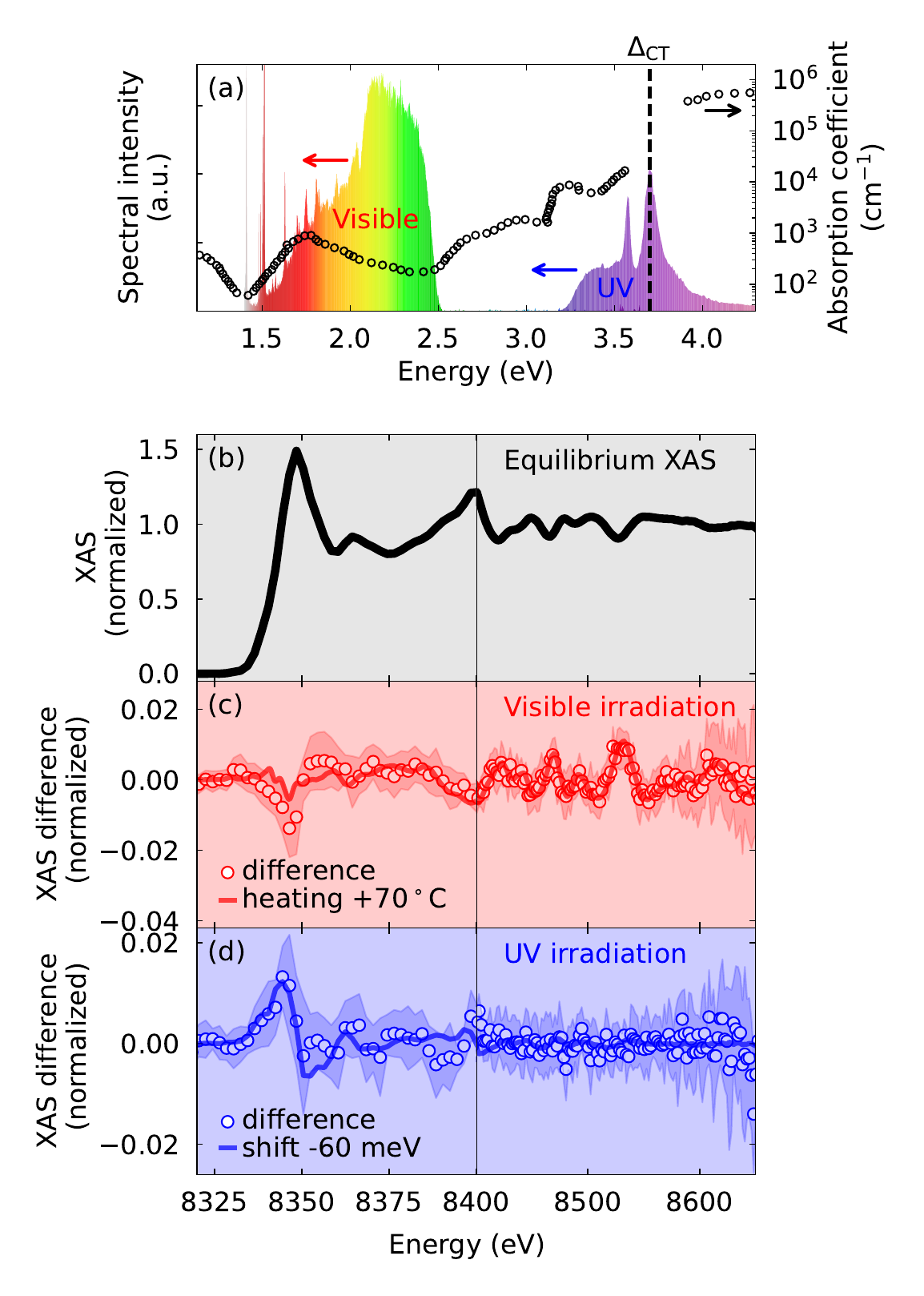}
    \caption{\textbf{Long-lived electronic excited state in NiO under continuous ultraviolet irradiation.} (a) Filtered spectral density of a xenon arc lamp used for the excitation of NiO in the visible and the UV (colored shaded areas, left axis). The absorption coefficient of NiO single crystal is shown for reference (black circles, right axis, adapted from reference \cite{Newman:1959gr}). The position of the charge-transfer gap ($\Delta_{\text{CT}}$) is shown with a vertical dashed line. (b) Normalized equilibrium Ni K-edge XAS spectrum of NiO. (c,d) Difference XAS spectra obtained by subtracting the dark spectrum in (b) from spectra recorded under continuous optical irradiation in the (c) visible and (d) ultraviolet spectral ranges (colored circles). Shaded areas represent the standard deviation between individual measurements. The difference XAS spectra are reproduced by simulations based on (c) a lattice heating of NiO by $+\SI{70\pm5}{\degreeCelsius}$ (red curve) and (d) a red shift of the equilibrium XAS spectrum by \SI{60}{\milli\electronvolt} (blue curve). The energy axis is splitted into two regions: the XANES region below \SI{8400}{\electronvolt} and the EXAFS region above \SI{8400}{\electronvolt}.}
    \label{fig:UV_Vis_irradiation}
\end{figure}

% Description of the XAS measurement at the Ni K-edge of NiO under Visible or UV irradiation from a polychromatic light source
We first investigate the effect of \emph{in situ} continuous optical irradiation on the Ni K-edge XAS spectra of NiO microcrystals (experimental details in SI~\S\ref{secSI:XAS_continuous_irradiation}; sample characterization in SI~\S\ref{secSI:material_characterization}). Figure~\ref{fig:UV_Vis_irradiation}a shows the selected spectral densities from a xenon arc lamp used to excite NiO below (red to green rainbow) or above (purple rainbow) its charge-transfer (CT) gap ($\Delta_{\text{CT}}$). The absorption coefficient of crystalline NiO is also shown on a logarithmic scale (black circles). In the visible part of the excitation spectrum, NiO features several resonances between \SI{1.5}{} and \SI{3.5}{\electronvolt} due to crystal-field multiplet excitations (dipole- and, potentially, spin-forbidden)~\cite{vanVeenendaal:2011hp} and a weak increasing absorption background with photon energy due to band gap transitions from Ni $3d$ to Ni $4s$-derived bands (dipole forbidden)~\cite{Gao2020:229963}. The ultraviolet excitation spectrum instead is in the region of the CT gap around \SI{3.7}{\electronvolt} (dipole allowed)~\cite{Newman:1959gr}, which corresponds to excitations between the O $2p$ and Ni $3d$-derived bands (purple arrow in the density of states of Figure~\ref{fig:experiment_illustration}a)~\cite{Gillen:2013bn}.

% Description of the equilibrium XAS spectrum and the in situ experiment
Figure~\ref{fig:UV_Vis_irradiation}b shows the equilibrium ground-state Ni K-edge XAS spectrum recorded without optical excitation. The spectrum is background subtracted and normalized to the edge jump (see SI~\S\ref{secSI:data_processing} for data-processing details). The effect of xenon lamp excitation on the XAS spectrum is investigated \emph{in situ}, with visible and ultraviolet excitation separately. The results are presented as difference XAS spectra, obtained by subtracting the edge-jump normalized dark spectrum from the corresponding spectrum measured under optical excitation (Figures~\ref{fig:UV_Vis_irradiation}c and~d).

% Effect of the Visible irradiation on the XAS spectrum: lattice heating
Under visible-light irradiation (Figure~\ref{fig:UV_Vis_irradiation}c), the difference XAS spectrum (red circles) shows pronounced oscillations with a maximum amplitude of~$\sim\SI{1}{\percent}$ at energies above~\SI{8400}{\electronvolt} in the extended X-ray absorption fine structure (EXAFS) region. These oscillations have maxima coinciding with the minima of the equilibrium XAS spectrum and vice versa (Figure~\ref{fig:UV_Vis_irradiation}b), which points to lattice heating~\cite{Rossi2021,Rossi2025:87275}. This behavior originates from an increase in the width of the distribution of atomic positions with rising lattice temperature. The resulting enhanced destructive interference of photoelectron scattering paths reduces the amplitude of EXAFS oscillations. This effect is commonly described as an increase in the Debye-Waller factor of the atomic shells surrounding the absorbing atom~\cite{Filez2025:127186,Espinosa2021:224748}. To quantify the degree of lattice heating, the difference XAS spectrum is compared to independent XAS spectra acquired at different temperatures using an external heat source (see SI~\S\ref{subsecSI:XAS_temperature_dependence}). The oscillations in the EXAFS are well reproduced by an effective increase of the lattice temperature by~$\SI{70\pm5}{\degreeCelsius}$, shown as a red curve in Figure~\ref{fig:UV_Vis_irradiation}c. In the X-ray absorption near-edge structure (XANES) region below~$\SI{8400}{\electronvolt}$, lattice heating accounts for the main spectral changes, except for a pronounced negative feature of amplitude~$\sim\SI{1.5}{\percent}$ at~$\SI{8350}{\electronvolt}$ and accompanying positive features on both sides. Lattice heating is likely due to the non-radiative decay of crystal-field multiplet excitations~\cite{Sachs2025:99143}. However, the discrepancy between the measured XANES difference spectrum and the lattice-heating simulation indicates an additional non-thermal response. One possible explanation is a change in the nickel oxidation state accompanied by an absorption-edge shift, as previously observed during the photocatalytic cycle of Ni complexes~\cite{Phelan2021:284520}. This explanation is unlikely, however, because the visible excitation energy lies below the charge-transfer (CT) gap. Oxygen-to-nickel charge transfer, and thus a change in the nickel oxidation state, is therefore not expected. Consistent with this interpretation, the difference XAS spectrum in Figure~\ref{fig:UV_Vis_irradiation}c does not exhibit the first-derivative lineshape near the absorption edge that would indicate an edge shift.
Instead, the excitation may populate antibonding Ni~$3d$ orbitals through crystal-field multiplet transitions. The resulting bond elongation should reduce the EXAFS oscillation frequency, but no such shift is observed, possibly because it is masked by the dominant lattice-heating contribution. Notably, the XANES difference spectrum contains a negative feature approximately \(\SI{4}{\electronvolt}\) below the absorption maximum, flanked by positive features, resembling spectral broadening. This response may arise from the simultaneous excitation of several crystal-field multiplets. The resulting incoherent local distortions could be indistinguishable from lattice heating in the EXAFS. In the XANES region, however, the combined electronic and structural changes at the Ni sites may produce a distinct response. In particular, the bond elongation and possible symmetry breaking induced by crystal-field multiplet excitations could account for the observed deviation from the lattice-heating spectrum~\cite{Liu2022:174651}.

% Effect of the UV irradiation on the XAS spectrum: energy renormalization
In contrast, under ultraviolet-light excitation with photon energies exceeding the NiO CT gap, no pronounced oscillations are observed in the EXAFS region of the difference XAS spectrum above \SI{8400}{\electronvolt} (blue circles in Figure~\ref{fig:UV_Vis_irradiation}d). This indicates that lattice heating is negligible under these conditions, despite the higher photon excitation energy. A first-derivative lineshape is observed near the absorption edge at~$\sim\SI{8350}{\electronvolt}$, which is well reproduced by a simulated red shift of the equilibrium XAS spectrum by~\SI{60}{\milli\electronvolt} (blue curve). The absence of EXAFS signatures of lattice heating, together with the presence of a pronounced near-edge spectral shift in the XANES, indicate that the observed response is predominantly electronic in origin. The lack of measurable lattice heating is likely a combination of: i) the limited dissipation of photoexcited carrier excess energy in correlated $3d$-bands~\cite{Kunes:2007dv}; ii) the small amount of excess energy in the excitation spectrum above the CT gap ($\sim\SI{0.1}{\electronvolt}$ for the most intense part of the spectral density, see Figure~\ref{fig:UV_Vis_irradiation}a); and iii) a large fraction of radiative carrier recombination~\cite{Gandhi2017:46928}. In band insulators, the red shift of the absorption edge would be assigned to a reduction of the Ni oxidation state. However, this would be combined with large structural changes due to the change of the local electronic density at the Ni sites~\cite{Phelan2021:284520}, which is not observed in the EXAFS region. Instead, NiO has a band structure strongly influenced by the strength of electron-electron repulsion in Ni~$3d$ orbitals (Figure~\ref{fig:experiment_illustration}a), meaning that the observed spectral shift likely originates from many-body interactions affecting the energy states, for instance through enhanced Coulomb screening by photoexcited carriers~\cite{Golez2019}.

% Comparison of the visible and UV excitation effects
Thus, the measurement of \emph{in situ} XAS spectra at the Ni K-edge of NiO shows that while visible excitation primarily induces thermal changes, UV excitation drives a non-thermal response characterized by an electronic redshift of the spectrum near the absorption edge. The observation of this shift under continuous irradiation demonstrates its persistent nature, which is uncommon in photoexcited materials with strong electron-electron repulsion, typically relaxing rapidly back to equilibrium through efficient energy-dissipation pathways enabled by strong coupling among electronic, lattice, and spin degrees of freedom~\cite{Giannetti:2016hp}. Our results show that a sustained modulation of the electronic structure can nevertheless be achieved in NiO under continuous UV irradiation.

% Interpretation of the red shift and comparison to the recent literature
The red shift confined to the near-edge region suggests that it arises from a light-induced renormalization of bound and weakly bound unoccupied electronic states. This interpretation is consistent with recent time-resolved XAS measurements performed at the Ni M$_{2,3}$~\cite{Cazali2025:166775} and Ni L$_{2,3}$~\cite{Lojewski2023} edges of NiO, where similar spectral red shifts were observed and attributed to a photoinduced screening of electron-electron repulsion, leading to a renormalization of the energy of the Ni $3d$ conduction band~\cite{Golez2019}. However, XAS spectra at the Ni K-edge are mostly sensitive to the unoccupied $p$ density of states, which means that a red shift of the spectrum cannot be directly linked to an energy change of the unoccupied Hubbard states near the NiO CT gap and is not analogous to previously reported red shifts~\cite{Cazali2025:166775,Lojewski2023}. Additionally, this effect is achieved at excitation densities several orders of magnitude lower than in previous pump-probe studies ($\sim\SI{1e13}{\per\cubic\centi\metre}$ in this work, see SI \S\ref{secSI:excitation_density}). Alongside its persistence, it raises the question of the nature and the microscopic mechanism responsible for stabilizing such an electronically excited state, which we address in the following section.

\section{Effect of pulsed excitation on the electronic and lattice structure of NiO: pump-probe XAS}

% There should probably be a Figure illustrating the setup somewhere here.

\begin{figure}[!ht]
    \centering
    \includegraphics[width=\linewidth]{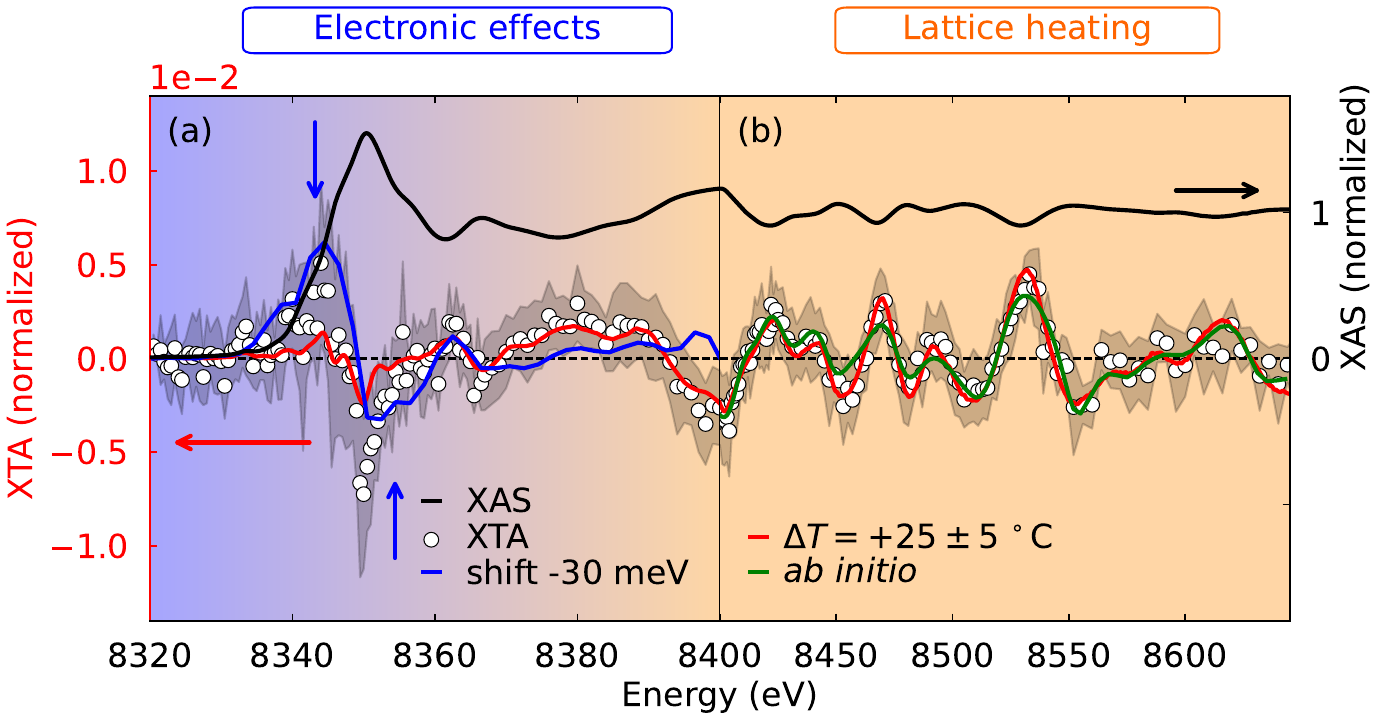}
    \caption{\textbf{Overlapping electronic and heating effects in the photoexcited state of NiO under pulsed laser excitation.} X-ray transient absorption (XTA) spectra at a pump-probe time delay of \SI{100}{\ps} (white circles) in the (a) XANES, and (b) EXAFS, normalized to the absorption edge jump of the equilibrium Ni K-edge XAS spectrum (black curve). The excitation fluence is \SI{24}{\milli\joule\per\centi\metre\squared} at a pump photon energy of \SI{3.49}{\electronvolt}. Shaded areas represent the standard deviation across individual measurements. Simulated XTA spectra in the EXAFS region are shown based on experimental temperature-dependent XAS (T-XAS) measurements upon lattice heating of $+$\SI{25\pm5}{\degreeCelsius} (continuous red curve), on \emph{ab initio} calculations incorporating both lattice expansion and increase in Debye--Waller factor (green curve), and upon a red shift of the XAS spectrum by \SI{30}{\milli\eV} in the excited state (blue curve). Blue and orange shaded background regions qualitatively indicate spectral ranges where electronic and lattice-heating contributions, respectively, dominate the XTA spectrum (see main text for details).}
    \label{fig:XTA}
\end{figure}

% Global look at the time-resolved XAS spectrum, separate analysis for the XANES and the EXAFS
We now investigate the picosecond response of NiO to pulsed ultraviolet (UV) excitation using pump-probe X-ray transient absorption (XTA) spectroscopy at the Ni K-edge. The pump photon energy is set near the CT gap at \(h\nu_{\text{pump}}=\SI{3.49}{\eV}\). The NiO crystals are suspended in a liquid water jet. Further details of the experimental setup are provided in SI~\S\ref{secSI:XTA_setup}. Figure~\ref{fig:XTA} shows the XTA spectrum measured at a pump–probe delay of \SI{100}{\ps} (white circles). The XTA spectrum is defined as the difference between the XAS spectra recorded with and without pump excitation. The transient responses in the XANES and EXAFS regions are analyzed separately in panels~(a) and~(b), respectively.

% Look at the EXAFS in particular: lattice heating
We first examine the EXAFS region (Figure~\ref{fig:XTA}b), which is sensitive to structural modifications in the excited state. The transient spectrum (white circles) displays oscillations whose maxima coincide with the minima of the equilibrium XAS spectrum (black curve), and vice versa, and closely resembles the difference spectrum under continuous visible-light irradiation (Figure~\ref{fig:UV_Vis_irradiation}c). This behavior indicates that lattice heating dominates the structural response at a delay of~\SI{100}{\ps} under UV pulsed excitation. At this time delay, the lattice and carrier temperatures are considered thermalized and in quasi-equilibrium at an elevated temperature, similarly to other photoexcited oxides with strong electron-electron repulsions~\cite{Konstantinova2018:191174}, with the heat transfer to the solvent taking place on timescales of hundreds of nanoseconds (\emph{vide infra}).

% Quantification of lattice heating
Although photoexcitation at the onset of the broad CT gap does not impart significant excess energy to the carriers, substantial lattice heating is observed at~\SI{100}{\ps}. This heating arises from the high density of photoexcited carriers ($\sim\SI{1.9e20}{\per\cubic\centi\metre}$), a fraction of which relaxes into crystal-field multiplets and subsequently undergoes non-radiative recombination. The excitation density is about 7 orders of magnitude higher than that achieved under the continuous UV excitation conditions shown in Figure~\ref{fig:UV_Vis_irradiation}d (see SI~\S\ref{secSI:excitation_density} for details of the calculation). The oscillatory features observed in the EXAFS region of the XTA spectrum are well captured by an optimization procedure that minimizes the difference between the XTA spectrum and independent temperature-dependent XAS (T-XAS) spectra (see SI~\S\ref{subsecSI:simulation_temperature_contribution_XTA} for a description of the optimization procedure). The best agreement is obtained for a temperature increase of~$\Delta\text{T}=+\SI{25\pm5}{\degreeCelsius}$ (red curve in Figure~\ref{fig:XTA}b), in good agreement with the maximum estimated temperature rise~$\Delta\text{T}=\SI{24}{\degreeCelsius}$ (SI~\S\ref{subsecSI:simulation_temperature_jump}). To further quantify the structural response associated with lattice heating, \emph{ab initio} calculations are performed to model the EXAFS signal (green curve in Figure~\ref{fig:XTA}b). These simulations show that the transient EXAFS response arises from a combination of an increased Debye--Waller factor, $\Delta\langle u^2\rangle=\SI{0.4\pm0.1}{\milli\angstrom\squared}$ (a~\SI{6}{\percent} increase \footnote{corresponding to~$\Delta B=8\pi^2\Delta\langle u^2\rangle=\SI{33\pm8}{\milli\angstrom\squared}$}), and a slight lattice expansion of~$\Delta\text{a}=\SI{1.7\pm0.1}{\milli\angstrom}$ (a relative change of~$\sim\SI{5e-4}{}$ from equilibrium; see SI~\S\ref{subsecSI:ab_initio_heat_calculations} for calculation details). Both the Debye--Waller factor increase and the lattice expansion are consistent with a lattice temperature rise of approximately~\SI{30}{}--\SI{40}{\degreeCelsius}, as inferred from T-XAS measurements (SI \S\ref{subsecSI:XAS_temperature_dependence}). The excellent agreement between the XTA spectrum in the EXAFS region, the \emph{ab initio} calculations, and the independent T-XAS measurements supports a single consistent interpretation: the structural response of photoexcited NiO with a UV light pulse at~\SI{100}{\ps} after excitation is dominated by lattice heating. This conclusion is in line with previous femtosecond electron diffraction~\cite{Windsor2021} and optical~\cite{Rossi2025:52718} studies, which report that lattice heating becomes the dominant contribution to the structural response within~$\sim\SI{20}{\pico\second}$ and at time delays exceeding~$\SI{100}{\pico\second}$, respectively.

% Look at the effect of lattice heating in the XANES
In contrast to the EXAFS region, the XTA spectrum in the XANES displays changes reflecting simultaneous dynamics in the lattice and the electronic structure (Figure~\ref{fig:XTA}a). The transient spectrum cannot be reproduced by simple spectral shifts or broadenings (see SI~\S\ref{secSI:simulation_shift_broadening}). The simulated contribution from lattice heating, using the same temperature increase extracted from the EXAFS analysis ($\Delta\text{T}=+\SI{25}{\degreeCelsius}$), fails to reproduce the experimental XTA spectrum in the~\SI{8330}{}--\SI{8360}{\eV} energy range. In particular, the measured XTA spectrum displays a pronounced positive signal on the rising edge and an enhanced negative feature near~$\sim\SI{8350}{\electronvolt}$ (marked with vertical blue arrows in Figure~\ref{fig:XTA}a), both of which are absent or underestimated in the simulation of a hot lattice. These spectral features are therefore attributed to a \emph{non-thermal} contribution and closely resemble the wavelet-like lineshape observed under continuous UV irradiation (Figure~\ref{fig:UV_Vis_irradiation}d), suggesting an electronic contribution in the transient XANES signal. A simulated XTA spectrum upon a red shift of the XAS spectrum by $\sim\SI{30}{\milli\electronvolt}$ reproduces the amplitude of the prominent positive signal in the rising edge (blue line in Figure~\ref{fig:XTA}a). The positive component of this wavelet persists for at least~\SI{10}{\nano\second} and decays on a timescale comparable to that of the prominent negative feature at~$\sim\SI{8350}{\electronvolt}$ (see SI~\S\ref{secSI:additional_XTA_spectra} Figure~\ref{figSI:XTA_spectra_time_delays}). A multi-exponential fit of a time trace at~\SI{8350}{\electronvolt} reveals two distinct decay components with time constants $\tau_1=\SI{600\pm200}{\pico\second}$ and $\tau_2=\SI{30\pm10}{\nano\second}$ with relative weights of~\SI{21}{} and~\SI{79}{\percent}, respectively (see SI~\S\ref{subsecSI:kinetics}). The shorter time constant, $\tau_1$, is similar to previously reported luminescence lifetimes in pristine NiO, which have been assigned to carrier recombination and is therefore consistent with the charge-carrier lifetime~\cite{Imran2021}. The longer time constant, $\tau_2$, is attributed to heat transfer from NiO to the solvent, as supported by simulations (see SI~\S\ref{subsecSI:heat_transfer_solvent}). Together, these observations demonstrate that the transient XANES response arises from a superposition of thermal and non-thermal (electronic) contributions with distinct temporal dynamics, which jointly shape the excited-state XAS spectrum. 

\begin{figure}[!ht]
    \centering
    \includegraphics[width=0.7\linewidth]{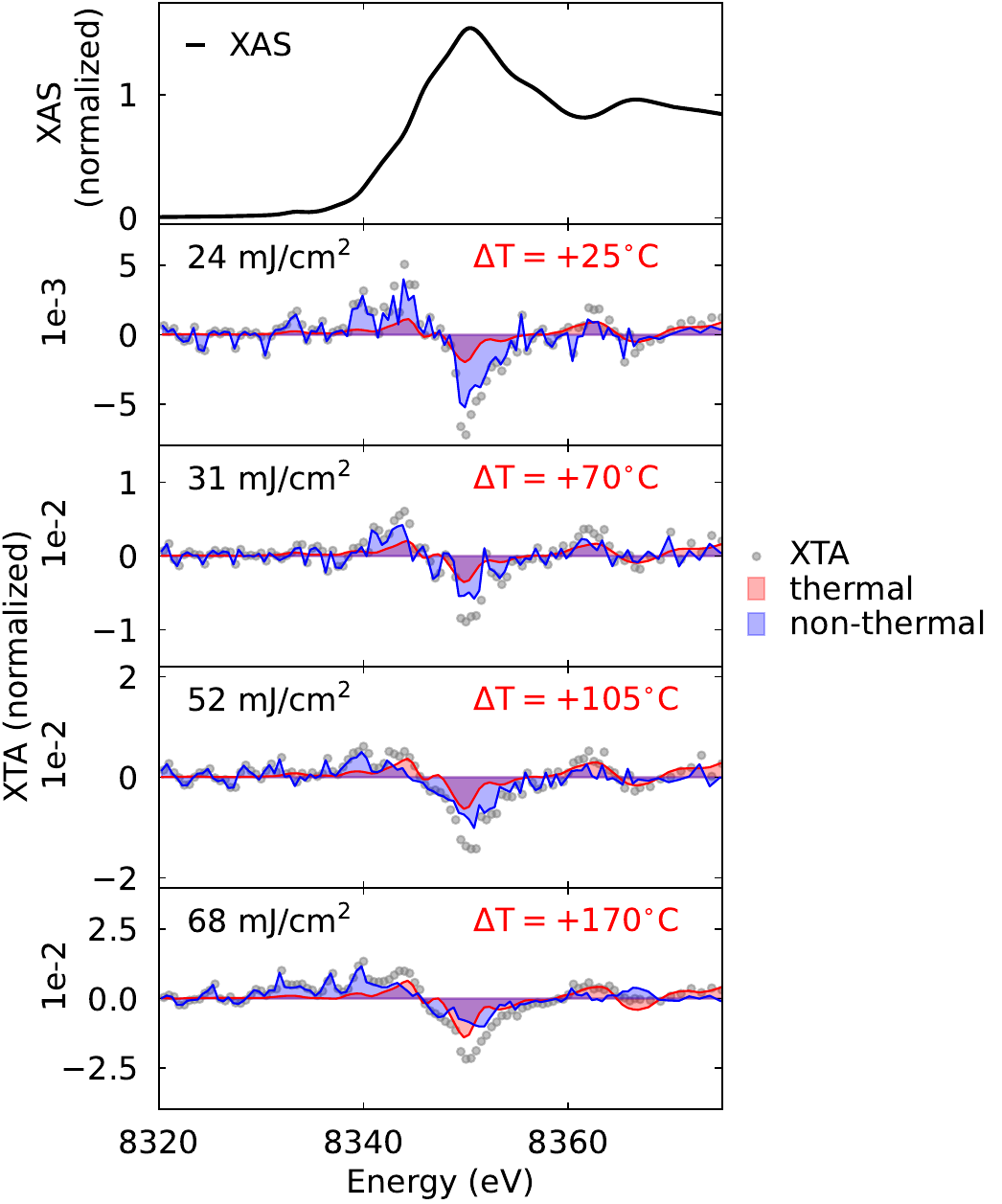}
    \caption{\textbf{Spectral signature of a non-thermal state in photoexcited NiO near the absorption edge.} Evolution of XTA spectra with the excitation fluence (grey circles). The equilibrium XAS spectrum is shown in black for reference (top panel). The XTA spectrum is decomposed into thermal (red shaded curves) and non-thermal (blue shaded curves) contributions based on the magnitude of the purely thermal contribution in the EXAFS region (lattice heating $\Delta\text{T}$ indicated in red, see main text).}
    \label{fig:non_thermal_XTA_evolution_fluence}
\end{figure}

% Principles of separation of thermal and non-thermal effects
To disentangle thermal and non-thermal contributions in the XTA spectrum, we employ a previously established analysis procedure in which the XTA spectrum is expressed as a linear combination of thermal and non-thermal components~\cite{Rossi2021,Rossi2025:87275}. At K-edges, this assumption is well justified because the core-hole lifetime is extremely short~\cite{Sipr2018:39343}, so that detected X-ray photons from the core-hole recombination effectively provide a snapshot of the electronic structure within a thermally disordered lattice. Electronic and lattice contributions to the XAS spectrum can therefore be treated as largely uncorrelated. 

% Applying the separation procedure
Applying this procedure, we find that the XTA spectrum in the EXAFS region is purely thermal (see SI~\S\ref{subsecSI:simulation_temperature_contribution_XTA}) and extract the \emph{non-thermal} component of the XTA spectrum at various pump fluences in the XANES, which are reported in Figure~\ref{fig:non_thermal_XTA_evolution_fluence} (blue shaded curves). The non-thermal signal is localized near the absorption edge and exhibits a characteristic wavelet-like lineshape, consisting of a positive feature in the rising edge ($\sim\SI{8340}{\electronvolt}$), and a negative feature at the main peak in the XAS spectrum ($\sim\SI{8350}{\electronvolt}$). This lineshape closely resembles that observed under continuous UV irradiation (Figure~\ref{fig:UV_Vis_irradiation}d), suggesting that the non-thermal response originates from a red shift of the absorption edge in the excited state. The non-thermal signal is assigned the lifetime $\tau_1$ ($\sim\SI{600}{\pico\second}$) based on the previous assignment that electronic effects must be represented by the shortest of the two observed decay times and that $\tau_1$ is similar to previously reported carrier lifetimes~\cite{Imran2021}. The amplitude of the non-thermal signal is between \SI{0.5}{\percent} and \SI{1}{\percent} of the absorption edge jump, which is less or comparable to the magnitude of the difference signal observed under continuous UV irradiation (Figure~\ref{fig:UV_Vis_irradiation}d), despite the excitation densities being 7 to 8 orders of magnitude higher with UV pulses (SI \S\ref{secSI:excitation_density} for detailed estimates). This observation points to the fact that the simultaneous presence of lattice heating and electronic effects reduces the magnitude of the non-thermal contribution for a given excitation density and underscores the need to mitigate lattice heating in photoexcited materials for enhanced electronic response. In the following, we perform simulations and \emph{ab initio} calculations to reproduce these non-thermal spectral features and elucidate their microscopic origin.

%%%%%%%%%%%%%%%%%%%%%%%%%%%%%%%%%%%%%%%%%%%%%%%%%%%%%%%%%%%%%%%%%%%%%%%%%%%
%%%%%%%%%%%%%%%%%%%%%%%%%%%%%%%%%%%%%%%%%%%%%%%%%%%%%%%%%%%%%%%%%%%%%%%%%%%
%%%%%%%%%%%%%%%%%%%%%%%%%%%%%%%%%%%%%%%%%%%%%%%%%%%%%%%%%%%%%%%%%%%%%%%%%%%
%%%%%%%%%%%%%%%%%%%%%%%%%%%%%%%%%%%%%%%%%%%%%%%%%%%%%%%%%%%%%%%%%%%%%%%%%%%

\section{Excited-state electronic effects: energy shift and broadening}

% Simulation of the excited-state XAS spectrum as a combination of an energy shift and a broadening of the equilibrium XAS spectrum

The XTA spectrum is obtained by subtracting the equilibrium XAS spectrum from the photoexcited XAS spectrum. We find that applying a rigid redshift and spectral broadening to the equilibrium spectrum to simulate the photoexcited state provides the best qualitative agreement with the observed non-thermal XTA response (SI~\S\ref{secSI:simulation_shift_broadening}, Figure~\ref{figSI:shift_broadening_XAS}). This finding is consistent with a previous dynamical mean-field theory (DMFT) study by Gole\u{z} and coworkers, who reported spectral shifts and broadening in the XAS spectra of photoexcited correlated materials~\cite{Golez2024:227781}. In that work, the energy shift arises from enhanced electronic screening in photoexcited NiO, which reduces the electron–electron repulsion, predominantly within the Ni $3d$ orbitals. This effect can be modeled as a renormalization of the effective Hubbard interaction $U$. In addition to this gap renormalization, DMFT predicts a concomitant broadening of the density of states due to coupling to low-energy plasmon modes~\cite{Golez2019a}. Similar energy shifts and broadenings have been reported in previous ultrafast XAS measurements of NiO~\cite{Lojewski2023}. The red shifts obtained here are between \SI{8}{} and \SI{17}{\milli\electronvolt}, consistent in magnitude with previously reported renormalizations of the CT gap~\cite{Lojewski2023,Rossi2025:52718}. However, we obtain several times larger broadenings of \SIrange{600}{800}{\milli\eV}, suggesting that this simple model does not accurately describe the photoexcited electronic state. In particular, the different electronic states probed at the Ni K-edge may experience shifts and broadenings of different magnitudes. This interpretation is consistent with the poor agreement previously obtained at the O K-edge of NiO using the same approach~\cite{Lojewski2023}. Nevertheless, the overall lineshape of the non-thermal XTA spectrum remains compatible with a red shift and broadening of the photoexcited-state XAS spectrum. This behavior points to dynamical screening of the electronic states in photoexcited NiO with no comparable effect previously observed in band insulators such as TiO$_2$ and ZnO. We therefore turn to \emph{ab initio} calculations to reproduce the non-thermal XTA spectrum and elucidate the microscopic origin of its first-derivative lineshape.

%%%%%%%%%%%%%%%%%%%%%%%%%%%%%%%%%%%%%%%%%%%%%%%%%%%%%%%%%%%%%%%%%%%%%%%%%%%
%%%%%%%%%%%%%%%%%%%%%%%%%%%%%%%%%%%%%%%%%%%%%%%%%%%%%%%%%%%%%%%%%%%%%%%%%%%
%%%%%%%%%%%%%%%%%%%%%%%%%%%%%%%%%%%%%%%%%%%%%%%%%%%%%%%%%%%%%%%%%%%%%%%%%%%
%%%%%%%%%%%%%%%%%%%%%%%%%%%%%%%%%%%%%%%%%%%%%%%%%%%%%%%%%%%%%%%%%%%%%%%%%%%

\section{Enhanced orbital hybridization in photoexcited NiO}

% Description of the calculations to simulate a renormalization of the Hubbard U and the screening of the core-hole interaction
In this section, we examine the electronic mechanisms responsible for the red shifts observed in the non-thermal XTA spectrum at the Ni K-edge of photoexcited NiO. These spectral shifts must arise without detectable structural changes, as the preceding analysis shows that no transient EXAFS signal is resolved once thermal contributions are removed (SI~\S\ref{subsecSI:simulation_temperature_contribution_XTA}). The absence of non-thermal structural dynamics suggests that polaron formation~\cite{Biswas:2018ec,Biswas:2018fv} and charge trapping~\cite{Janotti:2010cx,RittmannFrank:2014fu,Penfold:2018ie}, which are typically associated with local structural relaxations, are unlikely within the sensitivity of the present experiment. Accordingly, the non-thermal XTA signal observed here must be dominated by electronic effects.

% Full theoretical treatment of the effects currently not possible, description of the present approach
A comprehensive theoretical treatment of the several effects contributing to the photoexcited XAS spectrum of NiO would require DMFT combined with non-local many-body perturbative approaches to capture self-energy changes, excitonic interactions and dynamical spectral broadening (see SI~\S\ref{secSI:discussion_excited_XAS_contributions} for a discussion about the different contributions). At present, such calculations remain computationally demanding and typically rely on substantial simplifications of the electronic structure~\cite{Lojewski2023}. While such approaches are beyond the scope of the present work, they represent an important direction for achieving improved quantitative agreement between theory and experiments in future studies. Here, we instead use density functional theory (DFT) with an effective treatment of on-site electron correlations (Hubbard U) and inter-site Coulomb interactions (Hubbard V) to assess the influence of core-hole screening~\cite{Liu2021,Rossi2025:87275}, correlation renormalization~\cite{Tancogne-Dejean2020}, and orbital hybridization change~\cite{Cazali2025:166775} on the Ni K-edge XAS spectrum of photoexcited NiO. Computational details are provided in SI~\S\ref{secSI:ab_initio_calculations}. 

% Short description of the results of the effect of core-exciton screening and hybridization
We find that calculations incorporating an effective core-exciton screening and a reduction of the intersite screening between Ni $3d$ and O $2p$ states~\cite{Cazali2025:166775} lead to a blue shift of the Ni K-edge XAS spectrum (see SI~\S\ref{subsecSI:screening_core_potential} and~\S\ref{subsecSI:screening_Hubbard_V}). This behavior is inconsistent with the experimentally observed red shift. We therefore exclude both core-exciton screening and reduced Ni--O hybridization as the dominant mechanisms responsible for the red shift of the absorption edge in the non-thermal XTA spectra (Figure~\ref{fig:non_thermal_XTA_evolution_fluence}). 

% Description of the results from the renormalization of the Hubbard U
The screening of the Hubbard interaction $U$ by photoexcited carriers is modeled by computing the NiO XAS spectrum using a reduced effective Hubbard $U$. The corresponding simulated XTA spectra are obtained by taking the difference between XAS spectra calculated with the reduced and equilibrium values of $U$. The equilibrium value of the Hubbard $U$ is \SI{6.7}{\eV} with our set of pseudopotentials and functionals, determined by density functional perturbation theory (DFPT)~\cite{Timrov2022:301065}. Calculated XAS spectra for different values of $U$ show that a reduction of $U$ induces small red shifts of the absorption edge, with unphysical reductions of $U$ to match the magnitude of the red shift observed experimentally (SI \S\ref{subsecSI:screening_Hubbard_U}). Calculations were performed for reductions of the Hubbard interactions ranging from~\SI{100}{\milli\eV} to \SI{1}{\eV}, yielding red shifts between approximately a few meV to $\SI{40}{\milli\eV}$, i.e., more than two orders of magnitude smaller than the imposed reduction of $U$. This reduced sensitivity of the absorption-edge position with respect to variations of $U$ is expected because the unoccupied density of states at and above the Ni K-edge is dominated by $p$ states (see SI~\S\ref{subsubsecSI:XSpectra} Figure~\ref{figSI:pDOS_NiO}). These states are only indirectly affected by changes in electronic correlations through hybridization with Ni $3d$ orbitals. It points to the fact that a screening of electron correlations may be present in photoexcited NiO, as previously demonstrated~\cite{Lojewski2023,Cazali2025:166775,Rossi2025:52718}, but that another mechanism possibly intertwined with the dynamic screening of the Hubbard $U$ is likely responsible for the red shift of the XAS spectrum in the excited state. 

\begin{figure}
    \centering
    \includegraphics[width=0.7\linewidth]{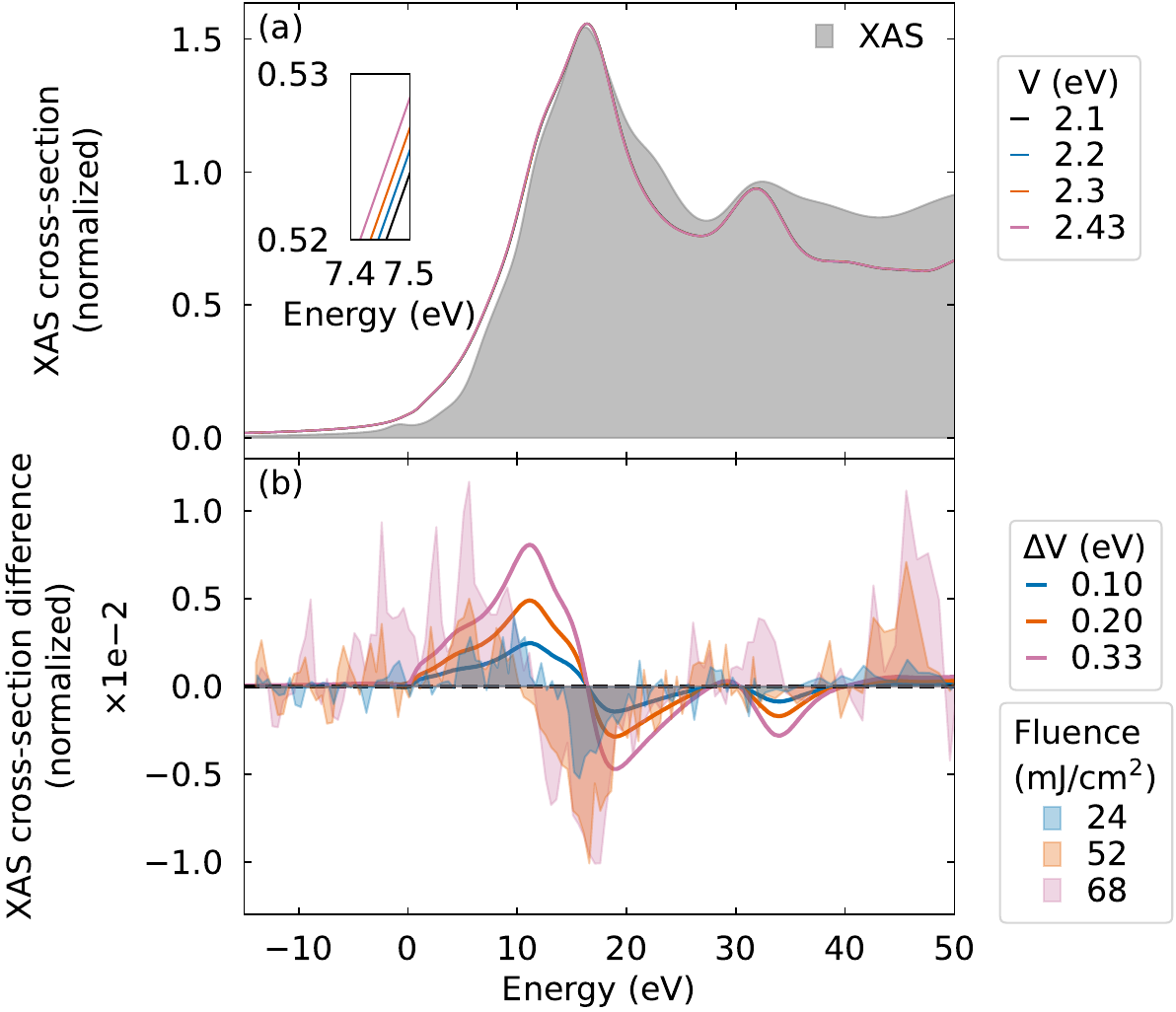}
    \caption{\textbf{Red shift of the XAS spectrum in the excited state interpreted as increase in intersite screening due to increased orbital hybridization.} Simulated (a) XAS and (b) XTA spectra upon increase of the intesite $V$ screening parameter between Ni $3d$ and O $2p$ orbitals using DFT+U+V (colored curves). The energy axis is relative to the Fermi level. The experimental XAS spectrum is shown for reference (grey shaded curve), red shifted by \SI{8334.2}{\eV} to match the energy scale of the calculation. Experimental non-thermal XTA spectra are shown with colored shaded areas for different excitation fluences. The energy alignment procedure between calculated spectra for different values of $V$ is detailed in SI~\S\ref{secSI:ab_initio_calculations}.}
    \label{fig:non_thermal_XTA_XANES_DFT}
\end{figure}

% Description of the results from the increase in the orbital hybridization
The Hubbard $V$ is used to model the screened interaction between electrons in orbitals on neighboring atomic sites. Although it is a Coulomb term, its role is to change the electronic ground state by altering the amount of mixing between orbitals, directly coupling to the degree of hybridization between neighboring localized orbitals~\cite{Campo2010:45010}. Calculations of XAS spectra upon increase of the $V$ parameter produce a more pronounced red shift of the XAS spectrum than with the Hubbard $U$ (Figure~\ref{fig:non_thermal_XTA_XANES_DFT}a), of the order of $\sim\SI{10}{\milli\eV}$ per \SI{100}{\milli\eV} change in $V$\footnote{The absence of the feature at \SI{20}{\eV} in the calculation is because it originates from a shake-up excitation, which cannot be modeled with the current level of theory~\cite{Calandra:2012kw}.}. The computed XTA spectrum upon this red shift is displayed with continuous lines together with non-thermal XTA spectra (shared area) in Figure~\ref{fig:non_thermal_XTA_XANES_DFT}b at three different excitation fluences. We note that there is a mismatch between the zero-crossing point of the signal between theory and experiment, which is interpreted as the absence of broadening of the excited-state spectrum in the simulation, which provided better agreement with transient spectra than pure energy shifts (SI \S\ref{secSI:simulation_shift_broadening}). At this theory level, it shows that the most likely explanation for the red shift of the XAS spectrum observed in the experiment is the increase in the $V$ parameter. A more quantitative treatment of photoexcited carriers and their effect on the screening of electron correlations and intersite screening would require the inclusion of dynamic correlations~\cite{Ren:2006fn,Miura:2008kt,Mocatti2026:192118}.

% Conclusion from the simulations
We therefore conclude that an increased Ni $3d$-O $2p$ orbital hybridization, modeled in the calculations by a change in the intersite $V$, provides a plausible explanation for the observed non-thermal red shift of the Ni K-edge in photoexcited NiO. This electronically excited state is likely connected to the screening of on-site electronic correlations reported previously~\cite{Lojewski2023,Rossi2025:52718,Cazali2025:166775}. Remarkably, this red shift is assigned the time constant $\tau_1=\SI{600\pm200}{\pico\second}$, indicating the formation of a long-lived, non-thermal metastable state characterized by these electronic changes.

%%%%%%%%%%%%%%%%%%%%%%%%%%%%%%%%%%%%%%%%%%%%%%%%%%%%%%%%%%%%%%%%%%%%%%%%%%%
%%%%%%%%%%%%%%%%%%%%%%%%%%%%%%%%%%%%%%%%%%%%%%%%%%%%%%%%%%%%%%%%%%%%%%%%%%%
%%%%%%%%%%%%%%%%%%%%%%%%%%%%%%%%%%%%%%%%%%%%%%%%%%%%%%%%%%%%%%%%%%%%%%%%%%%
%%%%%%%%%%%%%%%%%%%%%%%%%%%%%%%%%%%%%%%%%%%%%%%%%%%%%%%%%%%%%%%%%%%%%%%%%%%

\section{Discussion}

% Applications of the renormalization of electron correlations and dynamic screening in catalysis
Dynamic screening of electron–electron interactions can enhance superconductivity~\cite{Liu2021:149522} and drive photoinduced insulator-to-metal transitions~\cite{Werner2010:67889}. In molecular complexes, photoinduced changes in orbital hybridization alter the populations of bonding and antibonding orbitals~\cite{VanKuiken2016:67121}, enabling the activation of chemical reactions~\cite{Jay2023}. In solids, however, the potential of correlation screening and hybridization changes to improve carrier transport and catalytic efficiency remains largely unexplored. Growing evidence indicates that electronic correlations provide a powerful means of controlling catalytic functionality in correlated oxides. In many of these materials, weaker electronic correlations increase the covalency between metal centers and oxygen ligands. The resulting enhancement of charge-transfer character modifies the redox properties of the metal sites, which play a central role in catalytic performance. For example, chemical doping of NiO with different metals was recently shown to alter orbital hybridization and carrier mobility, thereby promoting charge delocalization and more favorable dynamics for the oxygen evolution reaction~\cite{Ostrom2025:101832}. Similarly, doping-induced reductions in electronic correlations have been linked to improved activity in the oxygen evolution and denitration reactions in iridates~\cite{Shang2019:32248} and vanadates~\cite{Suh2024:251340}, respectively. In perovskite oxides, correlation screening can reduce the charge-transfer (CT) gap and enhance catalytic efficiency by reshaping the energetic landscape of electron-transfer processes~\cite{Hong2017:102783,Miao2025:285895}. Building on evidence that photoexcitation screens electronic correlations in NiO on ultrafast timescales, the present work identifies a likely consequence: a change in Ni–O orbital hybridization without structural rearrangement. These results demonstrate that above-gap photoexcitation can control fundamental aspects of the electronic structure of correlated materials beyond simply generating charge carriers.

At present, establishing a direct connection electronic correlations or orbital hybridization to surface catalytic activity remains challenging, primarily because catalysis happens near material surfaces. Recent theoretical work on NiO has provided important insights in this direction, showing that different surface terminations can exhibit substantial gap reductions, along with changes in the nature of the insulating state in surface and subsurface layers driven by purely electronic effects~\cite{Leonov2021}. These findings highlight the crucial role of local electronic structure in determining surface reactivity, which motivates future studies to establish the extent to which orbital hybridization increase can be applied to catalytic applications. Probing such effects experimentally will require surface-sensitive and time-resolved techniques, such as time-resolved photoelectron spectroscopy, which can directly capture the evolution of electronic correlations under nonequilibrium conditions. Building on these considerations, we propose that UV light irradiation in NiO can serve as an effective external tuning parameter for electronic screening and enhanced performance. By dynamically tuning the electronic structure and the interaction between orbitals, it enables control over the energy and bandwidth of states near the Fermi level (Figure~\ref{fig:experiment_illustration}b) alongside improved carrier transport in the bulk (Figure~\ref{fig:experiment_illustration}c) which, together, lead to improved photocatalytic performance in correlated oxides (Figure~\ref{fig:experiment_illustration}d) using light excitation.

%Simultaneous contributions of lattice heating and renormalization of the correlation gap
The current work also shows that orbital hybridization changes can coexist with lattice heating. At a time delay of approximately \SI{100}{\pico\second} in the pump-probe measurements, the data reveal the coexistence of electronic changes and thermally induced structural changes in photoexcited NiO. The dual sensitivity of transient X-ray absorption spectroscopy allows us to disentangle these contributions. It highlights that both lattice heating and many-body electron interactions could potentially lead to a red shift of the XAS spectrum. By independently determining the lattice temperature from the EXAFS signal, the purely thermal contribution can be subtracted from the response in the XANES, including its potential effect on the unoccupied density of states near the absorption edge. Thermal lattice expansion is expected to reduce the magnitude of the Hubbard $U$~\cite{Campo2010:45010}. Using the temperature-dependent lattice parameters obtained from X-ray diffraction (SI~\S\ref{subsecSI:XAS_temperature_dependence}), we calculated the effect of thermal expansion on the Hubbard parameter $U$ using DFPT. Over the range of lattice parameters corresponding to temperatures between room temperature and \SI{150}{\degreeCelsius}, $U$ decreases linearly with increasing lattice parameter, with a slope of approximately \SI{-2}{\electronvolt\per\angstrom}. At an excitation fluence of \SI{24}{\milli\joule\per\centi\meter\squared}, the corresponding thermal expansion yields a calculated decrease in $U$ of \SI{4}{\milli\electronvolt}. This decrease is approximately half the magnitude of the red shift obtained by applying the shift-and-broadening model to simulate the non-thermal XTA spectrum (SI Figure~\ref{figSI:shift_broadening_XAS}b).
These results indicate that the thermal and purely electronic modifications of the energy levels induced by intense UV pulses are comparable in magnitude. Mitigating thermal effects may therefore enable stronger electronic modifications and potentially extend the lifetime of the associated excited state, as illustrated by the \emph{in situ} experiment under continuous irradiation (Figure~\ref{fig:UV_Vis_irradiation}d).

\section*{Conclusion}

%The Hubbard V models the inter-site Coulomb interaction between neighboring localized orbitals

This work reveals a long-lived metastable electronic state in photoexcited NiO, identified using \emph{in situ} XAS and X-ray transient absorption (XTA) spectroscopy at the Ni K-edge. The state is characterized by an increase in intersite screening compatible with a strengthening of orbital hybridization between Ni $3d$ and O $2p$ orbitals, consistent with DFT+U+V calculations. Based on previous works, this electronic change occurs simultaneously with the screening of electron correlations~\cite{Lojewski2023,Cazali2025:166775,Rossi2025:52718}. Our analysis further shows that, at high excitation densities, this electronic state coexists with substantial lattice heating, showing its robustness. However, it can also be generated under continuous UV irradiation with a xenon lamp without appreciable lattice heating, which makes the magnitude of the effect larger with orders of magnitude lower excitation density. This demonstrates that electronic changes achieved in NiO upon UV excitation are not merely secondary thermal effects, but an intrinsic electronic response. 

Building on these findings, we outline how photoinduced renormalization of electronic correlations and increased orbital hybridization could be exploited to tune the catalytic activity of correlated oxides and enhance carrier transport. In devices incorporating NiO, models aiming at simulating electronic performance should account for the dynamic modification of energy levels and orbital hybridization induced by photoexcited carriers -- an effect that is particularly pronounced in strongly correlated systems.

Beyond this, these results open the door to systematic investigations of screening-induced renormalization of electronic interactions in charge-transfer insulators~\cite{Cazali2025:166775}. Looking ahead, X-ray free electron lasers (XFELs) will enable direct correlation of gap renormalization with transient changes in $d$-state occupancy and electronic screening via time-resolved RIXS~\cite{Merzoni2025:187430}. Such experiments will provide microscopic insight into nonequilibrium correlation dynamics and establish NiO as a benchmark system for studying light-driven tunability of the electronic structure in strongly correlated materials.

%%%%%%%%%%%%%%%%%%%%%%%%%%%%%%%%%%%%%%%%%%%%
%%%%%%%%%%%%%%%%%%%%%%%%%%%%%%%%%%%%%%%%%%%%
%%%%%%%%%%%%%%%%%%%%%%%%%%%%%%%%%%%%%%%%%%%%
%%%%%%%%%%%%%%%%%%%%%%%%%%%%%%%%%%%%%%%%%%%%
%%%%%%%%%%%%%%%%%%%%%%%%%%%%%%%%%%%%%%%%%%%%

\section*{Acknowledgments}

This work was supported by the ERC Advanced Grant DYNAMOX (no.\ 695197) and CHIRAX (no.\ 101095012). Views and opinions expressed are however those of the authors only and do not necessarily reflect those of the European Union or the European Research Council Executive Agency. Neither the European Union nor the granting authority can be held responsible for them. This work was also supported by NCCR:MUST grant no.\ 51NF40-183615 of the Swiss SNF. We acknowledge DESY (Hamburg, Germany), a member of the Helmholtz Association HGF, for the provision of experimental facilities in the framework of the long-term project proposal LTP (II-20210010) and regular proposal (I-20220620). We acknowledge the Paul Scherrer Institut, Villigen, Switzerland for provision of synchrotron radiation beamtime at beamline SuperXAS and MicroXAS of the SLS. AW acknowledges the partial support by the Polish Ministry and Higher Education project (no.~1/SOL/2021/2). We thank the MicroXAS beamline staff and Chris Milne for support during the measurement. We thank Matteo Calandra and Matteo Cococcioni for fruitful discussions.

%%%%%%%%%%%%%%%%%%%%%%%%%%%%%%%%%%%%%%%%%%%%%%%%%%%%%%%%%%%%%%%%%
%%%%%%%%%%%%%%%%%%%%%%%%%%%%%%%%%%%%%%%%%%%%%%%%%%%%%%%%%%%%%%%%%
%%%%%%%%%%%%%%%%%%%%%%%%%%%%%%%%%%%%%%%%%%%%%%%%%%%%%%%%%%%%%%%%%
%%%%%%%%%%%%%%%%%%%%%%%%%%%%%%%%%%%%%%%%%%%%%%%%%%%%%%%%%%%%%%%%%

\cleardoublepage

\clearpage
\section*{Supporting Information}
\addcontentsline{toc}{section}{Supporting Information}

\setcounter{section}{0}
\setcounter{subsection}{0}
\setcounter{subsubsection}{0}
\setcounter{figure}{0}
\setcounter{table}{0}
\setcounter{equation}{0}

\renewcommand{\thesection}{S\arabic{section}}
\renewcommand{\thesubsection}{S\arabic{section}.\arabic{subsection}}
\renewcommand{\thesubsubsection}{S\arabic{section}.\arabic{subsection}.\arabic{subsubsection}}
\renewcommand{\thefigure}{S\arabic{figure}}
\renewcommand{\thetable}{S\arabic{table}}
\renewcommand{\theequation}{S\arabic{equation}}

% Give hyperref destinations SI-specific names after resetting the counters.
\renewcommand{\theHsection}{SI.\arabic{section}}
\renewcommand{\theHsubsection}{SI.\arabic{section}.\arabic{subsection}}
\renewcommand{\theHsubsubsection}{SI.\arabic{section}.\arabic{subsection}.\arabic{subsubsection}}
\renewcommand{\theHfigure}{SI.\arabic{figure}}
\renewcommand{\theHtable}{SI.\arabic{table}}
\renewcommand{\theHequation}{SI.\arabic{equation}}

\begin{center}
\small
\textit{Corresponding authors:}\\
Thomas C. Rossi (\texttt{thomas.rossi@helmholtz-berlin.de})\\
Majed Chergui (\texttt{majed.chergui@elettra.eu})
\end{center}

\cleardoublepage

\tableofcontents

\cleardoublepage

%%%%%%%%%%%%%%%%%%%%%%%%%%%%%%%%%%%%%%%%%%%%%%%%%%%%%%%%%%%%%%%%%%%%%%%
%%%%%%%%%%%%%%%%%%%%%%%%%%%%%%%%%%%%%%%%%%%%%%%%%%%%%%%%%%%%%%%%%%%%%%%
%%%%%%%%%%%%%%%%%%%%%%%%%%%%%%%%%%%%%%%%%%%%%%%%%%%%%%%%%%%%%%%%%%%%%%%
%%%%%%%%%%%%%%%%%%%%%%%%%%%%%%%%%%%%%%%%%%%%%%%%%%%%%%%%%%%%%%%%%%%%%%%

\section{\emph{In situ} XAS measurements under continuous light irradiation\label{secSI:XAS_continuous_irradiation}}

The experimental von H\'amos setup has been described elsewhere~\cite{Schlesiger2015,Schlesiger2020}. Briefly, the setup is based on a polychromatic micro focus X-ray tube dispersed in transmission of the sample by HAPG crystal (highly annealed pyrolytic graphite) with a radius of curvature of \SI{30}{\centi\metre}. The X-ray tube was operating at \SI{15.90}{\kilo\volt} with a current of \SI{1.88}{\milli\ampere}. The dispersed X-ray photons were collected on an Eiger 2K (Dectris) with low and high-energy thresholds to avoid the detection of inelastically scattered photons in air (\SI{5000}{} and \SI{12000}{\electronvolt}, respectively). Energy calibration was performed with a \SI{2}{\micro\metre} nickel foil. The NiO pellet was prepared by mixing \SI{53}{\milli\gram} of NiO powder (325 mesh, Sigma-Aldrich) and \SI{255}{\milli\gram} of silicon nitride powder (Sigma-Aldrich). To minimize inhomogeneity effects in the measured spectra, the sample was moved continuously about \SI{2}{\milli\meter} back and forth perpendicular to the X-ray beam.

\begin{figure}[!ht]
    \centering
    \includegraphics[width=\linewidth]{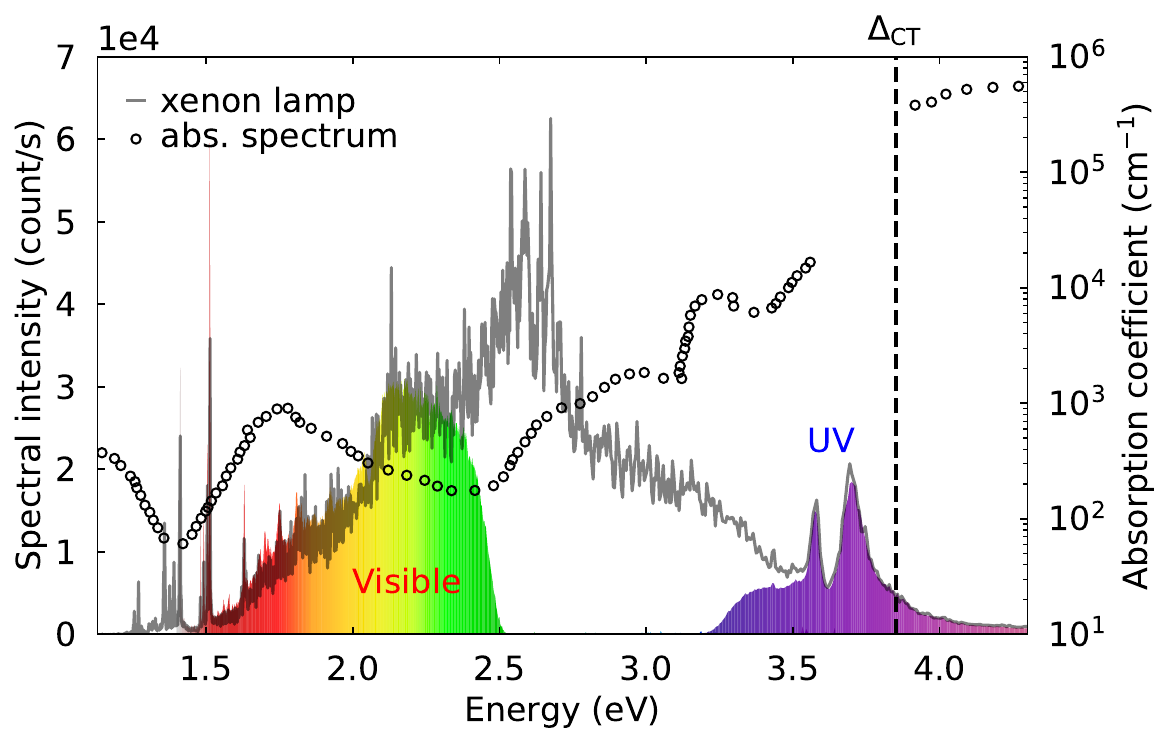}
    \vspace{-9mm}
    \caption{\textbf{Spectral densities used for \emph{in situ} excitation of NiO.} Emission spectrum of a xenon arc lamp (grey curve, left axis) with filtered spectral densities for measurements under visible irradiation (red to green rainbow) and UV irradiation (purple rainbow). The absorption coefficient of NiO single crystal is shown for reference (black circles, right axis, adapted from reference~\cite{Newman:1959gr}). The position of the charge-transfer gap ($\Delta_{\text{CT}}$) is shown on the energy axis (\SI{3.7}{\electronvolt}).}
    \label{figSI:xenon_lamp_spectrum}
\end{figure}

Light irradiation is performed with a Xenon arc lamp (SLS401, Thorlabs) focused on the pellet to a diameter of \SI{3}{\milli\metre} with focusing lenses. An OG515 filter is used to deliver a visible spectrum while a FGUV11M filter is used to deliver a UV spectrum (see the corresponding spectral densities in Figure \ref{figSI:xenon_lamp_spectrum}).

For \emph{in situ} measurements, the sample was measured alternately for \SI{10}{minutes} under light irradiation and without light irradiation. This alternate measurement procedure was done to minimize the influence of spectral shifts due to the slow temperature variations of the instrument, and to resolve differences of $<\SI{1}{\percent}$ between XAS spectra recorded with and without optical irradiation.

%%%%%%%%%%%%%%%%%%%%%%%%%%%%%%%%%%%%%%%%%%%%%%%%%%%%%%%%%%%%%%%%%%%%%%%%%%%%
%%%%%%%%%%%%%%%%%%%%%%%%%%%%%%%%%%%%%%%%%%%%%%%%%%%%%%%%%%%%%%%%%%%%%%%%%%%%
%%%%%%%%%%%%%%%%%%%%%%%%%%%%%%%%%%%%%%%%%%%%%%%%%%%%%%%%%%%%%%%%%%%%%%%%%%%%
%%%%%%%%%%%%%%%%%%%%%%%%%%%%%%%%%%%%%%%%%%%%%%%%%%%%%%%%%%%%%%%%%%%%%%%%%%%%

\section{Material characterizations\label{secSI:material_characterization}}

\begin{figure}[!ht]
    \centering
    \includegraphics[width=0.6\linewidth]{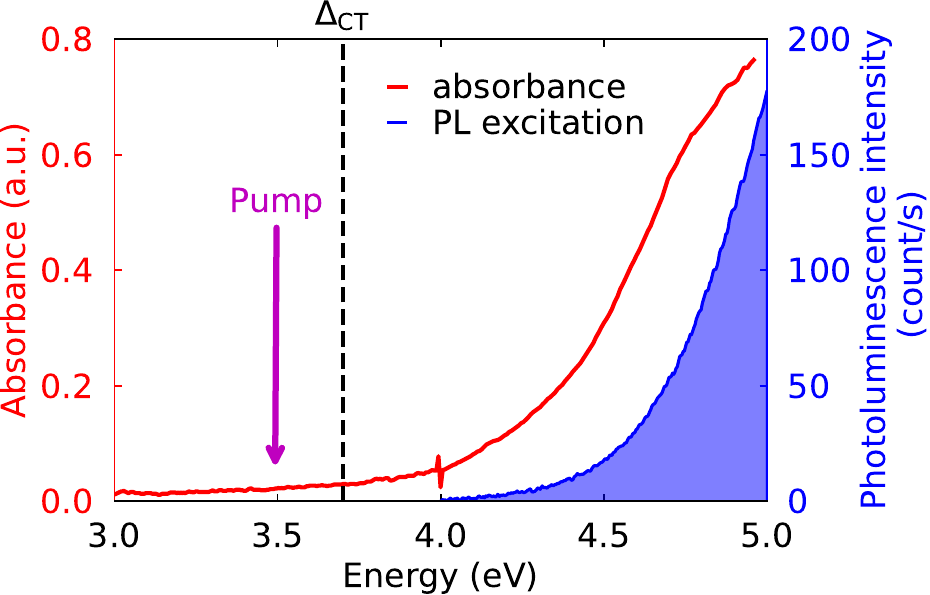}
    \vspace{-4mm}
    \caption{\textbf{Optical properties of NiO powder at the charge-transfer gap.} Absorption (red curve) and photoluminescence excitation spectrum (blue shaded curve, detection at \SI{3.5}{\electronvolt}). The photon energy of the pump laser used for XTA measurements is shown for reference (purple arrow), alongside the position of the charge-transfer gap in single crystal NiO ($\Delta_{\text{CT}}$, vertical dashed line).}
    \label{figSI:optical_spectra}
\end{figure}

The sample used through this work is a NiO powder (325 mesh, \SI{99}{\percent}, Sigma-Aldrich). The powder was further size filtered for XAS and XTA measurements with a 635 mesh to select particles with diameter $\leq\SI{20}{\micro\meter}$, which brings the average particle size to about \SI{10}{\micro\meter}. Absorption and photoluminescence excitation spectra are shown in Figure \ref{figSI:optical_spectra}. The pump excitation for the X-ray transient absorption (XTA) measurements is at \SI{3.49}{\electronvolt}, at the onset of the optical absorption. This photon energy is selected to achieve homogeneous excitation through the large NiO particles in the particle suspension. SEM images, XRD patterns, and FT-IR spectra of the particles were previously reported in~\cite{Ye2026:191622}.

%%%%%%%%%%%%%%%%%%%%%%%%%%%%%%%%%%%%%%%%%%%%%%%%%%%%%%%%%%%%%%%%%%%%%%%%%%%%
%%%%%%%%%%%%%%%%%%%%%%%%%%%%%%%%%%%%%%%%%%%%%%%%%%%%%%%%%%%%%%%%%%%%%%%%%%%%
%%%%%%%%%%%%%%%%%%%%%%%%%%%%%%%%%%%%%%%%%%%%%%%%%%%%%%%%%%%%%%%%%%%%%%%%%%%%
%%%%%%%%%%%%%%%%%%%%%%%%%%%%%%%%%%%%%%%%%%%%%%%%%%%%%%%%%%%%%%%%%%%%%%%%%%%%

\section{X-ray absorption data processing\label{secSI:data_processing}}

X-ray absorption spectra (XAS) are background subtracted, edge-jump normalized and flattened in the post-edge region of the spectrum using \texttt{Larch}~\cite{Newville:2013go}. The ionization potential was set at the maximum of the first derivative of the XAS spectrum. The pre-edge energy range and post-edge energy ranges (relative to the ionization potential) were set between \SI{-50}{} and \SI{-30}{\eV}  and between \SI{35} and \SI{250}{\eV}. The pre-edge and post-edge background were both subtracted with a linear function, and the spectra were normalized to the edge jump.

\cleardoublepage

%%%%%%%%%%%%%%%%%%%%%%%%%%%%%%%%%%%%%%%%%%%%%%%%%%%%%%%%%%%%%%%%%%%%%%%%%%%%
%%%%%%%%%%%%%%%%%%%%%%%%%%%%%%%%%%%%%%%%%%%%%%%%%%%%%%%%%%%%%%%%%%%%%%%%%%%%
%%%%%%%%%%%%%%%%%%%%%%%%%%%%%%%%%%%%%%%%%%%%%%%%%%%%%%%%%%%%%%%%%%%%%%%%%%%%
%%%%%%%%%%%%%%%%%%%%%%%%%%%%%%%%%%%%%%%%%%%%%%%%%%%%%%%%%%%%%%%%%%%%%%%%%%%%

\section{Temperature-dependent X-ray absorption spectroscopy and diffraction}

\subsection{X-ray absorption spectroscopy\label{subsecSI:XAS_temperature_dependence}}

XAS spectra as a function of lattice temperature were recorded at the Ni K-edge of NiO at the SuperXAS (X10DA) \SI{2.9}{\tesla} super-bending magnet beamline of the Swiss Light Source (SLS). The measurement was performed with a Si(111) double crystal monochromator oscillating at \SI{1}{\hertz} in quick-scanning XAS mode with a sampling rate of \SI{2}{\mega\hertz}. XAS spectra in total fluorescence yield were measured with a PIPS detector from Mirion Technologies (details of the setup in~\cite{Clark2020}). XAS spectra in transmission were measured with two ion chambers for the incident and transmitted X-ray flux. A reference spectrum of a nickel foil was measured simultaneously to the sample spectrum with another ionization chamber downstream the sample. The sample was placed in a quartz capillary with \SI{2}{\milli\metre} outer diameter and \SI{10}{\micro\metre} wall thickness (HR6-150, Hampton Research). It was composed of grinded and compacted powder mixture with \SI{494.2}{\milli\gram} (\SI{19.9}{\milli\mol}) of boron nitride (Sigma-Aldrich) and \SI{14.3}{\milli\gram} (\SI{191}{\micro\mol}) of NiO particles (325 mesh, Sigma-Aldrich). The edge jump was $\sim\SI{0.2}{}$ in transmission and $\sim\SI{0.66}{}$ in fluorescence. The sample was heated in a capillary reactor setup equipped with two infrared heaters and a thermocouple positioned inside the capillary in direct contact with the powder for temperature monitoring and control. A temperature ramp rate of \SI{1}{\degreeCelsius}/min was applied. Data analysis was performed with the ProQEXAFS software version 2.51~\cite{Clark2020a}. After monochromator encoding and energy calibration with respect to the nickel foil spectrum measured in transmission (performed on individual spectra), background subtraction was performed with a second order polynomial function before the edge and a fourth order polynomial function above the edge. High-frequency noise in the spectra are filtered out with a Savitzky-Golay filter as implemented in the python \texttt{scipy} library, with a window of 200 data points and a third order polynomial interpolation. Spectra were averaged together in bins of \SI{5}{\degreeCelsius}. An energy binning of \SI{0.5}{\electronvolt} was applied to increase the statistics at each energy point, which was necessary to achieve a signal-to-noise ratio comparable with the magnitude of the signals in the pump-probe measurements. Since the temperature and the energy binning were performed on independent datasets, the covariance matrix is zero and the standard deviation propagation was calculated according to the sum of the squares of individual standard deviations. Difference XAS spectra were computed by difference between the XAS spectrum at the temperature bin $T$ and the XAS spectrum measured at room temperature (\SI{25}{\degreeCelsius}). Standard deviation propagation for the difference XAS spectrum was performed following,
\begin{equation}
\sigma_{\text{diff}}(E_{\text{bin}},T)=\sqrt{\sigma^2(E_{\text{bin}},T)+\sigma^2(E_{\text{bin}},25 ^\circ\text{C})}.
\end{equation}

\begin{figure}
    \centering
    \includegraphics[width=\linewidth]{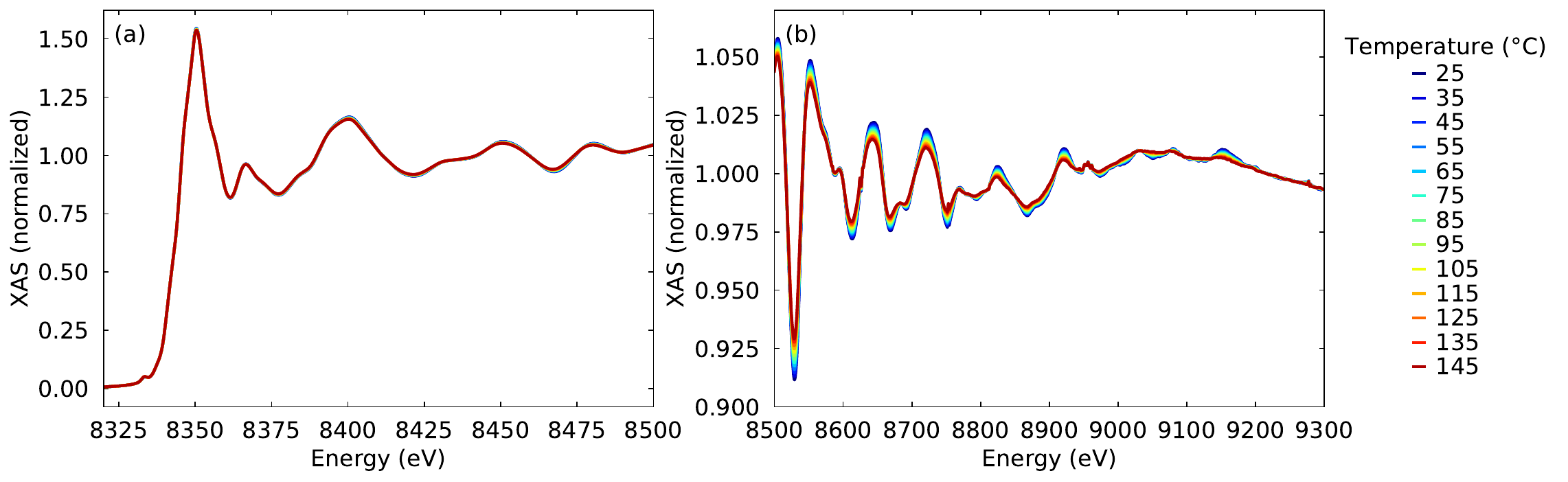}
    \vspace{-7mm}
    \caption{\textbf{Temperature-dependence of XAS spectra at the Ni K-edge of NiO.} Temperature-dependent XAS spectra at the Ni K-edge of NiO between room temperature (\SI{25}{\degreeCelsius}) and \SI{145}{\degreeCelsius} in the (a) XANES, and (b) EXAFS (detection in fluorescence).}
    \label{figSI:XAS_temperature_dependence}
\end{figure}

\begin{figure}
    \centering
    \includegraphics[width=\linewidth]{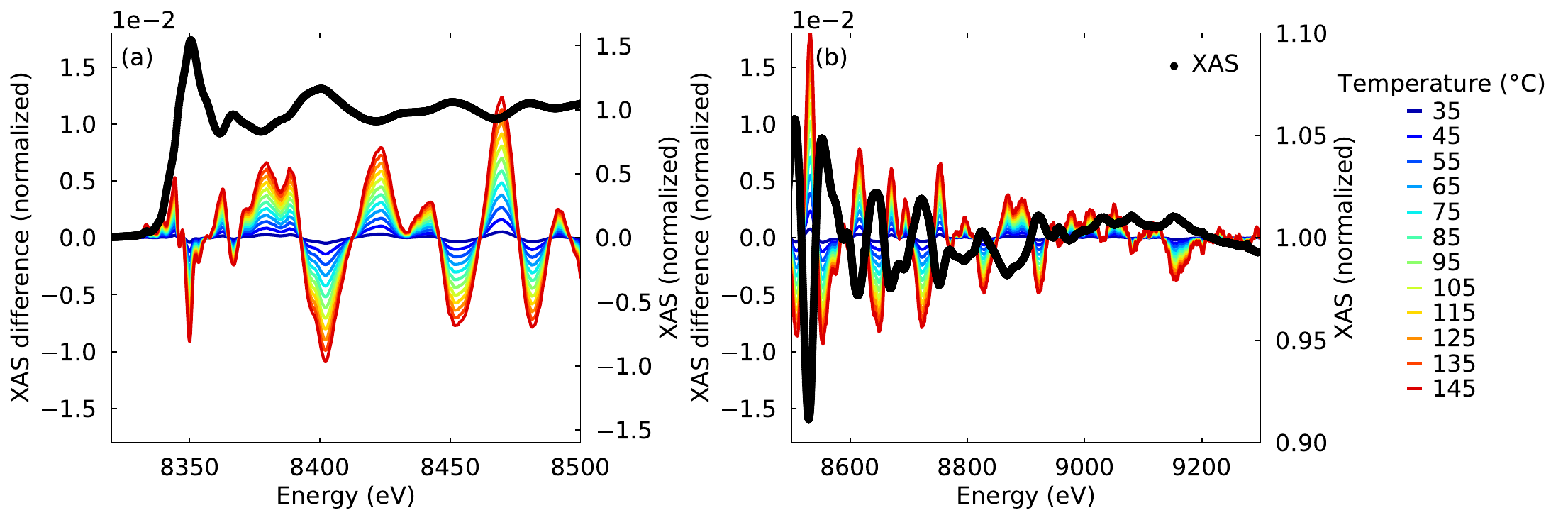}
    \vspace{-7mm}
    \caption{\textbf{Temperature-dependent XAS difference spectra at the Ni K-edge of NiO.} Evolution of XAS difference spectra between a XAS spectrum measured at a hot lattice temperature and a XAS spectrum at \SI{25}{\degreeCelsius} (colored curves, left axis) in the (a) XANES, and (b) EXAFS. The room temperature XAS spectrum (\SI{25}{\degreeCelsius}) is shown for reference (grey circles, right axis). The detection is performed in fluorescence.}
    \label{figSI:XAS_difference_temperature_dependence}
\end{figure}

Figure \ref{figSI:XAS_temperature_dependence} shows the evolution of normalized XAS spectra between room temperature (\SI{25}{\degreeCelsius}) and \SI{145}{\degreeCelsius}. In the EXAFS (Figure \ref{figSI:XAS_temperature_dependence}b), oscillations damp as the temperature increases, which is due to the increase of the Debye-Waller factor. Figure \ref{figSI:XAS_difference_temperature_dependence} displays difference XAS spectra computed from the difference between a XAS spectrum at a hot lattice temperature and a XAS spectrum measured at room temperature. The difference spectra were used to simulate the lattice heating contribution in excited state XAS spectra of NiO.

Background subtraction, normalization and flattening of the XAS spectra was performed with the python implementation of \texttt{Larch} (version 0.9.80)~\cite{Newville:2013go}. The evolution of the bond distances and the Debye-Waller factor as a function of lattice temperature were obtained by fitting EXAFS oscillations at each lattice temperature using the python implementation of \texttt{Larch}~\cite{Newville:2013go}. The pre-edge background subtraction was performed with a linear function between \SI{-145.5}{} and \SI{-80.0}{\electronvolt} with respect to $E_0$ (set to \SI{8345.5}{\electronvolt}) and the post-edge background subtraction was performed with a linear function between \SI{150}{} and \SI{854}{\electronvolt} with respect to $E_0$ with no spline clamp on the low energy side and a strong clamp on the high energy side. Fourier transform of the EXAFS oscillations was performed between \SI{3.3}{} and \SI{15}{\per\angstrom} (with a Hanning window of slope \SI{1}{\per\angstrom}). Fitting of the EXAFS oscillations was performed by including the shortest four scattering paths (two single scattering and two multiple scattering paths, detail in Table \ref{tab:scattering_paths}). A radial space background subtractions was performed below \SI{1}{\angstrom} ($R_{bckg}$ parameter). The $k$-weighting of power 1, 2, and 3 were used simultaneously during the fitting, which was performed in real space between \SI{1.0}{} and \SI{4.0}{\angstrom}. The Debye-Waller factors (DWFs) of the nickel and the oxygen atoms in the single scattering paths were considered independent and follow an Einstein model with independent temperatures $\theta_{E,O}$ and $\theta_{E,Ni}$ for the single scattering paths to the closest oxygen and nickel atoms to the absorber, respectively. The expression of the DWF in the Einstein model was given in reference~\cite{Sevillano1979}, the Einstein temperatures of oxygen ($\theta_{E,O}$) and nickel atoms ($\theta_{E,Ni}$) were fitted globally for all the datasets. Individual DWFs were used for the multiple scattering paths ($\sigma^2_{MS1}$, $\sigma^2_{MS2}$). The EXAFS parameters $S_0^2$, $E_0$ were fixed to the same value for all the datasets. The coordination numbers were fixed to the nominal stoichiometric values. The bond distances were fitted such that the variation of bond distance $\Delta R$ with respect to the input value is given by $R=\Delta R_{offset}+\alpha TR_{eff}$ with $T$ the lattice temperature in kelvin. The scatterings paths included in the fit are given in Table~\ref{tab:scattering_paths}.

\begin{figure}
    \centering
    \includegraphics[width=\linewidth]{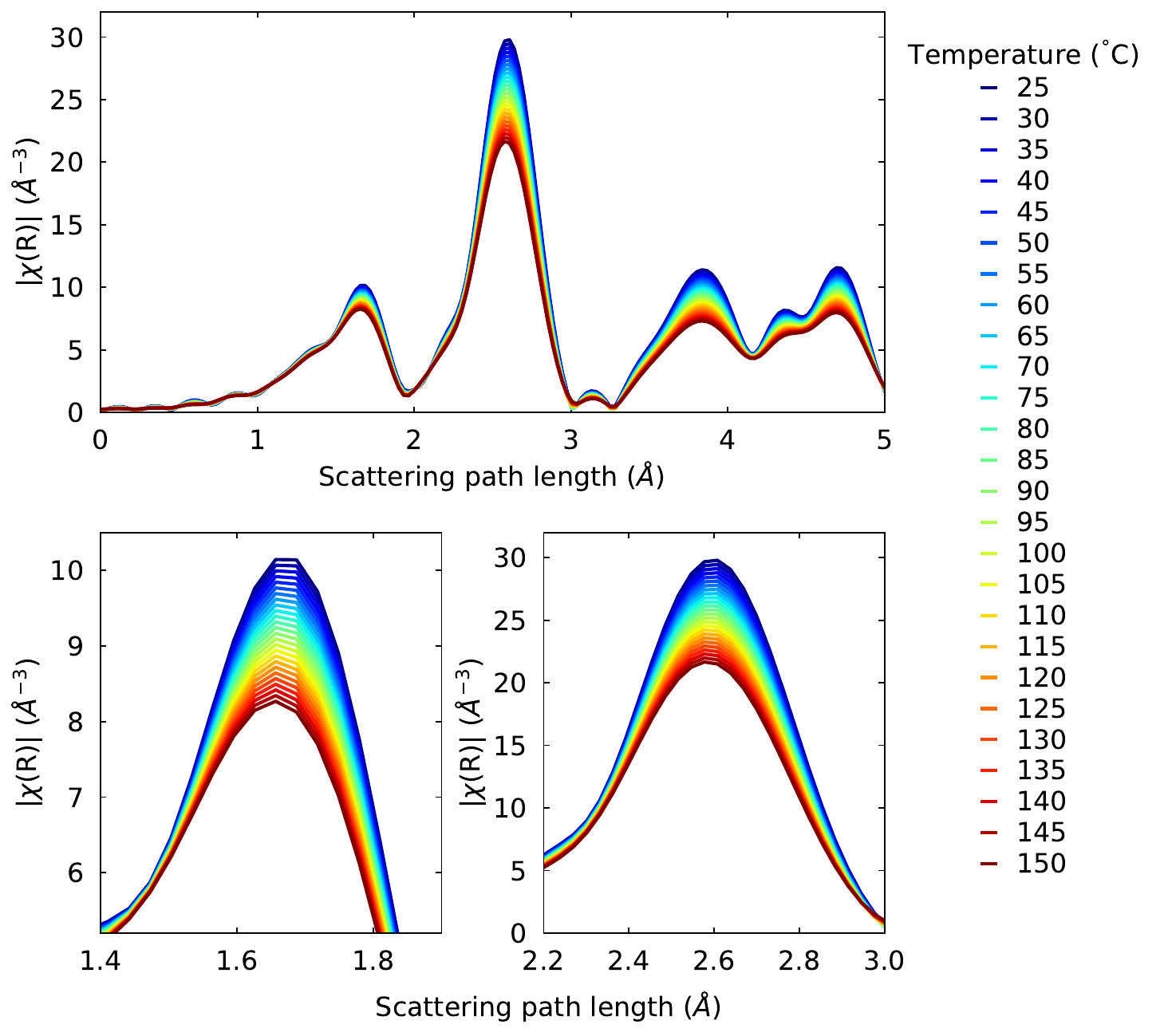}
    \caption{\textbf{Real-space scattering amplitude as a function of lattice temperature.} Evolution of the scattering cross-section along the path length with lattice temperature.}
    \vspace{-12mm}
    \label{figSI:radial_distribution_temperature_dependence}
\end{figure}

\begin{figure}
    \centering
    \includegraphics[width=0.24\linewidth]{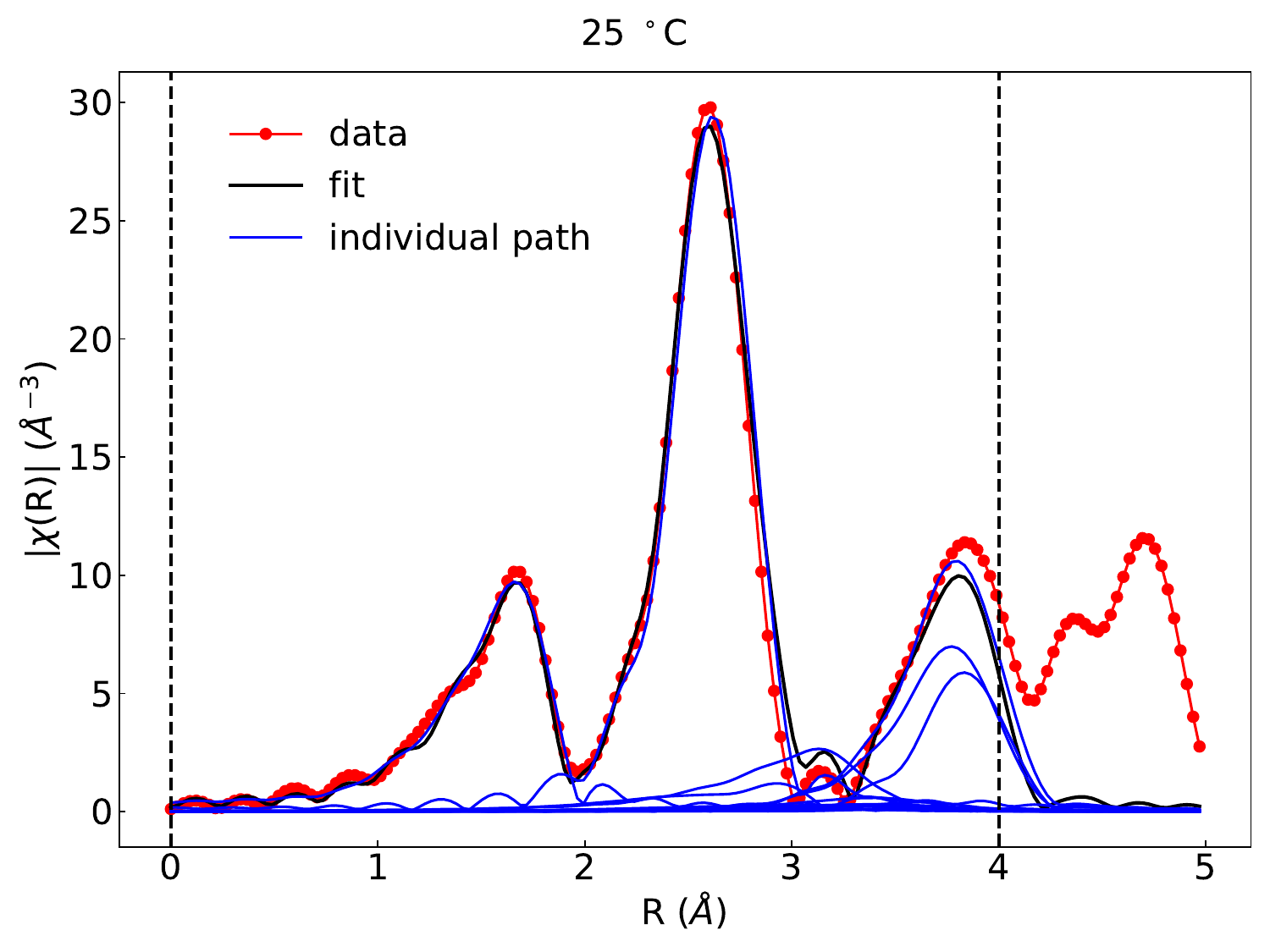}
    \includegraphics[width=0.24\linewidth]{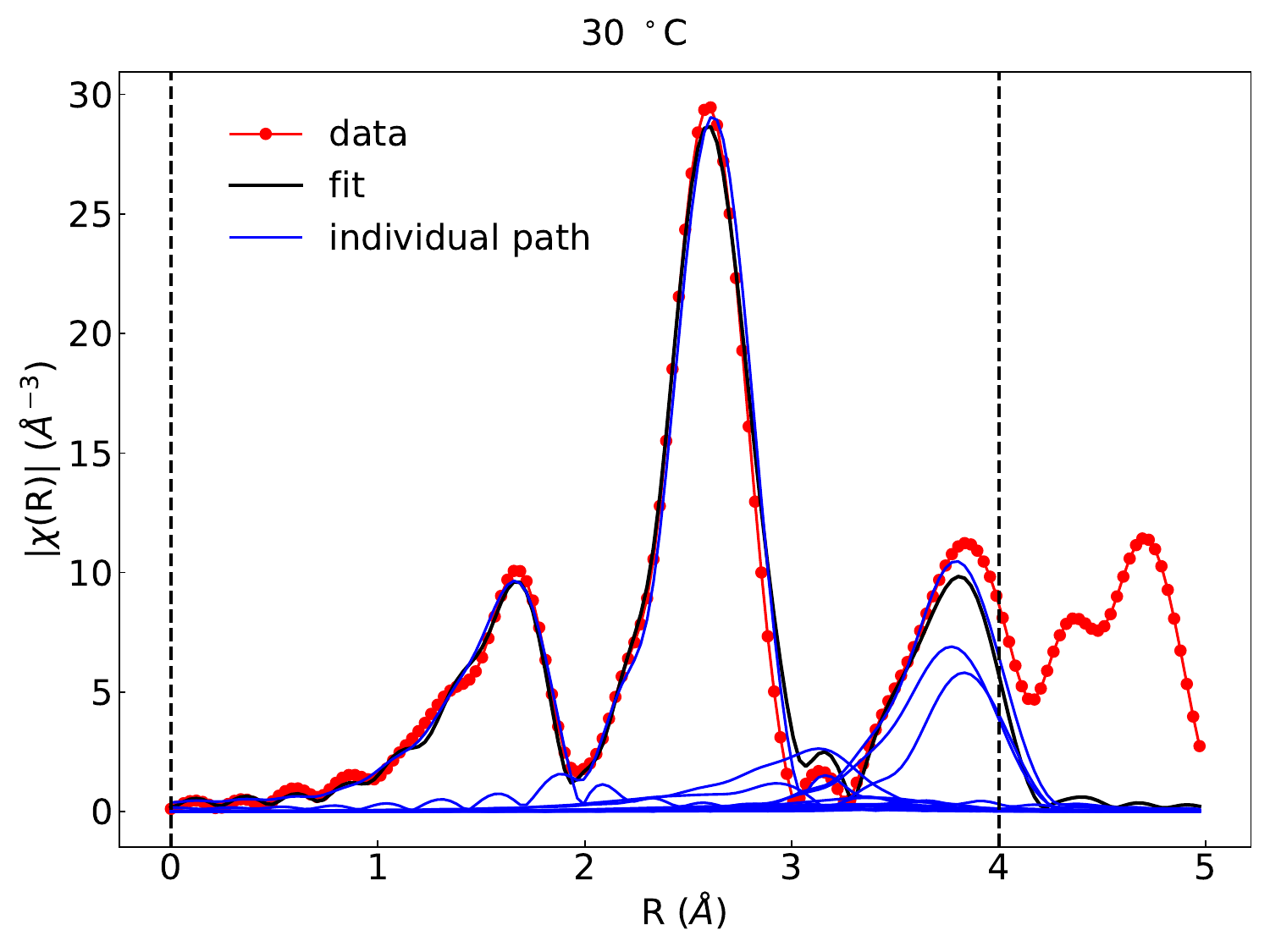}
    \includegraphics[width=0.24\linewidth]{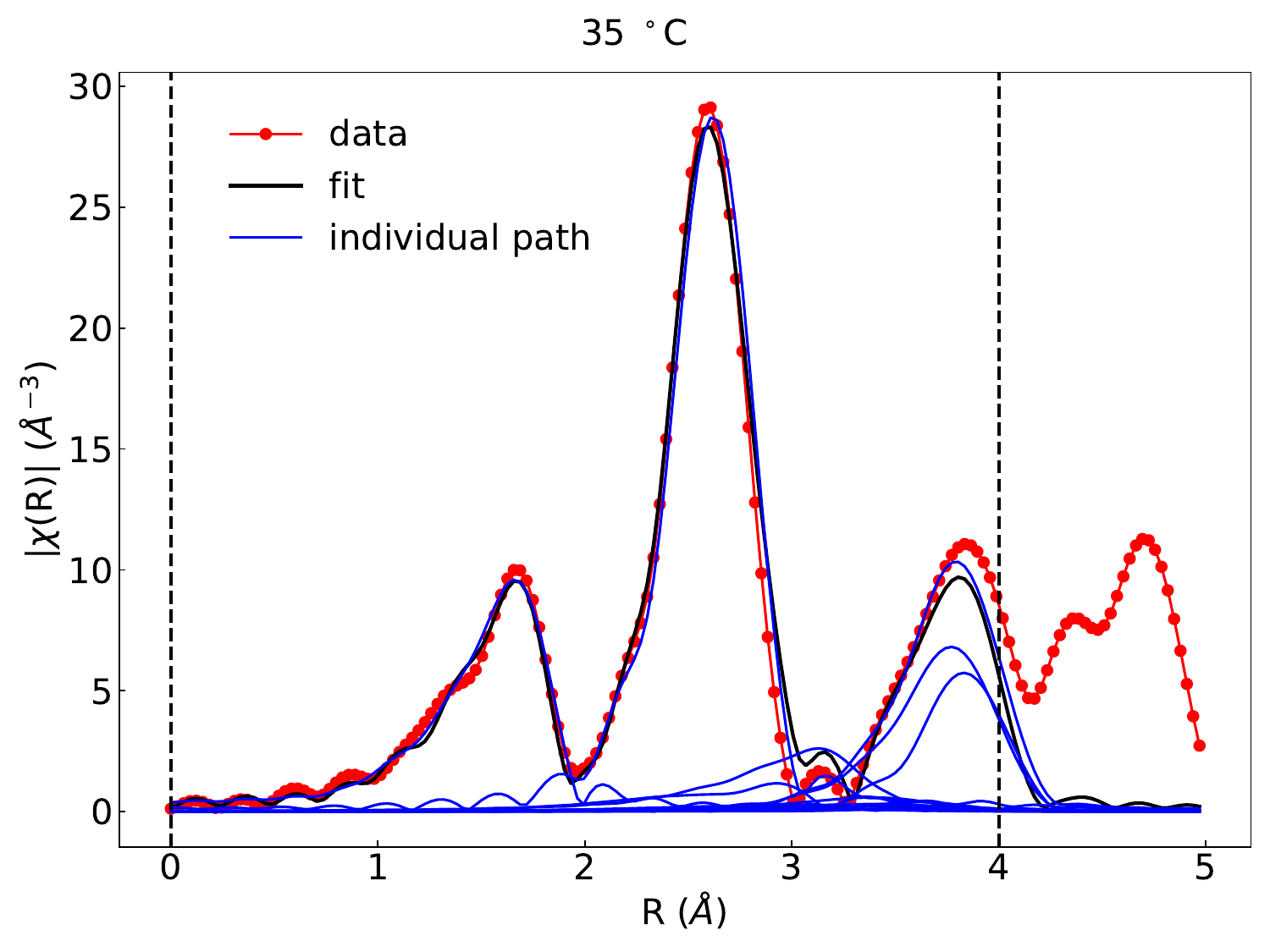}
    \includegraphics[width=0.24\linewidth]{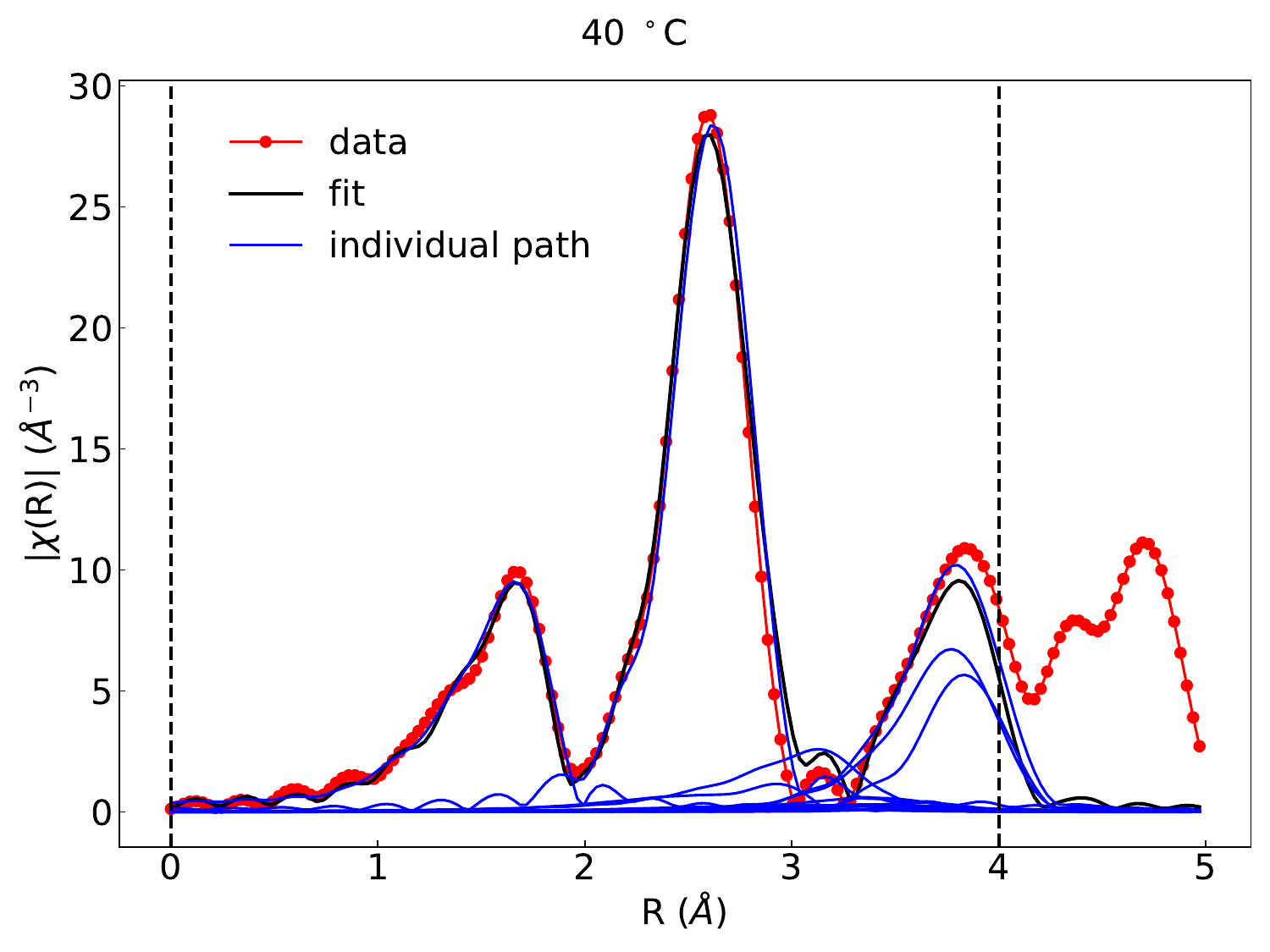}
    
    \includegraphics[width=0.24\linewidth]{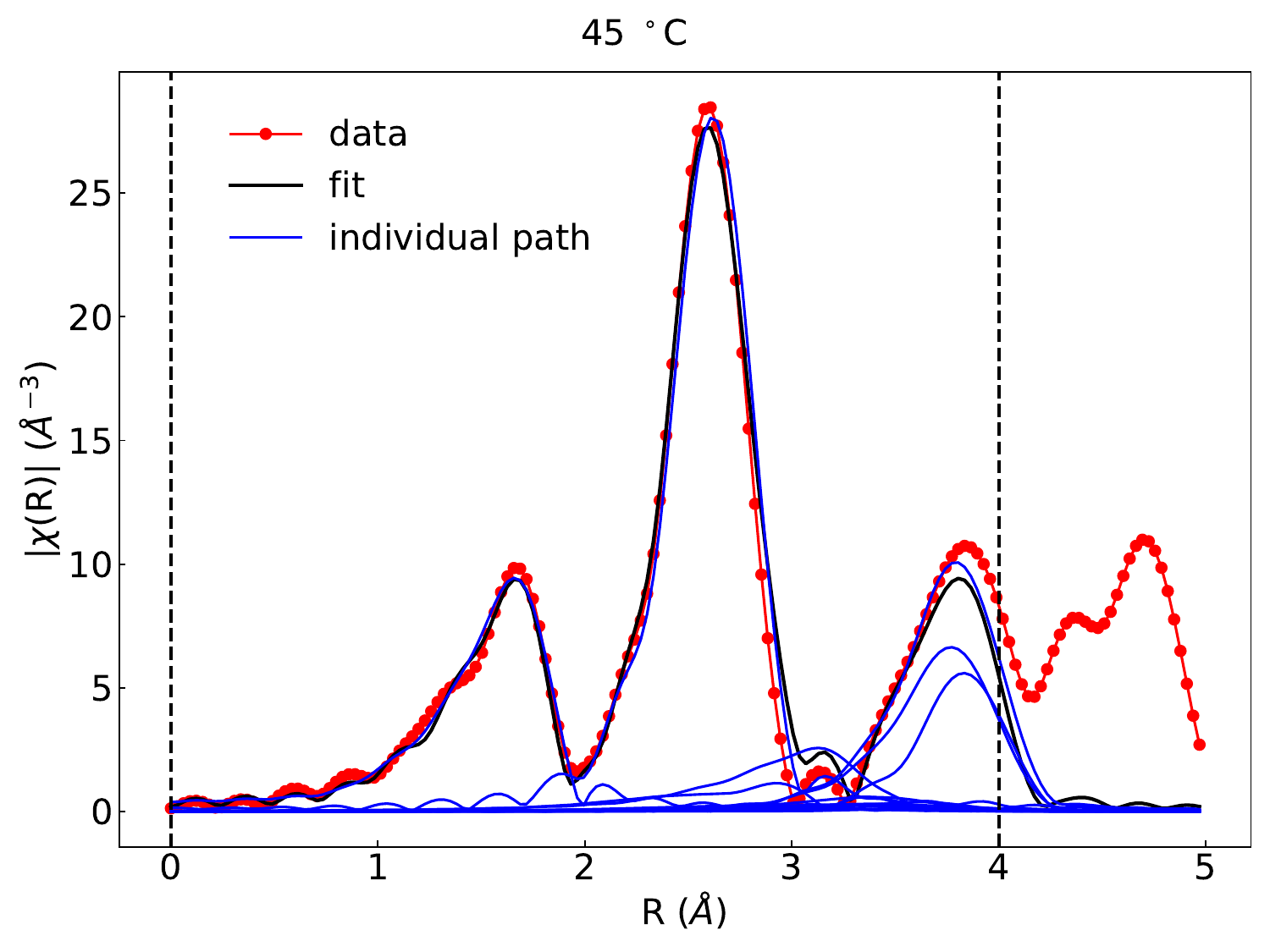}
    \includegraphics[width=0.24\linewidth]{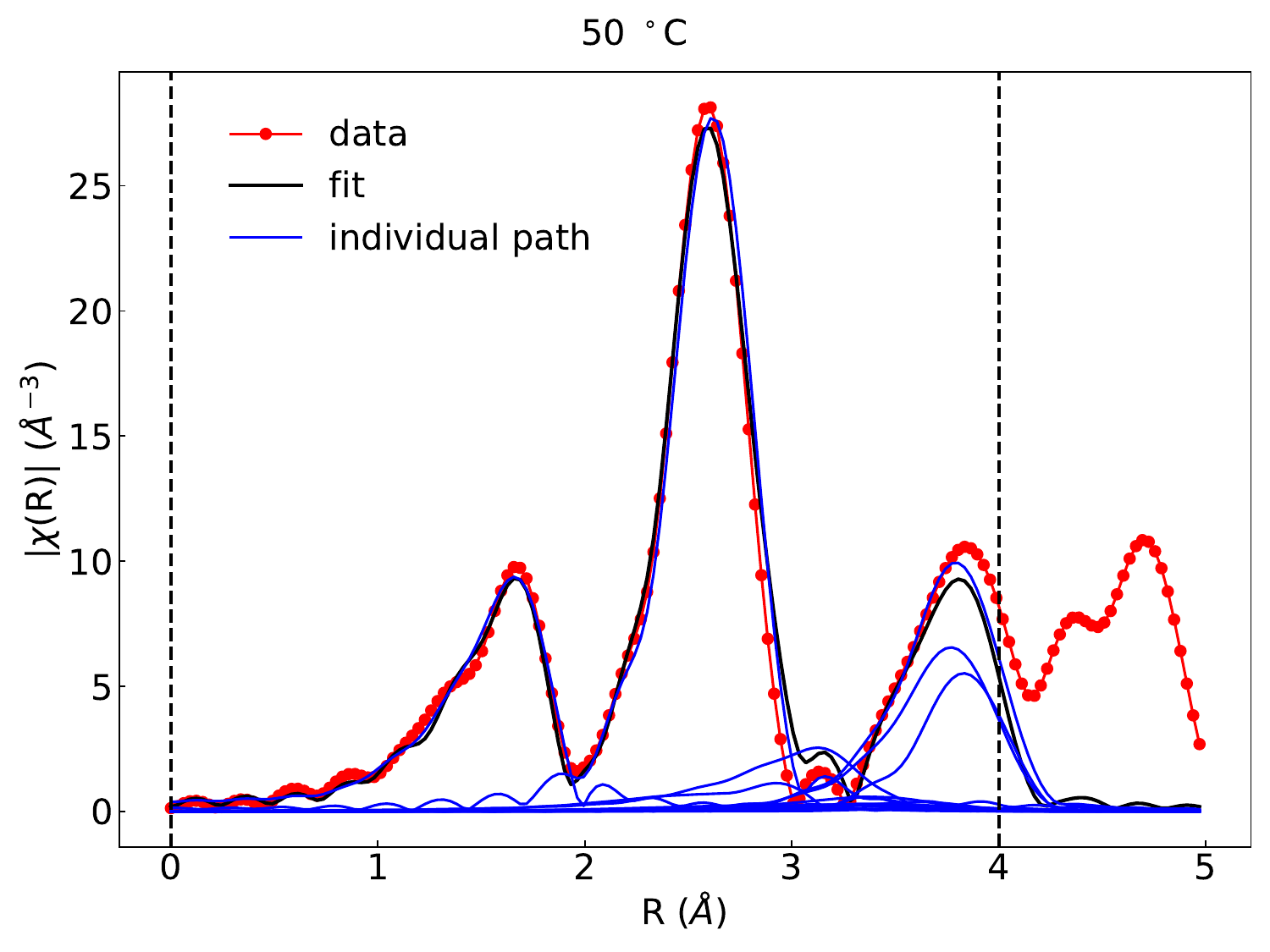}
    \includegraphics[width=0.24\linewidth]{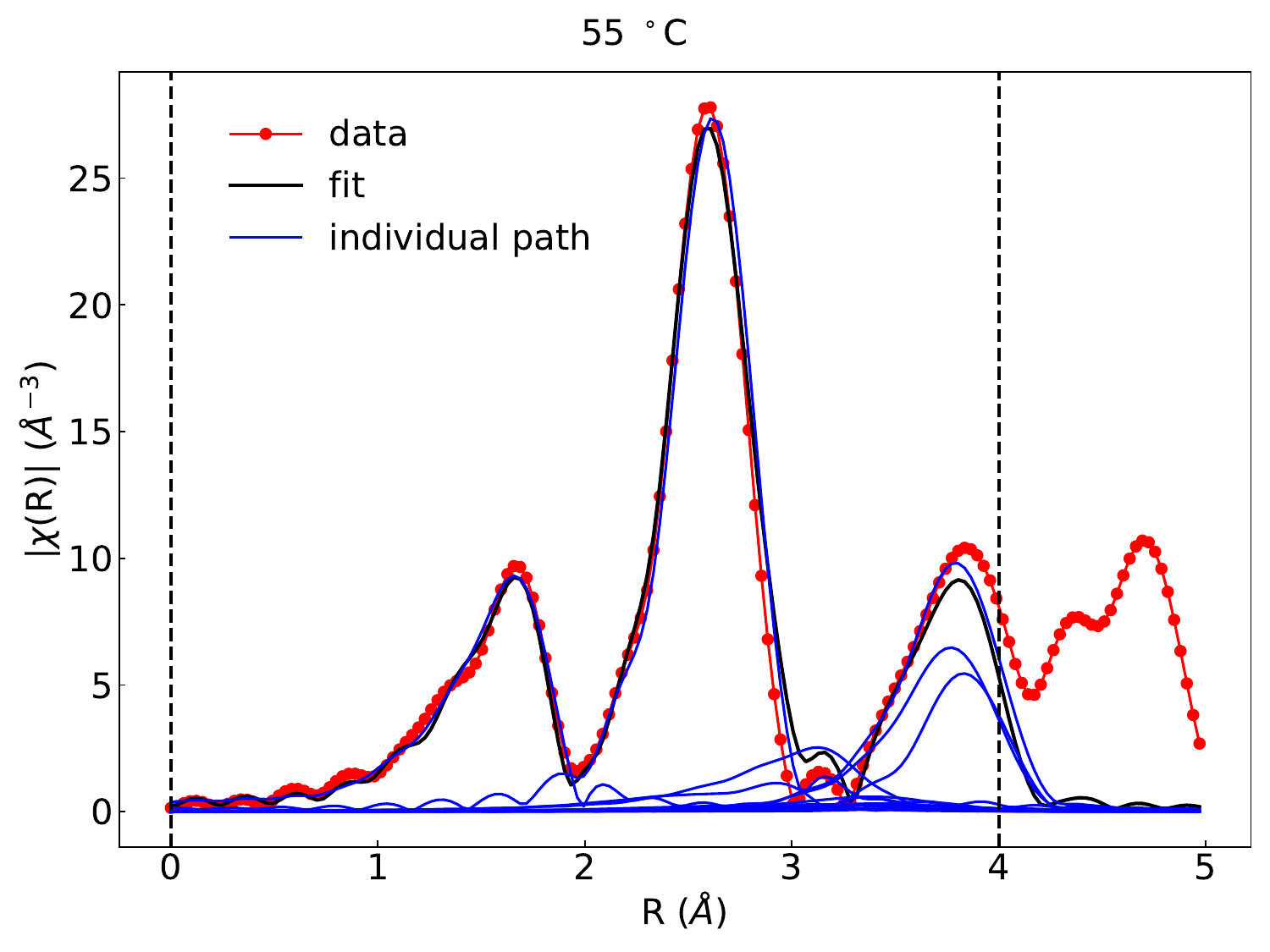}
    \includegraphics[width=0.24\linewidth]{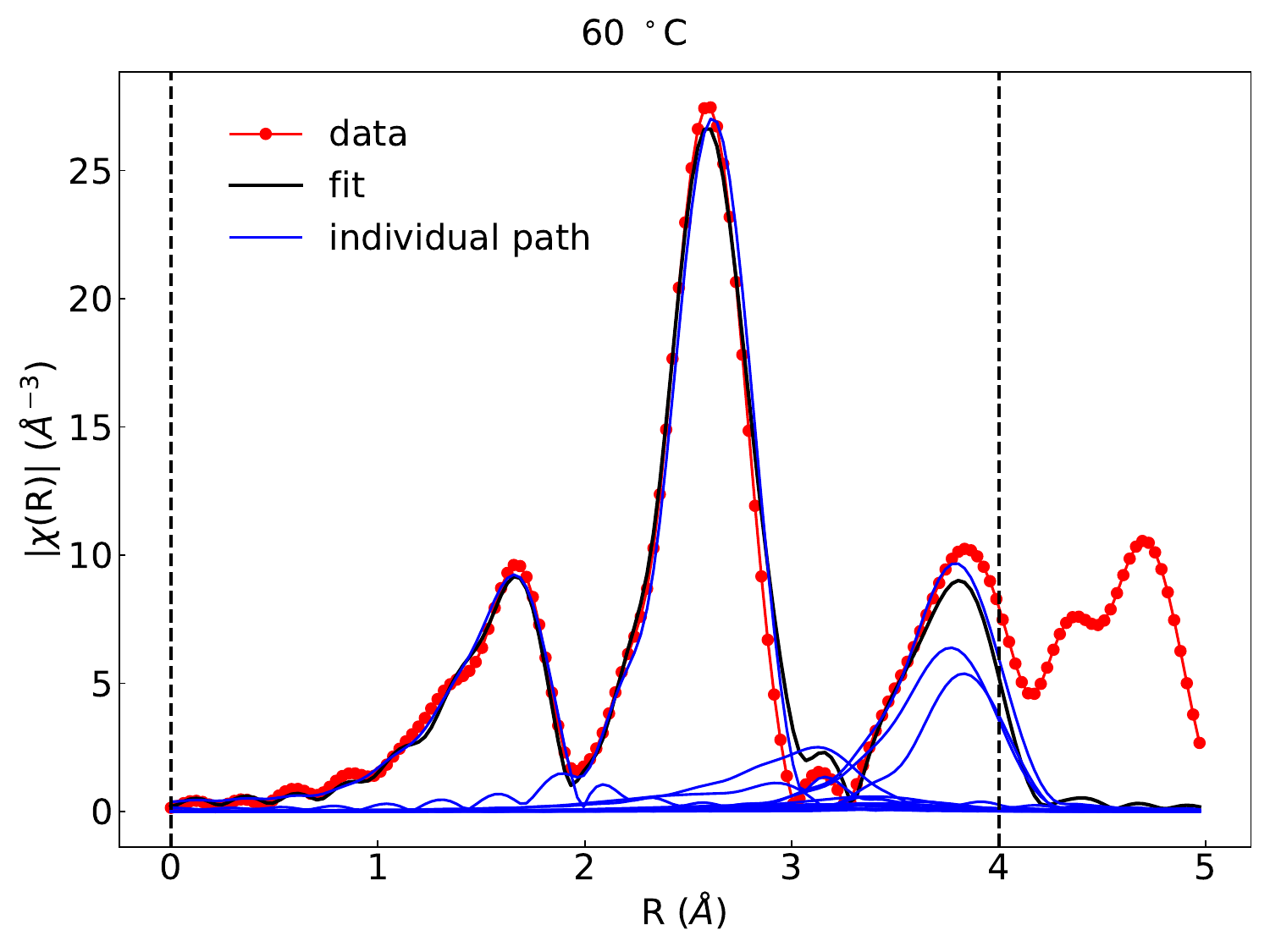}
    
    \includegraphics[width=0.24\linewidth]{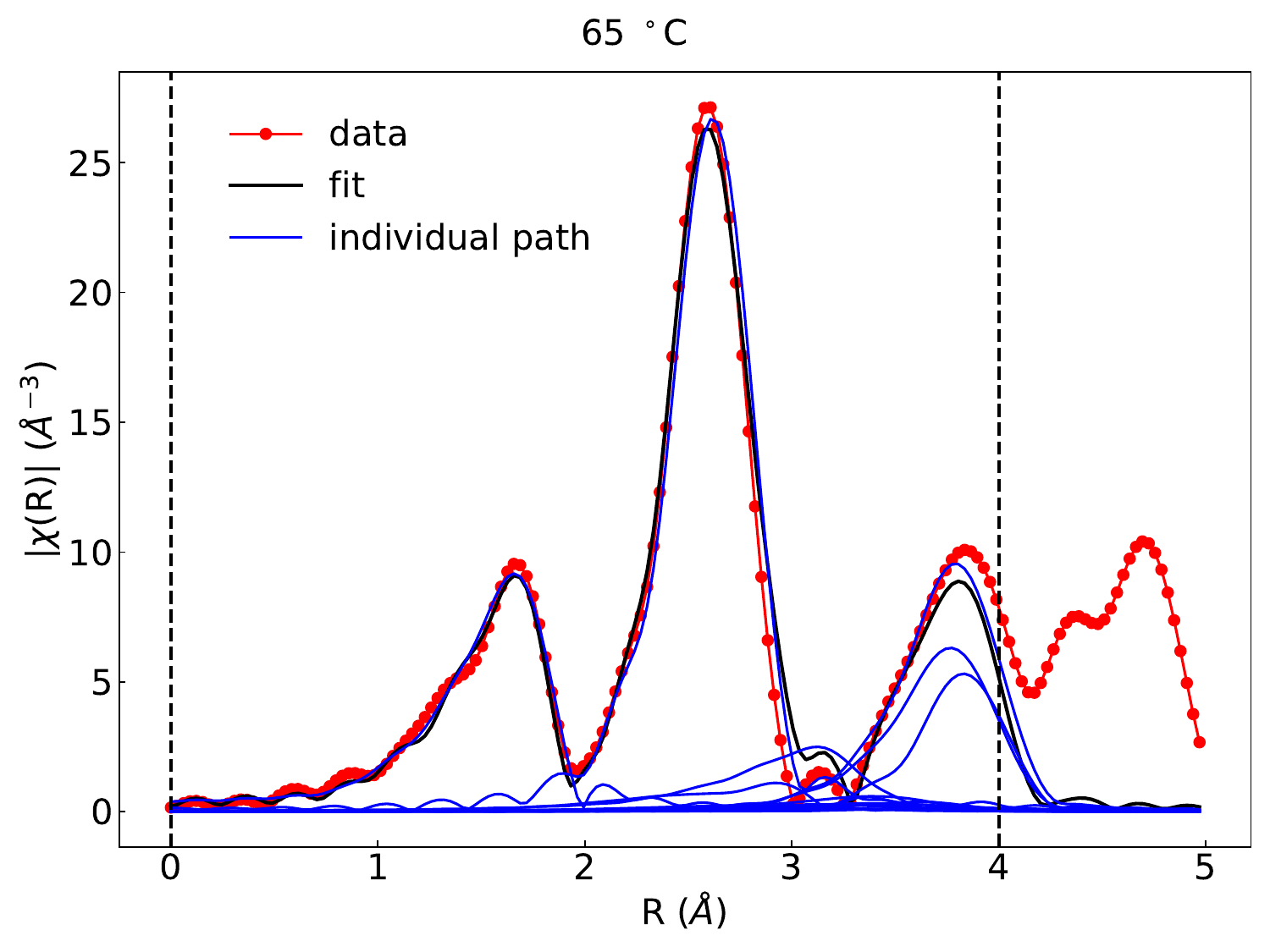}
    \includegraphics[width=0.24\linewidth]{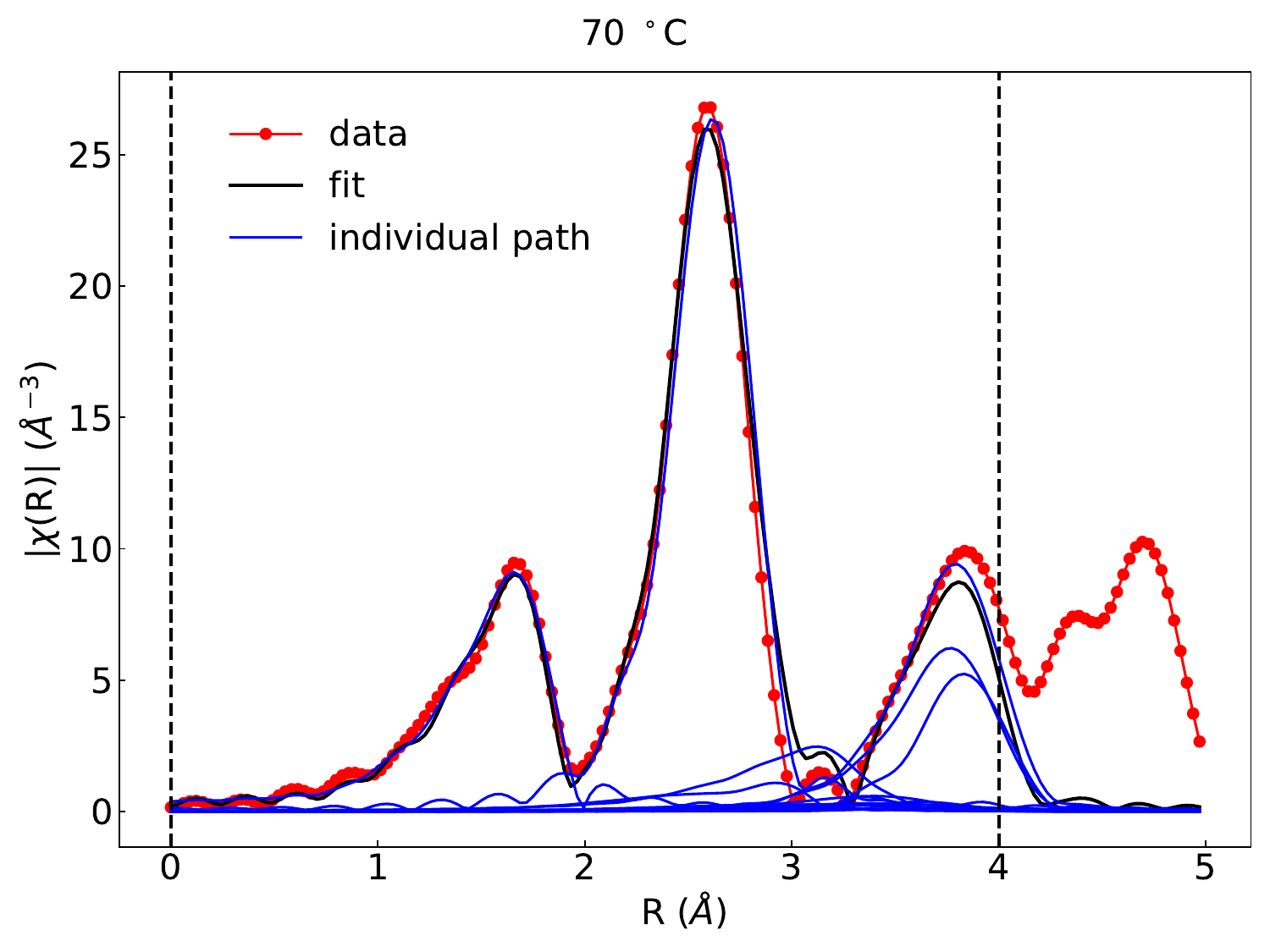}
    \includegraphics[width=0.24\linewidth]{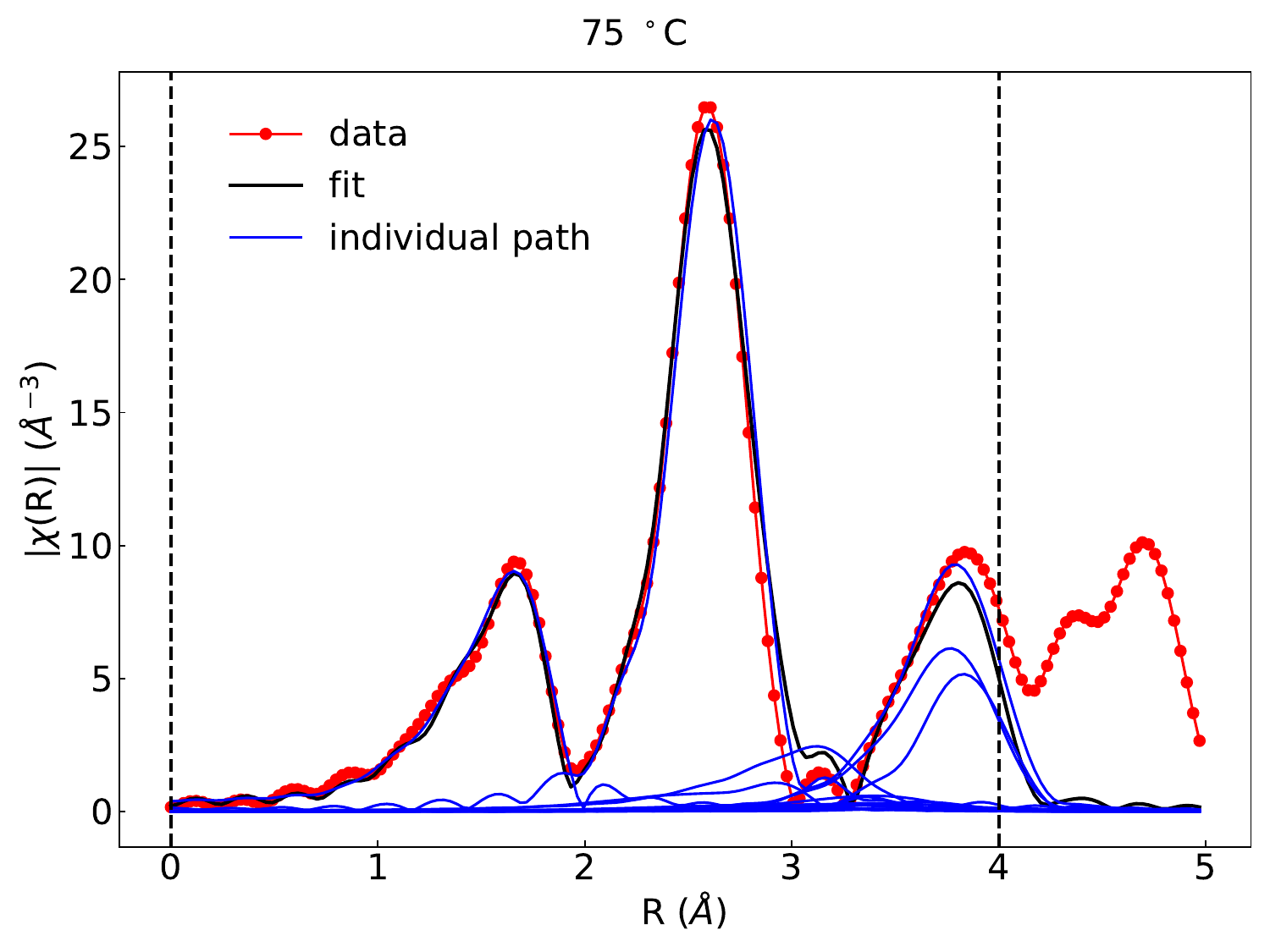}
    \includegraphics[width=0.24\linewidth]{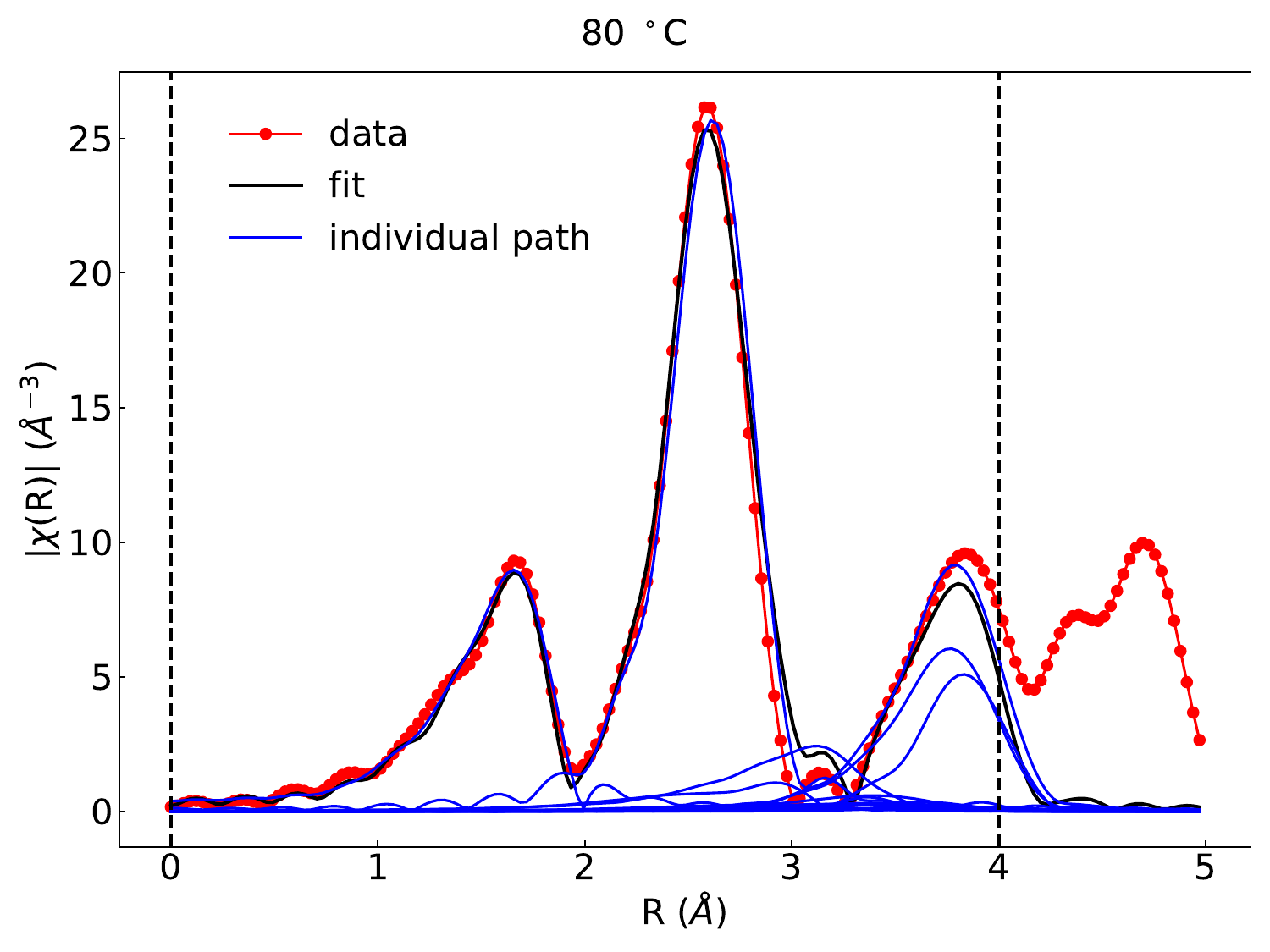}
    
    \includegraphics[width=0.24\linewidth]{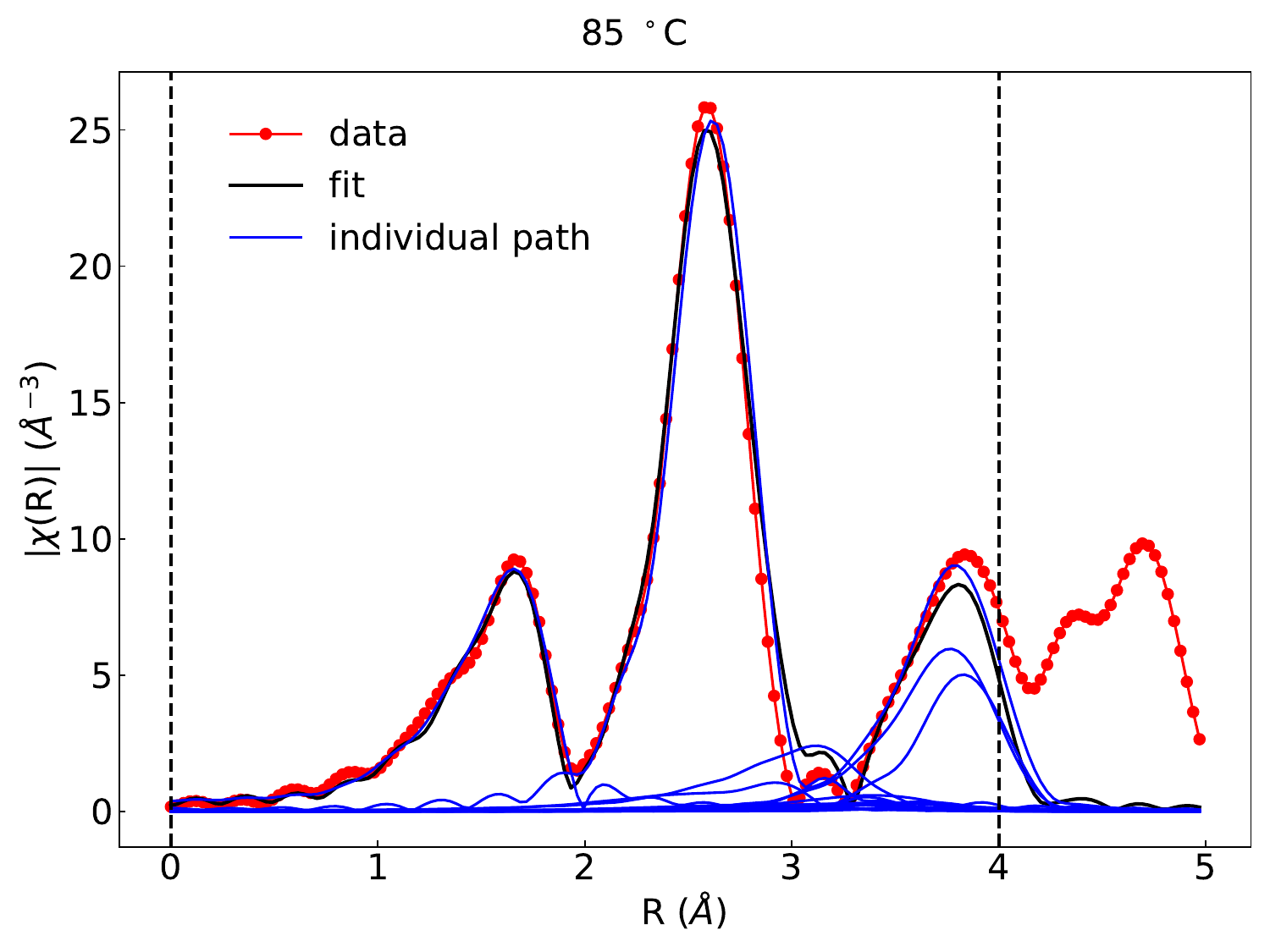}
    \includegraphics[width=0.24\linewidth]{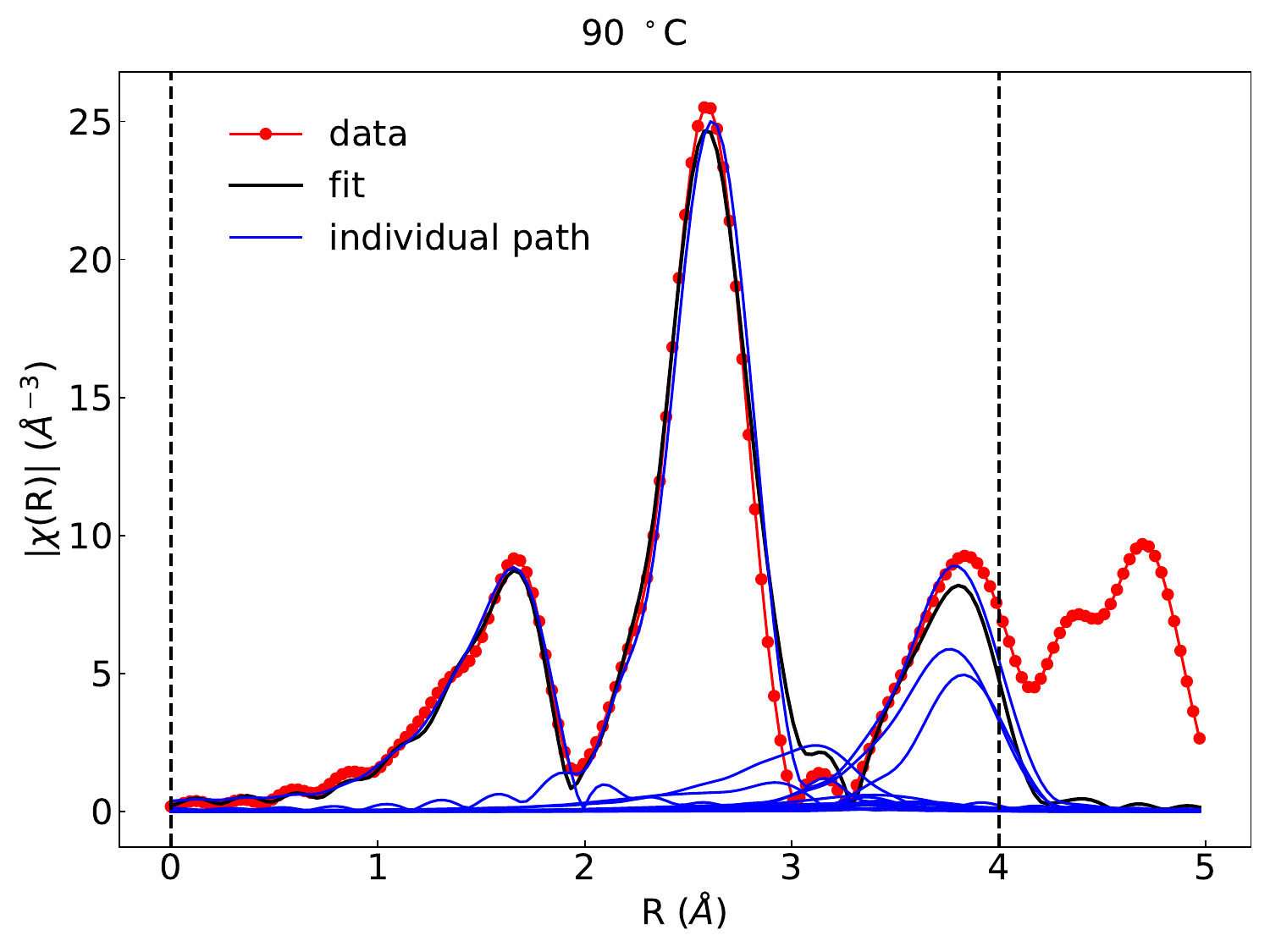}
    \includegraphics[width=0.24\linewidth]{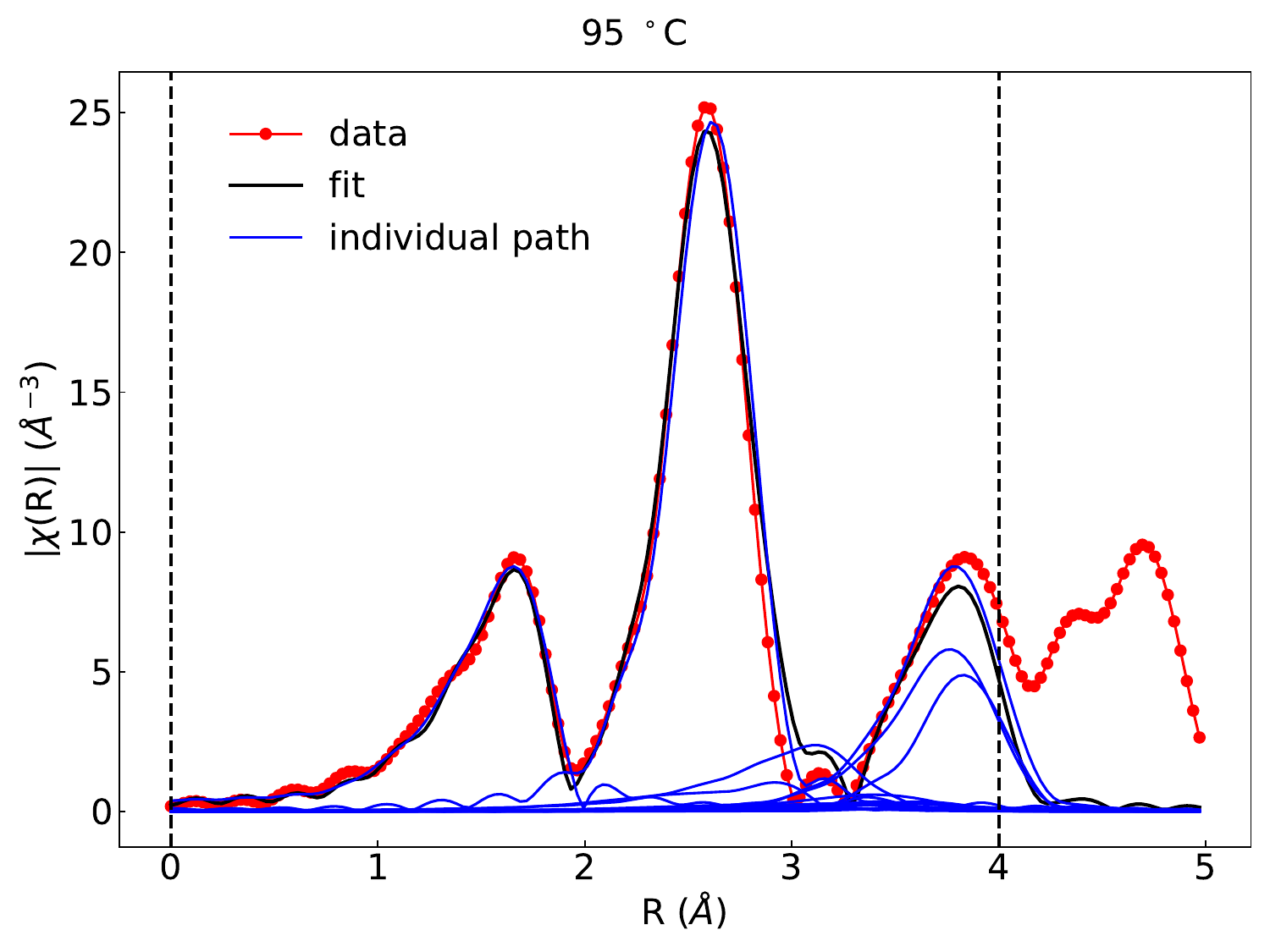}
    \includegraphics[width=0.24\linewidth]{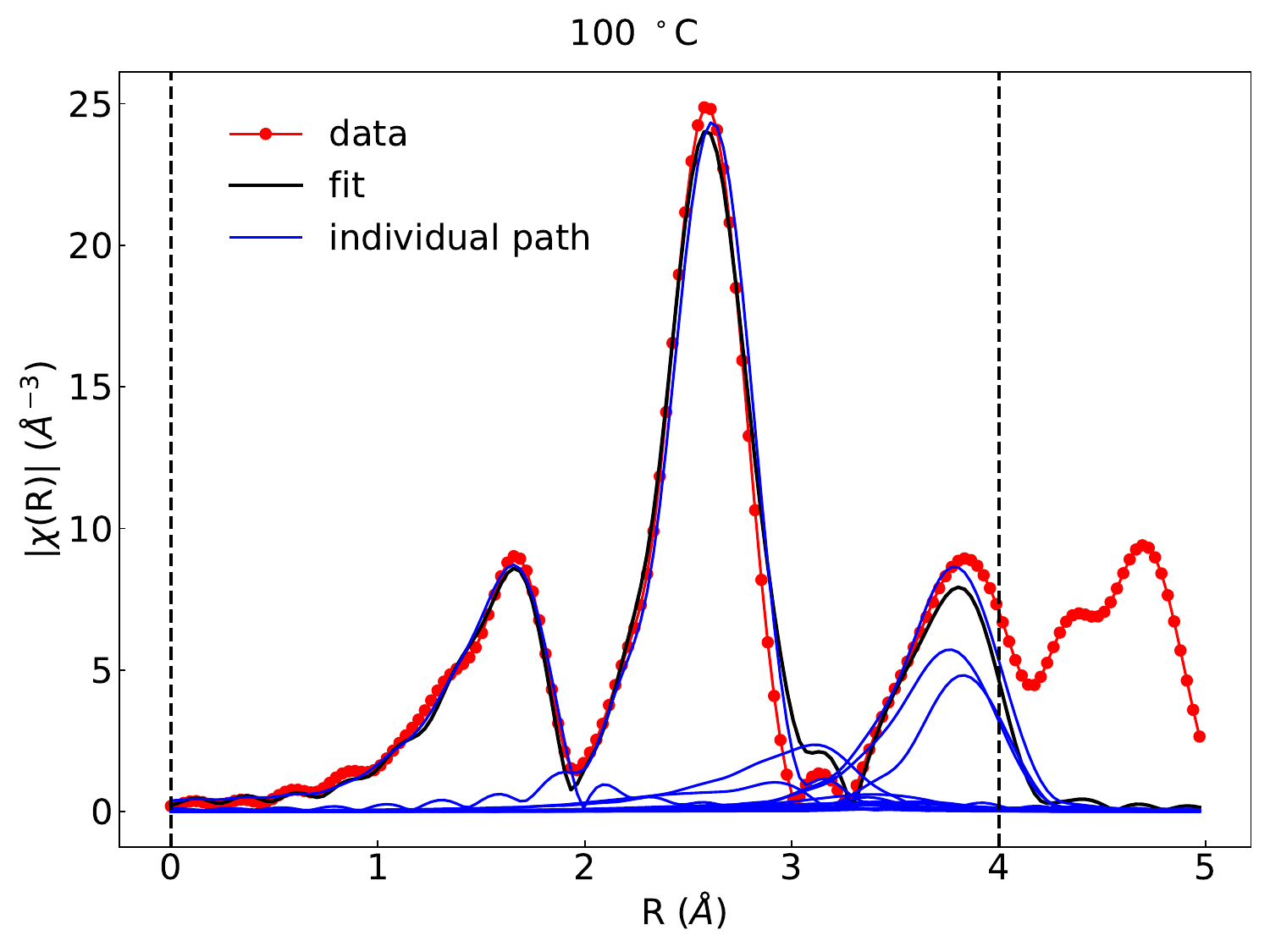}
    
    \includegraphics[width=0.24\linewidth]{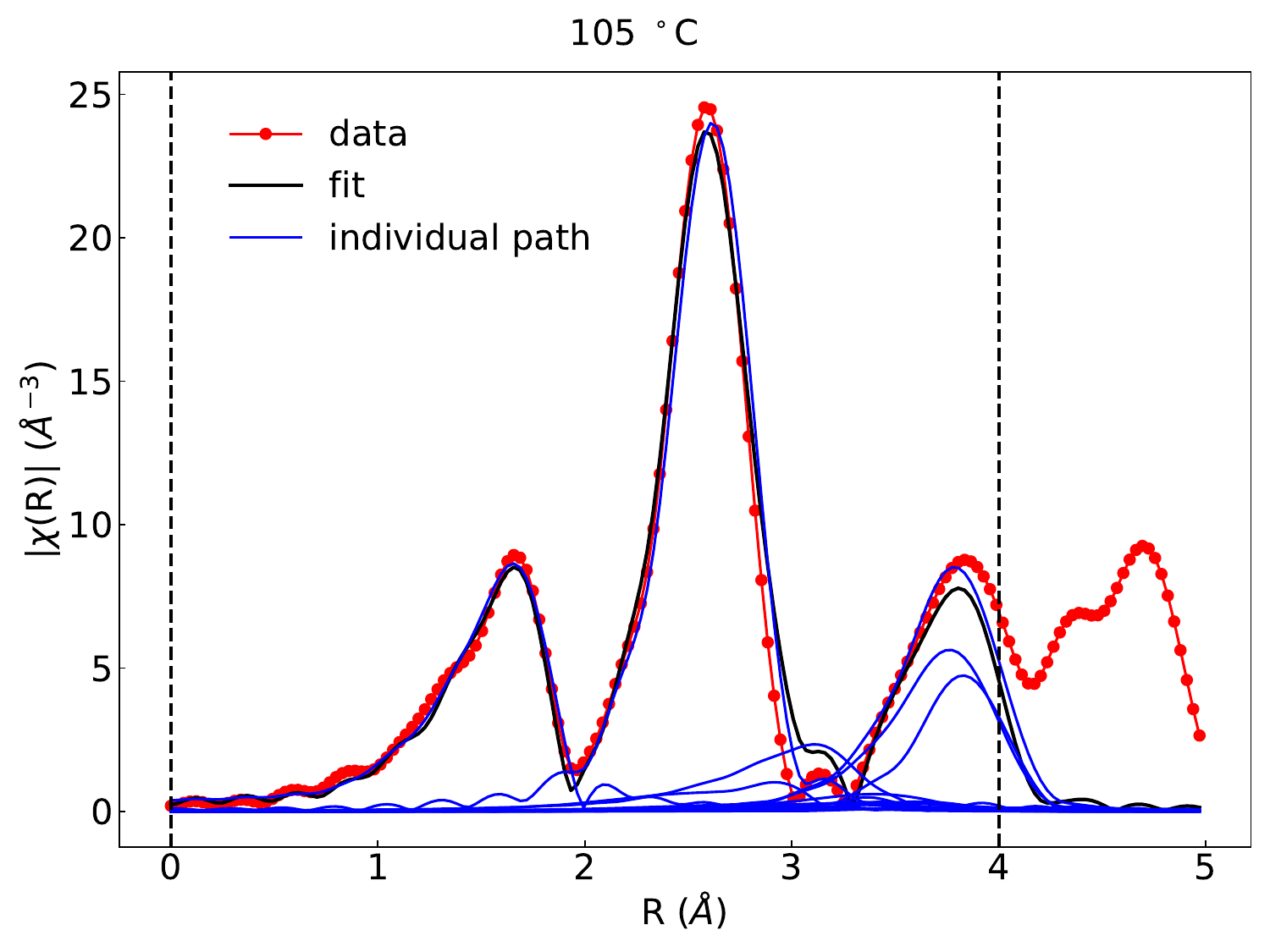}
    \includegraphics[width=0.24\linewidth]{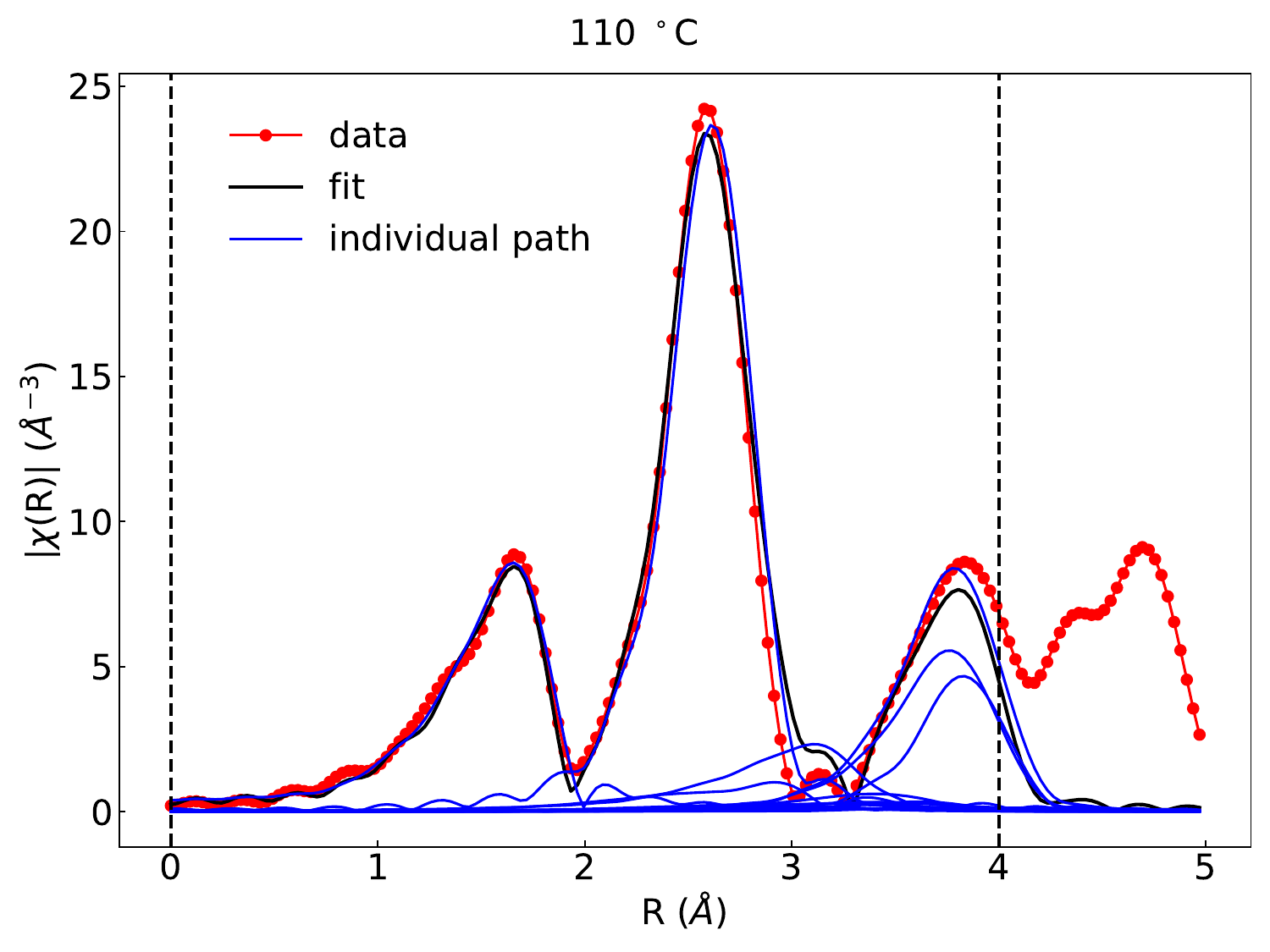}
    \includegraphics[width=0.24\linewidth]{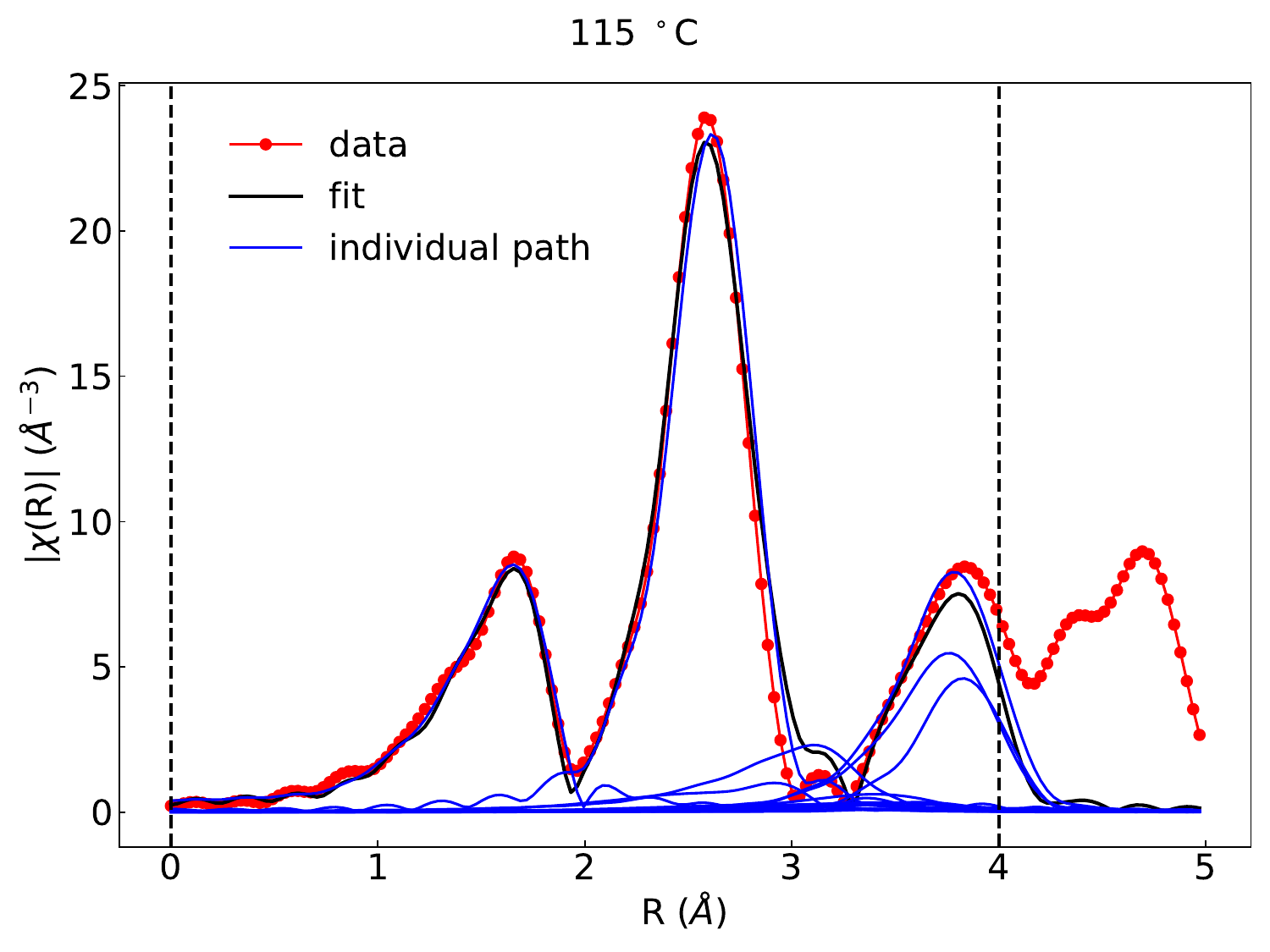}
    \includegraphics[width=0.24\linewidth]{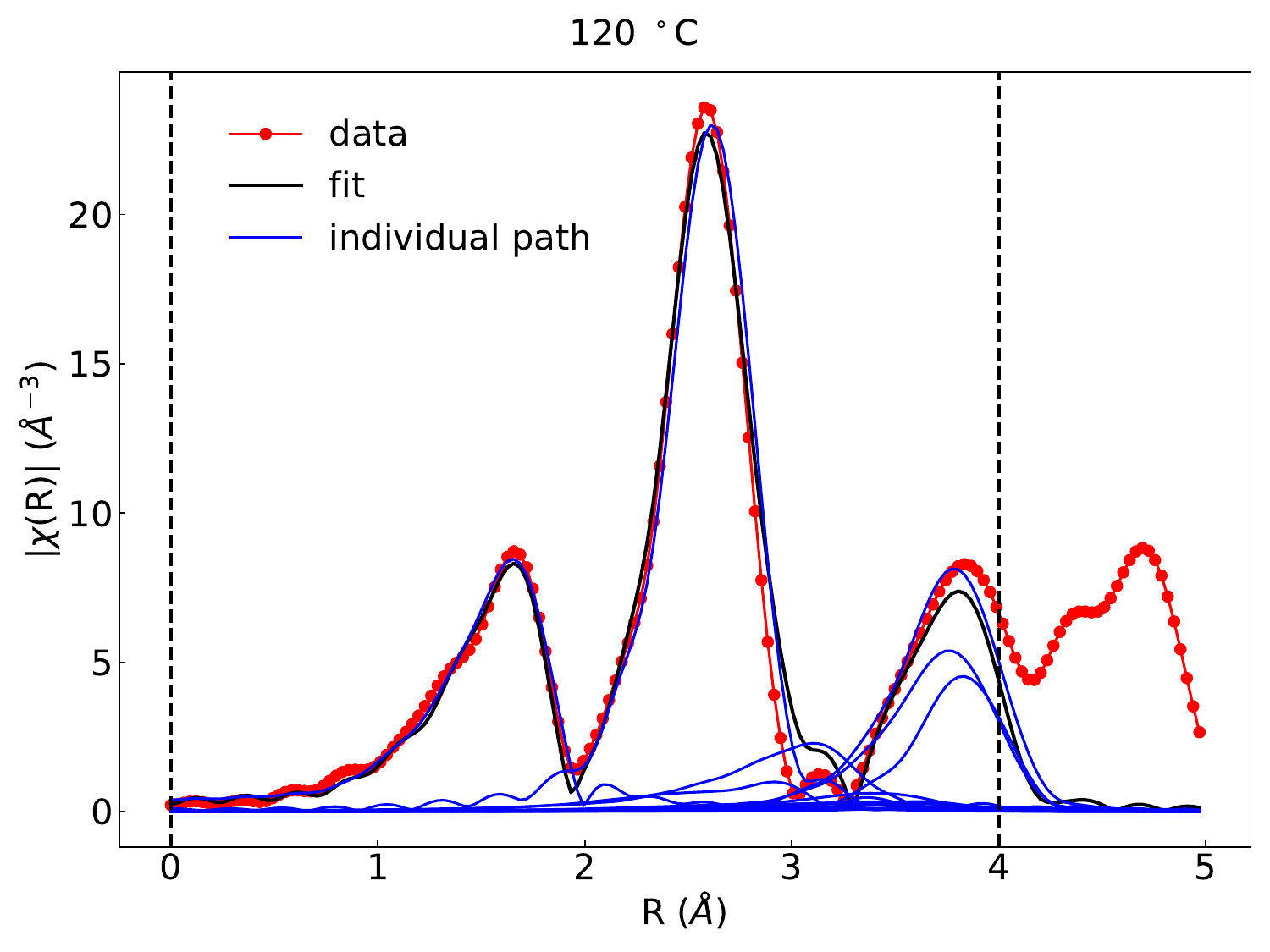}
    
    \includegraphics[width=0.24\linewidth]{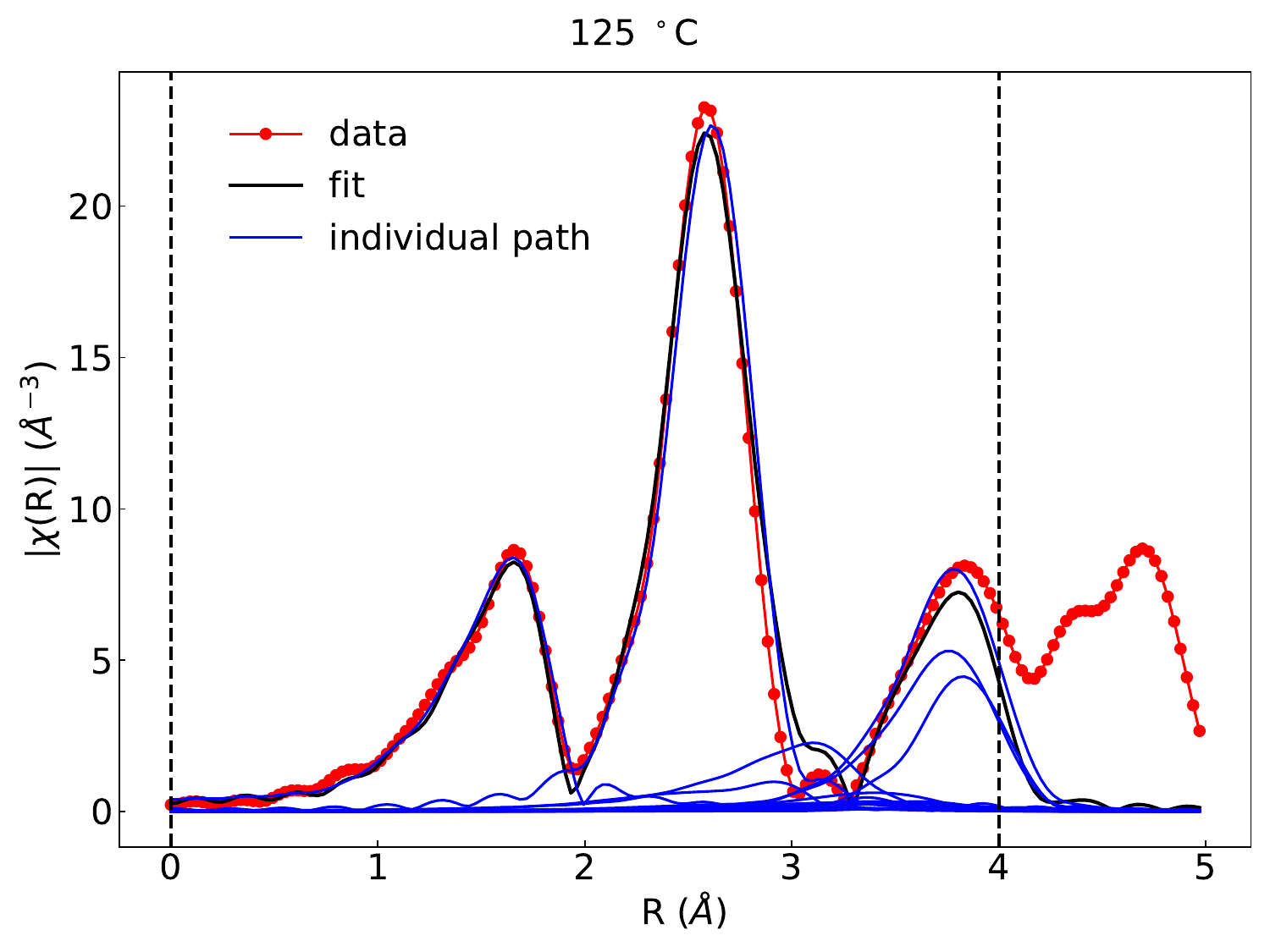}
    \includegraphics[width=0.24\linewidth]{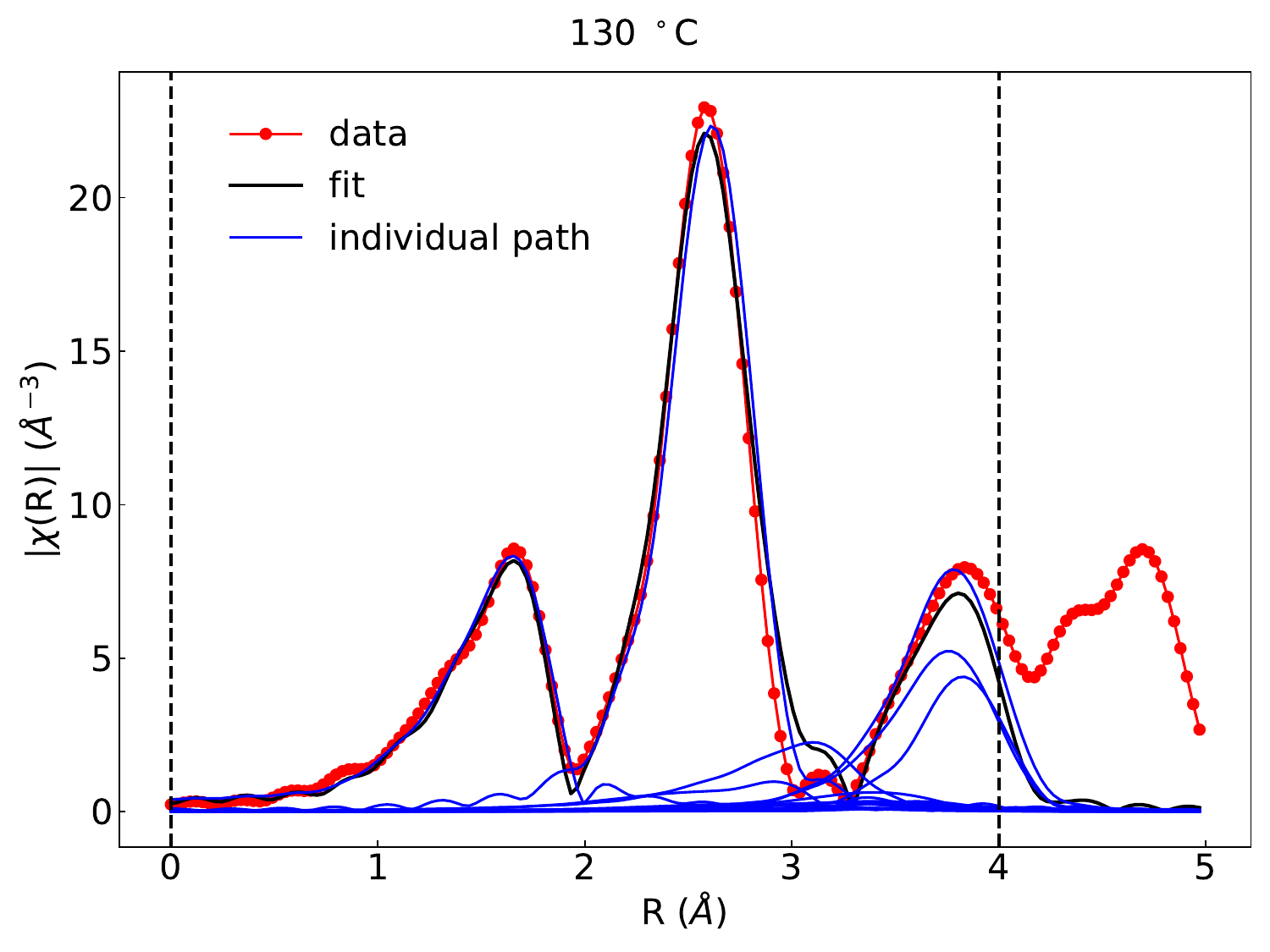}
    \includegraphics[width=0.24\linewidth]{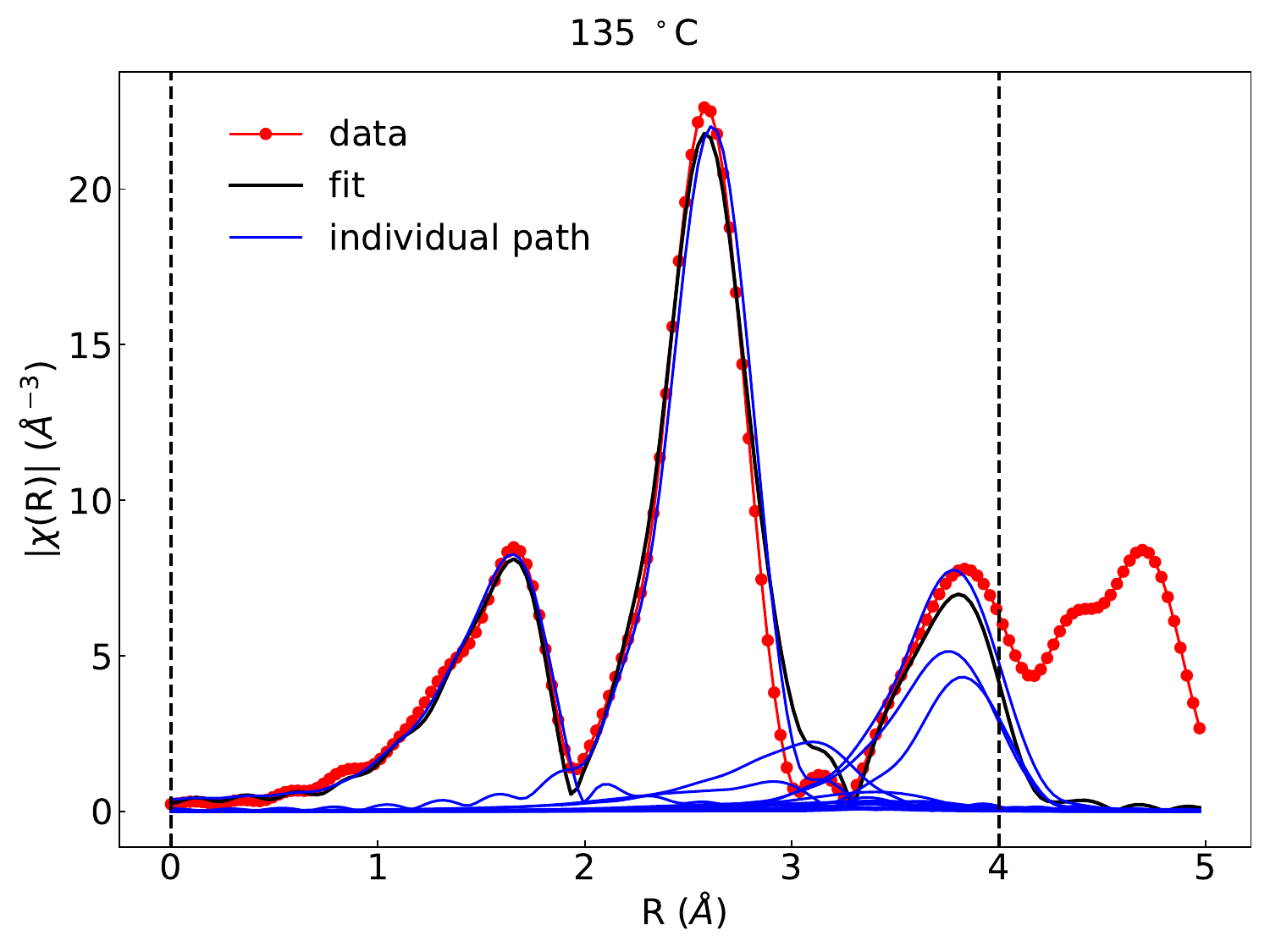}
    \includegraphics[width=0.24\linewidth]{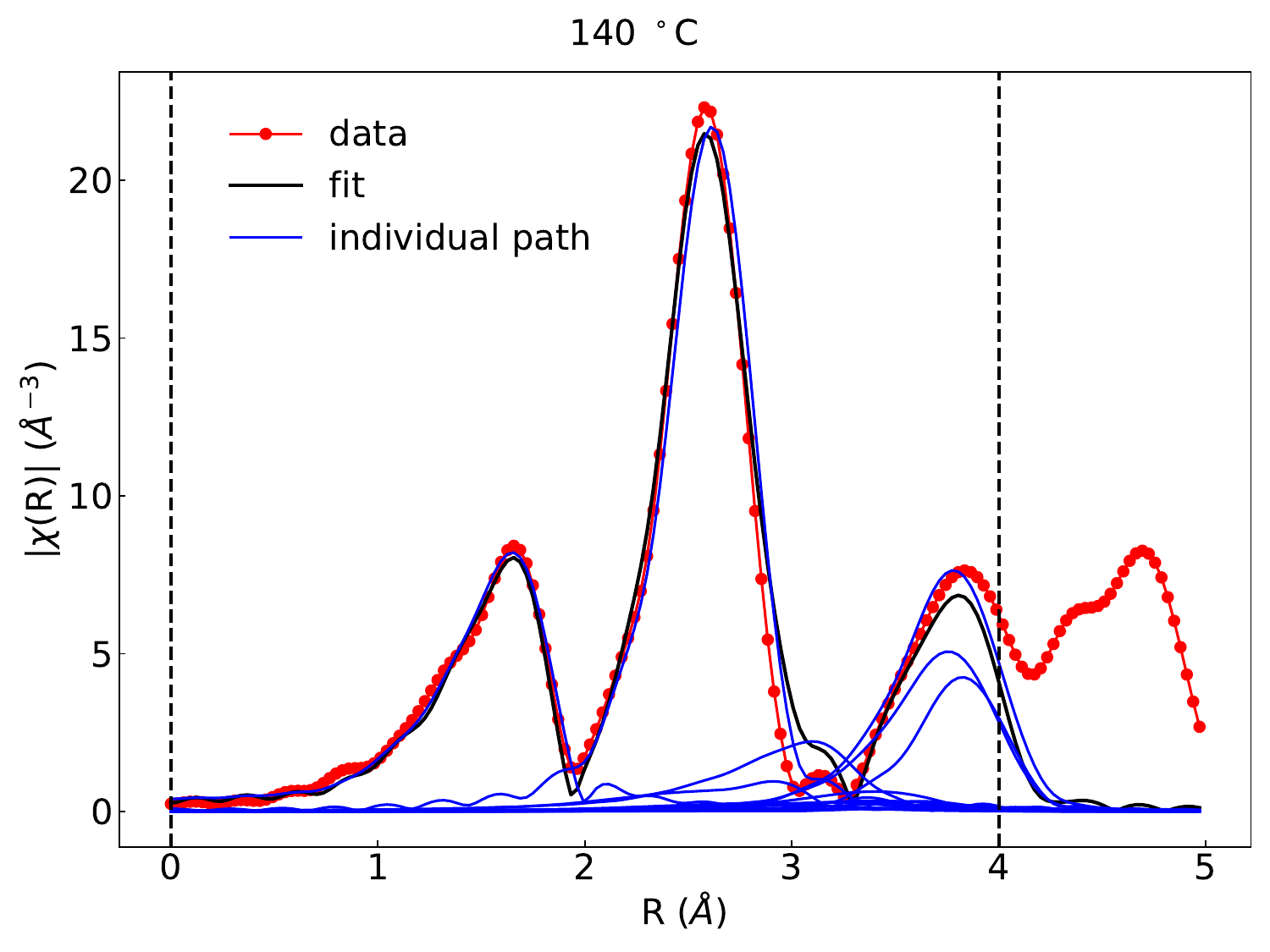}
    
    \includegraphics[width=0.24\linewidth]{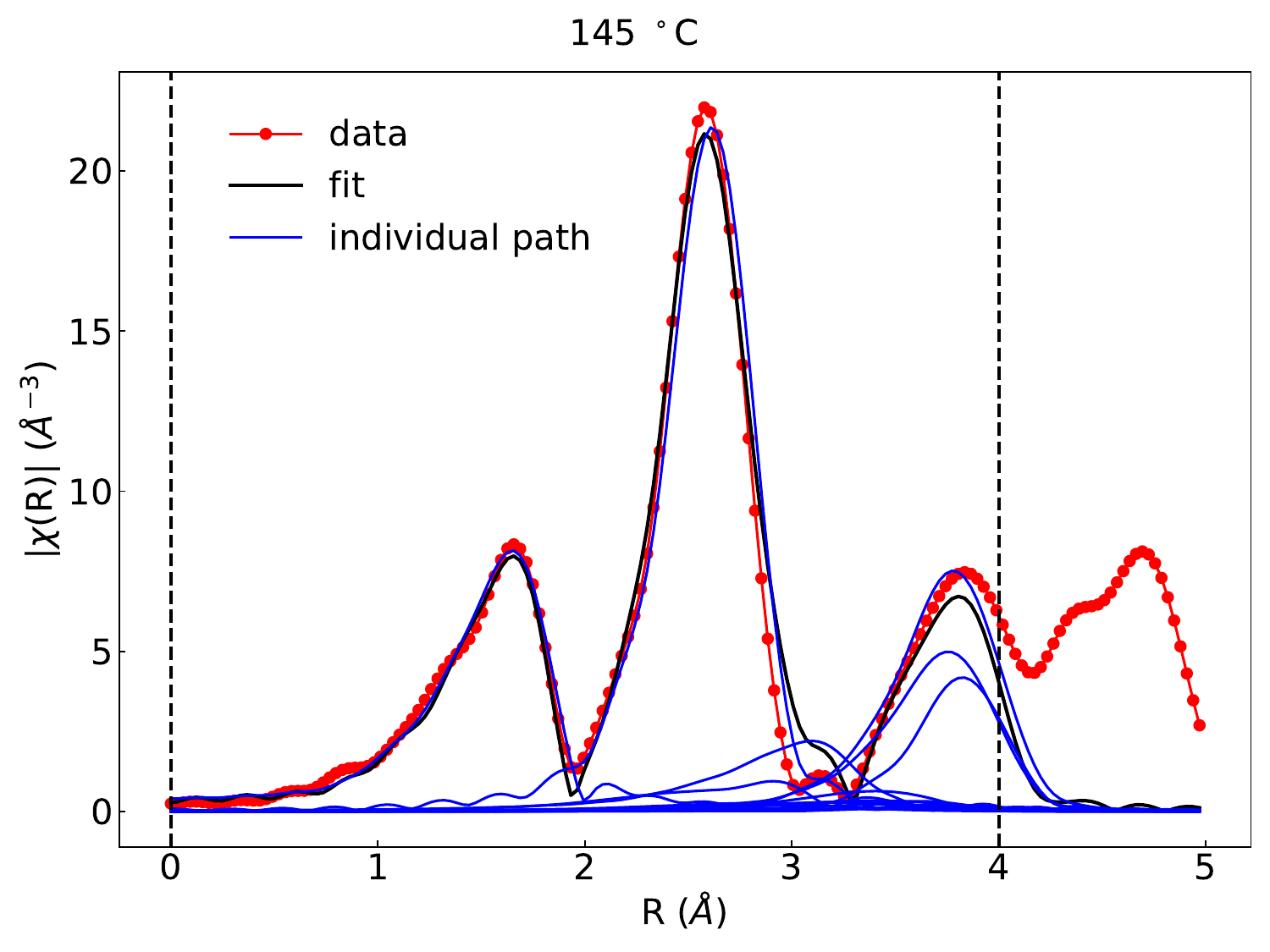}
    \includegraphics[width=0.24\linewidth]{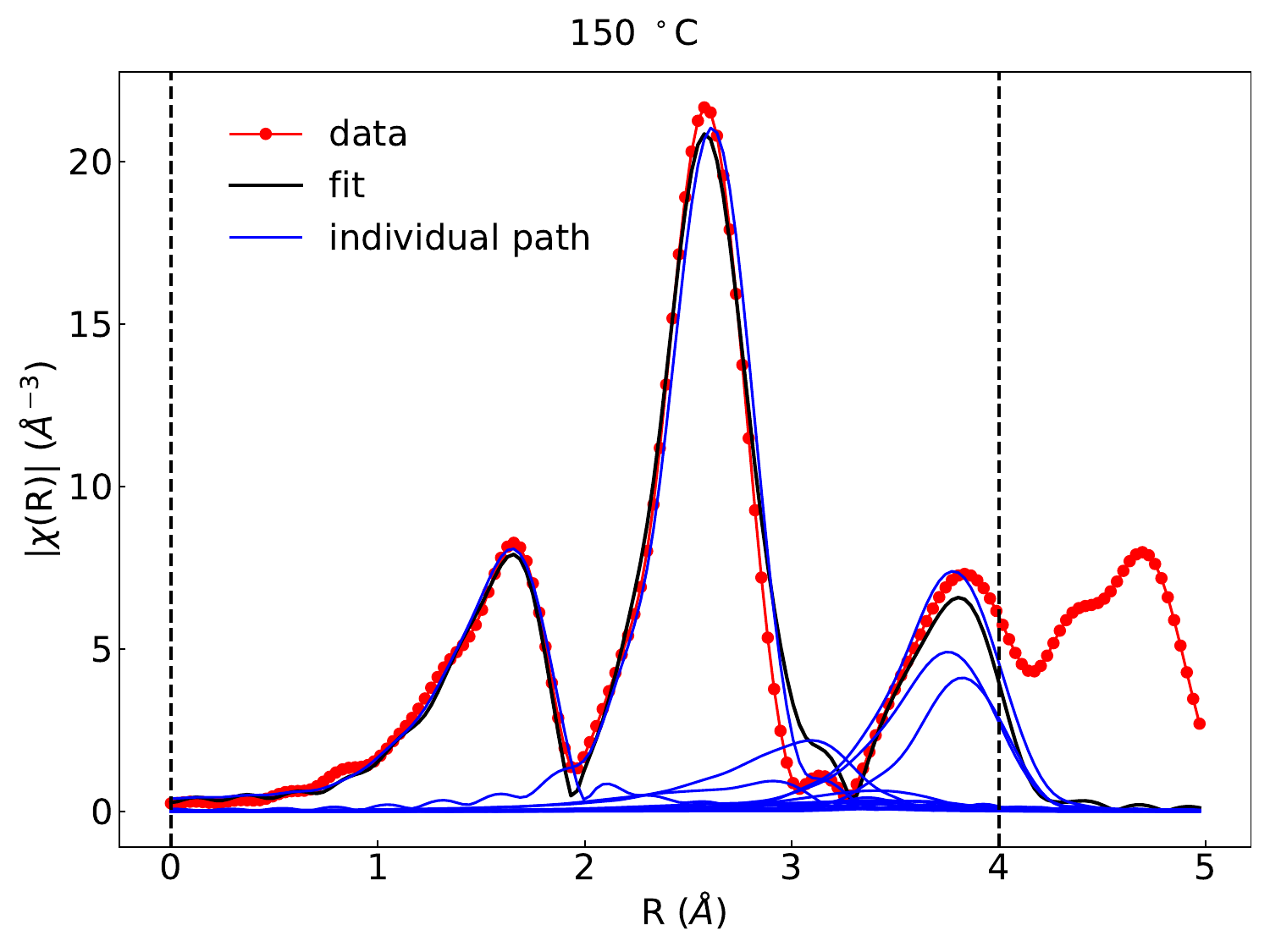}
    \vspace{-4mm}
    \caption{\textbf{Fitting of scattering amplitude in real space.} Fitting of the scattering amplitude including 4 scattering paths with mean free paths between \SI{0}{} and \SI{4}{\angstrom}.}
    \label{figSI:fit_radial_distribution}
\end{figure}

The evolution of the scattering amplitude as a function of lattice temperature is shown in Figure~\ref{figSI:radial_distribution_temperature_dependence}. The distribution is in good agreement with previously reported distributions from the Fourier transform of the EXAFS at the Ni K-edge of NiO~\cite{Anspoks:2011im,Anspoks:2012cx,Anspoks:2014hr}. Fitting results of the scattering amplitude in real space are given in Figure~\ref{figSI:fit_radial_distribution}. The fitting model was a global model in which scattering amplitudes over the full temperature range were fitted simultaneously. With this model, the fitted Einstein temperatures is $\theta_{E,O}=\SI{377\pm5}{\kelvin}$ and $\theta_{E,Ni}=\SI{389\pm2}{\kelvin}$ for the oxygen and nickel atoms, respectively. The lattice thermal expansion coefficient is $\SI{4.5\pm0.9e-5}{\angstrom\per\kelvin}$, which compares well with previously published values (\SI{4.444e-5}{\angstrom\per\kelvin}~\cite{PSrivastava:1977fv}) and the Ni--O bond distance at zero kelvin is $\SI{2.09\pm0.01}{\angstrom}$. The amplitude and energy offset values are $S_0^2=\SI{0.82\pm0.01}{}$ and $\Delta E_0=\SI{-1.8\pm0.1}{\electronvolt}$, respectively. The evolution of the Ni--O bond distance and the isotropic Debye-Waller factor as a function of the lattice temperature is shown in Figure~\ref{figSI:XAS_XRD_fitting_parameters} (blue lines). The results individual fittings at each temperature instead of a global fitting are also shown (blue circles).

\begin{table}
\centering
\begin{tabular}{ccccc}
\toprule
Scattering path index & path length (\AA) & scattering path & degeneracy & weight \\
\midrule
1 & 2.1083 & O (SS) & 6 & 100.00 \\
2 & 2.9816 & Ni (SS) & 12 & 87.05 \\
3 & 3.5991 & O-O (MS)& 24 & 13.91 \\
4 & 3.5991 & Ni-O (MS) & 48 & 19.79 \\
\bottomrule
\end{tabular}
\vspace{-2mm}
\caption{\textbf{List of photoelectron scattering paths.} Characteristics of the scattering paths included in the fitting of $|\chi(R)|$ with lattice temperature. SS: single scattering, MS: multiple scattering.}
\label{tab:scattering_paths}
\end{table}

\subsection{X-ray diffraction}

% Description of the in-situ XRD setup
\emph{In situ} temperature-dependent X-ray diffraction (XRD) was performed at the P02.1 beamline of the Petra III synchrotron in Hamburg, Germany (DESY). The details of the setup have been described elsewhere~\cite{Bischoff2024:55910}. Briefly, around \SI{1}{\milli\gram} of the crushed NiO powder was introduced inside an amorphous quartz capillary with \SI{1}{\milli\metre} internal diameter, packed in between quartz wool. The quartz capillay acts as a flow setup, often used as a fixed bed plug flow reactor for catalytic studies. An external and concentrical SiC tube acts as a furnace, heated with 2 IR lamps. A k-type thermocouple downstream the sample allows for temperature control. An inert gas flow is controlled with a set of mass flow controllers (Bronkhorst). The pressure in the cell is regulated using a pressure regulator system (Equilibar and pressure regulator from Bronkhorst). The incident X-ray wavelength was \SI{0.207327}{\angstrom} ($\sim\SI{59.8}{\kilo\electronvolt}$). 

\begin{figure}
    \centering
    \includegraphics[width=0.8\linewidth]{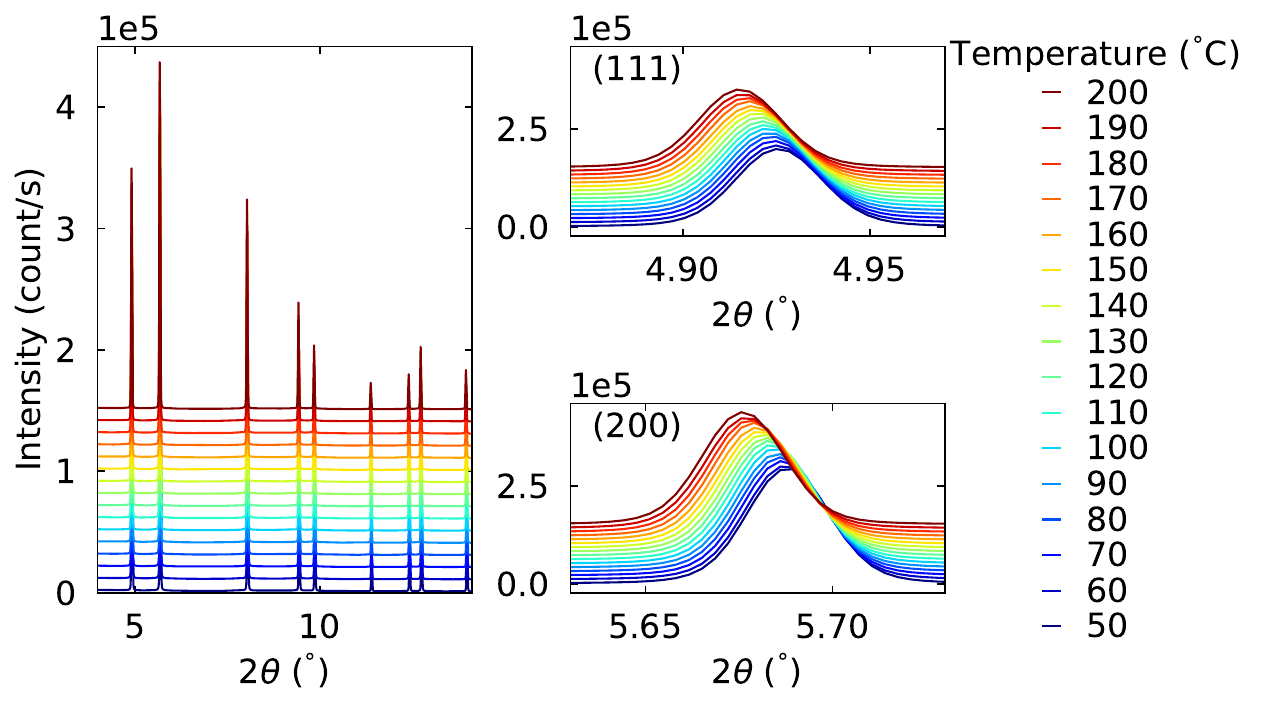}
    \vspace{-4mm}
    \caption{\textbf{Temperature-dependent X-ray diffraction (XRD).} XRD pattern with lattice temperature between \SI{50}{} (purple) and \SI{200}{\degreeCelsius} (red). Insets: magnifications over the (111) and (200) Bragg reflections.}
    \label{fig:XRD_temperature}
\end{figure}

\begin{figure}
    \centering
    \includegraphics[width=\linewidth]{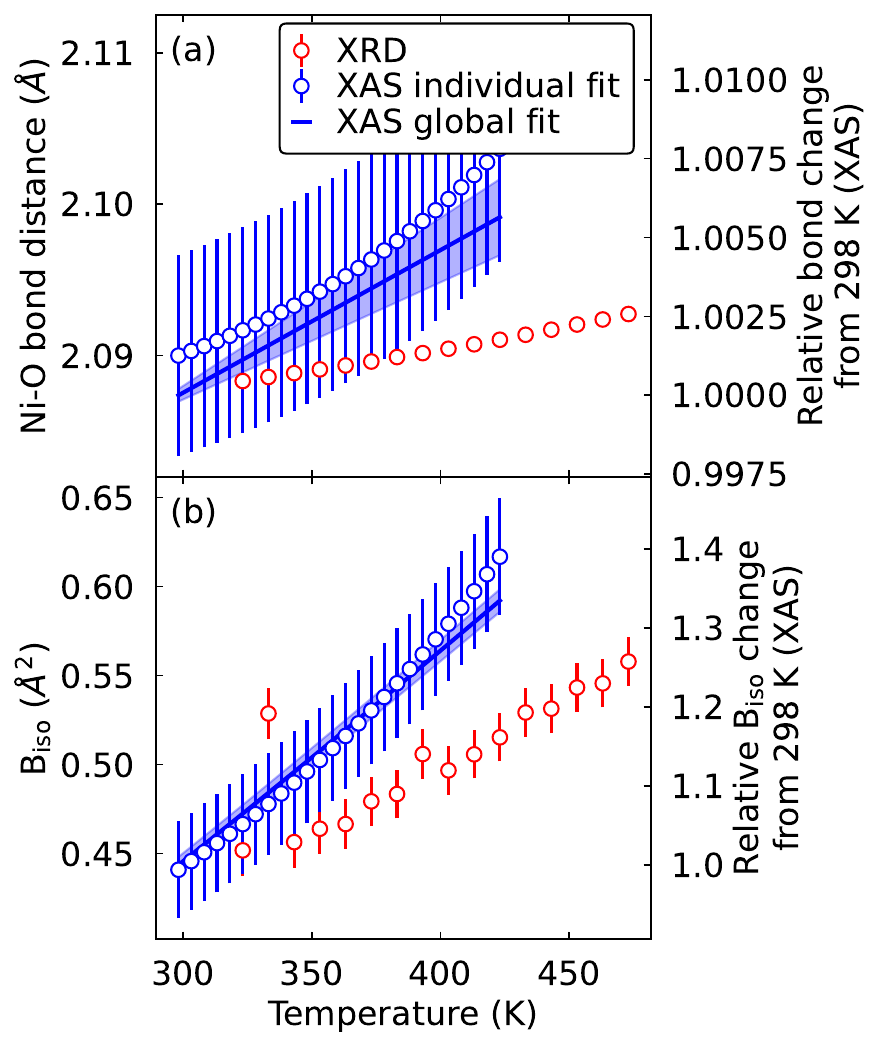}
    \vspace{-4mm}
    \caption{\textbf{Temperature-dependent bond distances and Debye-Waller factor.} Evolution of (a) the Ni--O bond distance, and (b) Debye-Waller factors measured by XRD (red circles) and XAS (blue circles, first coordination shell). Error bars are standard errors computed from the fittings, represented either as vertical sticks or shaded areas.}
    \label{figSI:XAS_XRD_fitting_parameters}
\end{figure}

% Description of the XRD setup
Figure~\ref{fig:XRD_temperature} shows the evolution of the XRD pattern with the lattice temperature between \SI{50}{} and \SI{200}{\degreeCelsius}. A clear shift to lower scattering angles is observed as the temperature increases, due to the thermal expansion of the lattice. Analysis of the lattice parameter and Debye-Waller factors was performed by Rietveld refinement with the \texttt{FullProf} software~\cite{Rodriguez-Carvajal1993:207992}. Background curves were generated with the automatic background algorithm in the WinPlotr module~\cite{Roisnel2001:140190} of FullProf (3 iterations with a background threshold at \SI{0.05}{}). The background function covering the entire angular range of the measurement was then generated from a cubic spline interpolation between the hereobtained background points. The "Special Polarization Correction (Synchrotron)" was specified to allow for polarization correction of the diffraction peak intensities. The diffraction peaks were fitted with a Thompson-Cox-Hastings pseudo-Voigt function with axial divergence asymmetry. The Lorentz polarization correction was applied to account the polarization of the incident X-rays with $K=\SI{0.1}{}$ and with the monochromator polarization correction $\cos^2(2\theta_{mono})=\SI{0.9855}{}$. The input structure for the Rietveld refinement was from bunsenite (space group F$_{m\overline{3}m}$ \#225). The nickel and oxygen atoms were assumed to occupy fixed positions during the fitting at (0.0000, 0.0000, 0.0000) and (0.5000, 0.5000, 0.5000) in units of fractional coordinates, respectively. A single free isotropic Debye-Waller factor was used for all atoms ($B_{iso}$). Parameters U and Y contributing to the broadening of the diffraction peaks were fixed to \SI{-0.003468}{\degree} and \SI{0.032373} respectively, given by the instrument response, established from the scattering of a known compound. Other parameters for the diffraction peaks were fitted through the $V$ parameter (contributing as $V\tan\theta$ to the squared Gaussian FWHM) and $X$ for the Lorentzian strain parameter (contributing as $X\tan\theta$ to the Lorentzian FWHM). Lattice parameters were constrained to $a=b=c$ (refined during the fitting) and $\alpha=\beta=\gamma=\SI{90}{\degree}$ (fixed). The least-squares refinement algorithm was used for the fitting with reflection ordering calculated at each cycle. All other parameters not explicitly mentioned were left as default. Figure~\ref{figSI:XAS_XRD_fitting_parameters} displays the increase of the Ni-O bond distance (panel a) and $B_{iso}$ (panel b) with the lattice temperature. The slope of both the Ni-O bond distance and $B_{iso}$ evolution with the temperature are compatible with the results from XAS measurements at the Ni K-edge.

\cleardoublepage

%%%%%%%%%%%%%%%%%%%%%%%%%%%%%%%%%%%%%%%%%%%%%%%%%%%%%%%%%%%%%%%%%
%%%%%%%%%%%%%%%%%%%%%%%%%%%%%%%%%%%%%%%%%%%%%%%%%%%%%%%%%%%%%%%%%
%%%%%%%%%%%%%%%%%%%%%%%%%%%%%%%%%%%%%%%%%%%%%%%%%%%%%%%%%%%%%%%%%
%%%%%%%%%%%%%%%%%%%%%%%%%%%%%%%%%%%%%%%%%%%%%%%%%%%%%%%%%%%%%%%%%

\section{Calculation of the excitation density\label{secSI:excitation_density}}

\subsection{Excitation density under continuous UV irradiation\label{subsecSI:excitation_density_continuous}}

For the calculation of the excitation density under continuous (\emph{in situ}) UV excitation at quasi-equilibrium, the calculation is in multiple steps.

\begin{enumerate}
	\item We account for the intensity of the light source filtered in the UV. The average power in this spectral region of the excitation spectrum is $\sim\SI{11}{\percent}$ of the unfiltered spectral power, based on the ratio between the integral of the spectral density filtered in the UV and the spectral density of the unfiltered lamp, which amounts to $P=\SI{137}{\milli\watt}$ (the nominal spectral power of the lamp is \SI{1.3}{\watt}).
	\item We calculate the spectral density $I(h\nu)$ (in W/m$^2$/eV) by solving the equation,
	\begin{equation}
	P=\int \frac{I(h\nu)}{A}d(h\nu),
	\end{equation}
	with $A$ the beam area (the beam waist is \SI{3}{\milli\metre} in diameter).
	\item We calculate the effective NiO optical path of the powder in the mixture with the silicon nitride powder. The NiO pellet is prepared to achieve \SI{6}{\percent} of absorption on top of the scattering background, which corresponds to an effective thickness of NiO of $L=\SI{450}{\nano\meter}$, taking into account the scattering background (source: CXRO database). The sample thickness is larger than the optical penetration depth of the photons in the region of the CT gap, meaning that we can consider that the photons are fully absorbed by NiO.
	\item We calculate the mean excitation rate, using the absorption coefficient $\alpha(h\nu)$ at the photon energies of the xenon lamp reported in reference~\cite{Newman:1959gr}. The mean excitation rate $<G>$ is given by,
	\begin{equation}
		<G>=\int \frac{I(h\nu)\times(1-e^{-\alpha(h\nu)L})}{h\nu\times e\times L}d(h\nu),
	\end{equation}
	yielding $<G>=\SI{5e22}{\per\cubic\centi\metre\per\second}$.
	\item We calculate a mean excitation density using as a lifetime of the photoexcited carrier the shortest time constant obtained from XTA measurements of $\tau=\SI{600}{\pico\second}$. The mean excitation density $\overline{n}$ at quasi-equilibrium is then given by,
	\begin{equation}
	\overline{n}=<G>\tau,
	\end{equation}
	giving an excitation density of \SI{3e13}{\per\cubic\centi\metre}.
\end{enumerate}
Hence, the excitation density is 7--8 orders of magnitude less under \emph{in situ} conditions at quasi-equilibrium than under laser pulse excitation in the pump-probe XTA measurements (Table~\ref{tabSI:excitation_density_100ps}). For reference, this excitation density is of the same order of magnitude as the quasiequilbrium carrier concentration under simulated solar illumination in lead halide perovskites~\cite{Wolff2021:301721}.

\subsection{Incident laser fluence in XTA}

The incident laser fluence $F$ is given by $F=P/(A\times R)$, where $P$ is the average laser power, $A$ the laser beam waist at the sample position (assuming $1/e^2$ beam diameter), and $R$ the laser repetition rate. Table~\ref{tabSI:initial_excitation_density} provides these experimental parameters for the four different excitation conditions reported throughout the manuscript.

\subsection{Initial excitation density in XTA}

The calculation of the initial excitation density is based on the incident laser fluence (Table~\ref{tabSI:initial_excitation_density}), the concentration of NiO microcrystals in the solution, and the optical path length through NiO in the liquid jet.

The molar concentration of NiO in the aqueous solution is $c=\SI{170}{\milli\mole\per\litre}$. With a molar mass of $M=\SI{74.69}{\gram\per\mole}$, this corresponds to a mass concentration of \SI{12.70}{\gram\per\litre}. Using the density of NiO, $\rho=\SI{6.67}{\gram\per\cubic\centi\metre}$, we obtain a total NiO volume of $V_{\text{NiO}}=\SI{1.90}{\cubic\centi\metre\per\litre}$ per unit volume of solution.

Assuming that the NiO particles are spherical and randomly distributed (homogeneous dispersion), the fraction of the optical path length within NiO is equal to the volume fraction, i.e., $l_{\text{NiO}}/l_{sol}=V_{\text{NiO}}/V_{sol}=0.0019$. For a liquid jet thickness of \SI{200}{\micro\metre}, this corresponds to an effective optical path length through NiO of \SI{0.38}{\micro\metre}.

We then apply the Lambert–Bouguer law to determine how much of the incident light is absorbed over this path length at the pump photon energy (\SI{3.49}{\electronvolt}), using an absorption coefficient of $\alpha=\SI{1.0e4}{\per\centi\metre}$~\cite{Newman:1959gr}. The resulting transmitted intensity ratio is \SI{0.68}{}.

The initial excitation density is calculated from the number of absorbed photons divided by the volume of NiO within the irradiated region. The irradiated solution volume is given by the product of the laser beam waist and the liquid jet thickness, and the corresponding NiO volume follows from the volume fraction determined above. The number of absorbed photons is obtained by dividing the pulse energy by the photon energy. The resulting excitation densities for the different experimental conditions are summarized in the last column of Table~\ref{tabSI:initial_excitation_density}.

\begin{table}[!ht]
    \centering
    \begin{tabular}{
        >{\centering\arraybackslash}p{2cm}
        >{\centering\arraybackslash}p{3.5cm}
        >{\centering\arraybackslash}p{3cm}
        >{\centering\arraybackslash}p{2cm}
        >{\centering\arraybackslash}p{3cm}
        }
    \toprule
    Average power (W) & laser beam diameter ($\mu$m, $1/e^2$) & laser fluence (mJ/cm$^2$) & pulse energy ($\mu$J) & excitation density (cm$^{-3}$) \\
    \midrule
    \SI{0.63}{} & $\SI{82\pm2}{}\times\SI{160\pm3}{\micro\metre}$ & \SI{24}{} & \SI{2.4}{} & \SI{3.5e20}{} \\
    \SI{1.24}{} & $\SI{212\pm3}{}\times\SI{93\pm2}{\micro\metre}$ & \SI{31}{} & \SI{4.8}{} & \SI{4.6e20}{} \\
    \SI{2.10}{} & $\SI{212\pm3}{}\times\SI{93\pm2}{\micro\metre}$ & \SI{52}{} & \SI{8.0}{} & \SI{7.8e20}{} \\
    \SI{3.35}{} & $\SI{212\pm3}{}\times\SI{93\pm2}{\micro\metre}$ & \SI{83}{} & \SI{13.0}{} & \SI{1.2e21}{} \\
    \bottomrule
    \end{tabular}
    \caption{\textbf{Initial excitation density in pump-probe XTA measurements of NiO.} The laser repetition rate is \SI{260}{\kilo\hertz} and the pump photon energy is $h\nu_{pump}=\SI{3.49}{\electronvolt}$.}
    \label{tabSI:initial_excitation_density}
\end{table}

\subsection{Excitation density at \SI{100}{\pico\second} in XTA}

For the calculation of the excitation density at the time delay of the pump-probe measurement (\SI{100}{\pico\second}), we rely on previously published carrier lifetimes through spontaneous recombination with a kinetic constant $k=\SI{0.006}{\per\pico\second}$~\cite{Biswas:2018ec}, hence a time constant $\tau=\SI{167}{\pico\second}$. The excitation density at \SI{100}{\pico\second} is given by $n_{\SI{100}{\pico\second}}=n_{t=0}\times e^{-\SI{100}{\pico\second}/\tau}$ with $n_{t=0}$ the initial excitation density given in the last column of Table \ref{tabSI:initial_excitation_density}. It gives the excitation density at \SI{100}{\pico\second} given in Table~\ref{tabSI:excitation_density_100ps}.

\begin{table}
	\centering
	\begin{tabular}{cc}
	\toprule
	Laser fluence (mJ/cm$^2$) & excitation density \SI{100}{\pico\second} (cm$^{-3}$) \\
	\midrule
	\SI{24}{} & \SI{1.9e20}{} \\
	\SI{31}{} & \SI{2.5e20}{} \\
	\SI{52}{} & \SI{4.3e20}{} \\
	\SI{68}{} & \SI{6.8e20}{} \\
	\bottomrule
	\end{tabular}
	\caption{\textbf{Excitation density at \SI{100}{\pico\second} time delay in pump-probe XTA measurements of NiO.} Calculated values of the excitation at the time delay of the XTA measurements assuming previously reported time constants of spontaneous carrier decay in photoexcited NiO~\cite{Biswas:2018ec}.}
	\label{tabSI:excitation_density_100ps}
\end{table}

%%%%%%%%%%%%%%%%%%%%%%%%%%%%%%%%%%%%%%%%%%%%%%%%%%%%%%%%%%%%%%%%%%%%%%%
%%%%%%%%%%%%%%%%%%%%%%%%%%%%%%%%%%%%%%%%%%%%%%%%%%%%%%%%%%%%%%%%%%%%%%%
%%%%%%%%%%%%%%%%%%%%%%%%%%%%%%%%%%%%%%%%%%%%%%%%%%%%%%%%%%%%%%%%%%%%%%%
%%%%%%%%%%%%%%%%%%%%%%%%%%%%%%%%%%%%%%%%%%%%%%%%%%%%%%%%%%%%%%%%%%%%%%%

\section{Pump-probe X-ray transient absorption (XTA) spectroscopy\label{secSI:XTA_setup}}

X-ray transient absorption (XTA) measurements were performed at the MicroXAS and SuperXAS beamlines of the Swiss Light Source (SLS). The setup has been extensively described in reference~\cite{Lima:2011dya}. The sample was made of NiO microparticles (325 mesh, Sigma-Aldrich), downsize filtered (see \S\ref{secSI:material_characterization}), and suspended in MilliQ water. Particle solutions were sonicated for half an hour sonication before XTA measurements. The solution was vigorously steered during the measurement to prevent agglomeration and deposition of NiO particles. The solutions were prepared at a concentration of \SI{170}{\milli\mol\per\litre} of NiO stoichiometric units per litre of solvent.

\begin{figure}
    \centering
    \includegraphics[width=\linewidth]{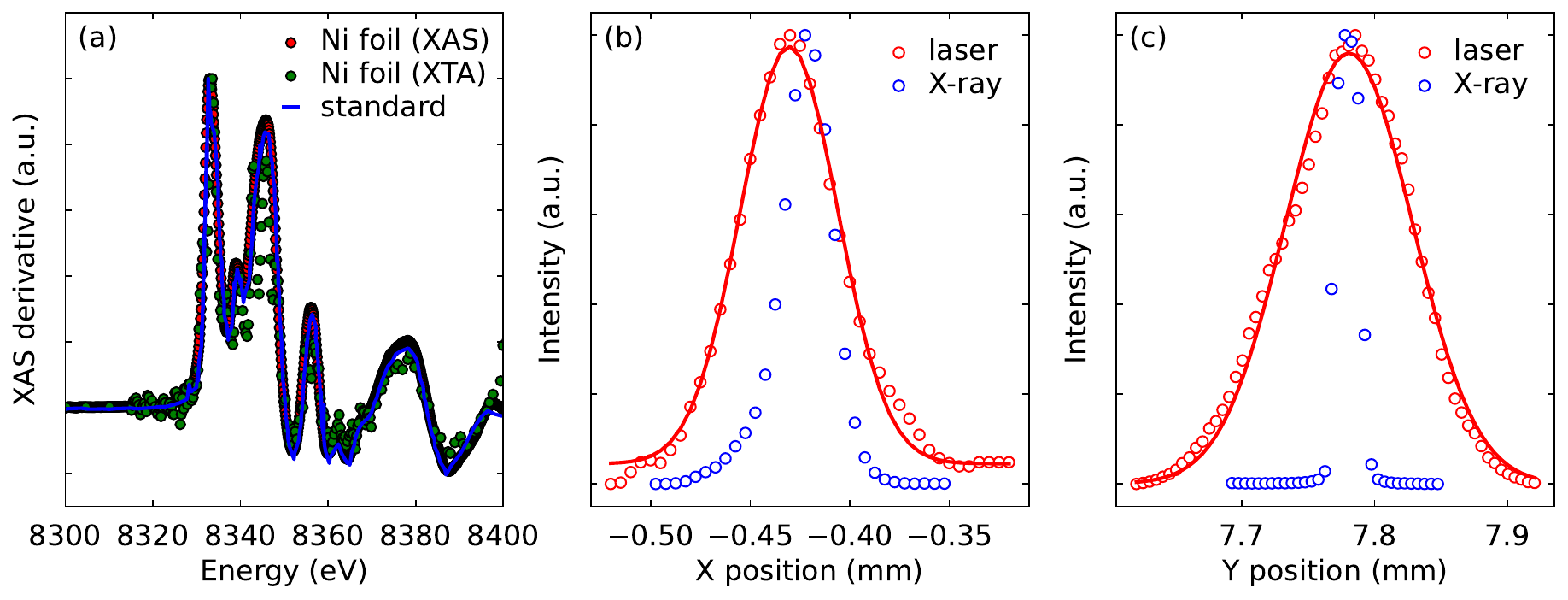}
    \vspace{-9mm}
    \caption{\textbf{XAS energy calibration and laser/X-ray beam dimensions.} (a) First derivative of XAS spectra at the Ni K-edge in the EXAFS standard database (blue circle) and from the measurements of a nickel metal foil in transmission during \emph{in situ} XAS (red circles) and pump-probe XTA measurements (green circles). (b) Horizontal and (c) vertical X-ray (blue circles) and laser (red circles) beam dimensions at the spatial overlap. Continuous red curve is a gaussian fit to the laser beam profile.}
    \label{figSI:XAS_energy_calibration}
\end{figure}

The energy calibration was performed by measuring the XAS spectrum in transmission of a \SI{10}{\micro\metre} thin Ni foil with a Si diode coupled to a variable gain current amplifier (DHPCA-100 from FEMTO). Figure \ref{figSI:XAS_energy_calibration}a shows that the first derivative of the Ni K-edge XAS spectrum of the foil reproduces the EXAFS standard in the near-edge region.

In brief, the sample was excited at a repetition rate of \SI{260}{\kilo\hertz} with a laser spot diameter of \SI{32}{\micro\metre} (obtained from the fitting of the laser beam profile through a \SI{50}{\micro\metre} pinhole in Figure \ref{figSI:XAS_energy_calibration}b,c). The particles were flown though a liquid jet made of a sapphire nozzle with a thickness of \SI{200}{\micro\metre}. A summary of the pump excitation conditions for 4 experiments at increasing excitation fluences are provided in Table~\ref{tabSI:initial_excitation_density}.

\cleardoublepage

%%%%%%%%%%%%%%%%%%%%%%%%%%%%%%%%%%%%%%%%%%%%%%%%%%%%%%%%%%%%%%%%%
%%%%%%%%%%%%%%%%%%%%%%%%%%%%%%%%%%%%%%%%%%%%%%%%%%%%%%%%%%%%%%%%%
%%%%%%%%%%%%%%%%%%%%%%%%%%%%%%%%%%%%%%%%%%%%%%%%%%%%%%%%%%%%%%%%%
%%%%%%%%%%%%%%%%%%%%%%%%%%%%%%%%%%%%%%%%%%%%%%%%%%%%%%%%%%%%%%%%%

\section{Excited-state lattice heating\label{secSI:lattice_heating_simulation}}

\subsection{From temperature-dependent XAS spectra\label{subsecSI:simulation_temperature_contribution_XTA}}

\begin{figure}[!ht]
    \centering
    \includegraphics[width=\linewidth]{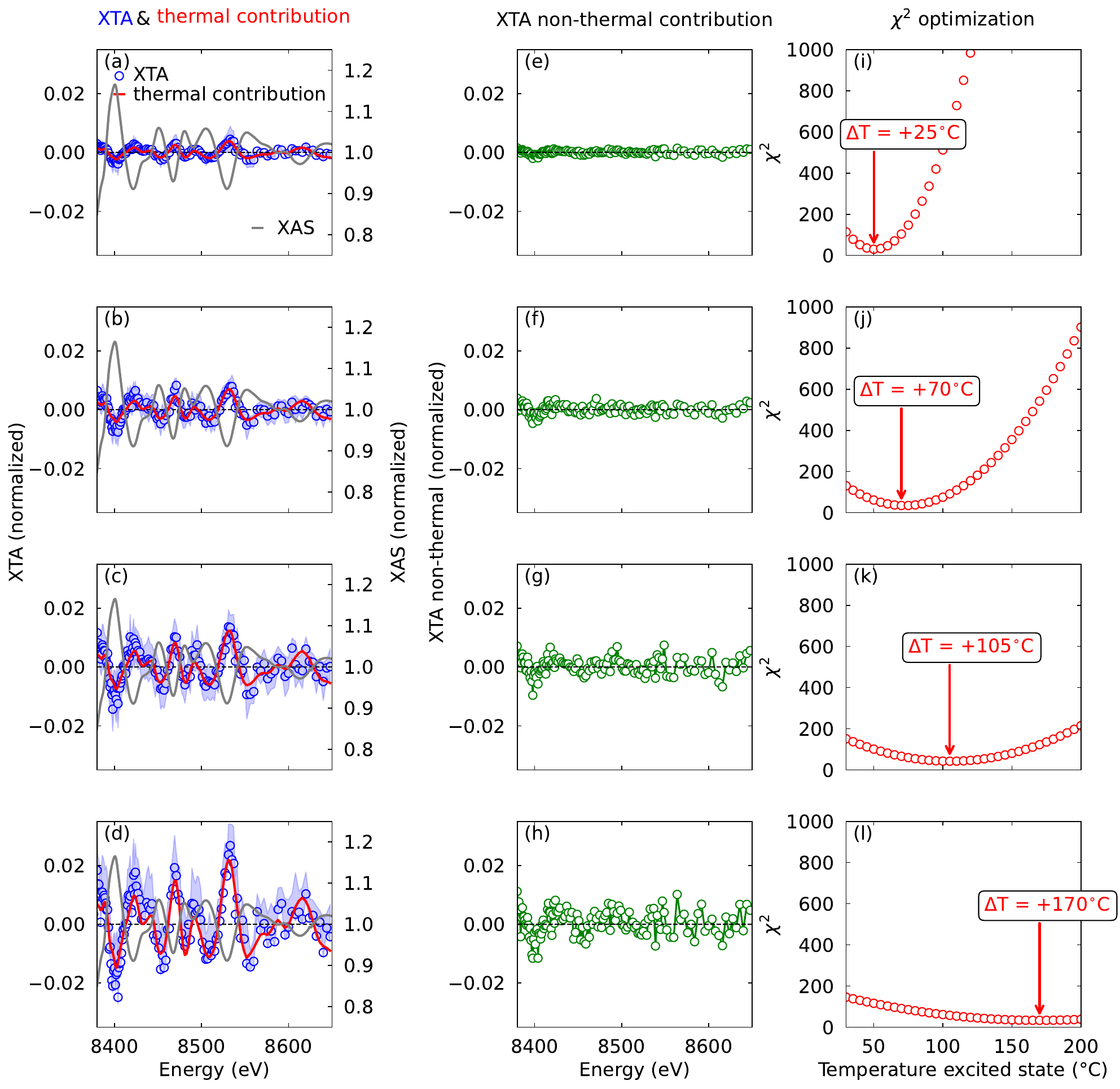}
    \caption{\textbf{Simulated lattice heating contribution to XTA spectra.} (a, b, c, d) Results of $\chi^2$-minimization between temperature-dependence difference XAS spectra (red curve) and XTA spectra (blue circles). The equilibirum XAS spectrum is shown for reference (grey curve). (e, f, g, h) Residuals between XTA spectra and best-matched difference XAS spectra (green circles). (i, j, k, l) $\chi^2$ value as a function of the lattice temperature in the excited state. The minimum of the $\chi^2$ distribution and the corresponding lattice temperature is shown with a vertical red arrow. The four rows of plots correspond to increasing excitation fluences: (first row) \SI{24}{\milli\joule\per\centi\metre\squared}, (second row) \SI{31}{\milli\joule\per\centi\metre\squared}, (third row) \SI{52}{\milli\joule\per\centi\metre\squared}, and (fourth row) \SI{68}{\milli\joule\per\centi\metre\squared}.}
    \label{figSI:simulation_temperature_contribution_XTA}
\end{figure}

Temperature-dependence of XAS difference spectra were used to model lattice heating contributions to XTA spectra. The procedure was previously described in references~\cite{Rossi2021,Rossi2025:52718}. Figure \ref{figSI:simulation_temperature_contribution_XTA} displays the results of a $\chi^2$-minimization between the XAS difference spectra and XTA spectra at increasing excitation fluences. The XTA spectra are shown in the first column (blue circles), together with the best fit (red curve). The residuals of the fitting are shown in the central column (green circles), showing that no pump-probe signal remains in the EXAFS after the subtraction of the lattice heating effect. The last column shows the evolution of $\chi^2$ as a function of the lattice temperature. It shows decreasing curvature for increasing excitation fluences (from top to bottom), meaning that the determination of the lattice temperature has a larger uncertainty with the excitation fluence. We assign this effect to potentially larger distribution of temperatures in the probed sample with increasing excitation density, which affects the accuracy of the model in which a single temperature is assumed to represent the hot lattice.

\subsection{Estimated heat jump under laser pulse excitation\label{subsecSI:simulation_temperature_jump}}

We evaluate the expected lattice heating upon absorption of the laser pump pulse. We assume that the pump excess energy is fully transferred to the lattice at \SI{100}{\pico\second} time delay and that heat diffusion is negligible due to the short time delay with respect to the heat diffusion time of \SI{3}{}-\SI{11}{\micro\second} (see section 8.1 of reference~\cite{Rossi2025:52718}). We compute the absorbed pump energy $E_{abs}$ as $E_{abs}=E_{pump}(1-e^{-t\lambda_p})$ with $E_{pump}$ the incident pump pulse energy and $\lambda_p$ the penetration depth at the pump photon energy ($h\nu_p$). The energy converted into heat is the excess energy given by the carriers, which is given by $E_{excess}=E_{abs}(h\nu_{pump}-E_g)/h\nu_{pump}$ with $E_g$ the value of the indirect band gap between the valence band at the T-point and the bottom of the conduction band derived from Ni 4s orbitals at the $\Gamma$-point~\cite{Gougoussis:2009dg}. Due to the attenuation of the pump beam along the depth direction of the film, the excited volume is computed as $V_{exc}=\pi\int_0^t(r_0e^{-z/\lambda_p})^2dz$ with $r_0$ the pump radius (derived from the FWHM), and $t$ the NiO crystal diameter. The heat jump $\Delta T$ is given by $\Delta T=E_{excess}/(C_VV_{exc})$ with $C_V=C_p\rho$ the volumic heat capacity and $\rho$ the density. All parameters used in the calculation are reported in Table~\ref{tabSI:parameters_calculation_heat_jump}, which provides a temperature jump $\Delta T=\SI{24}{\kelvin}$ for the excitation fluence of \SI{24}{\milli\joule\per\centi\metre\squared}, in good agreement with the jump experimentally observed of \SI{25}{\kelvin}.

\begin{table}[!ht]
    \centering
    \begin{tabular}{cccc}
    \toprule
    parameter & value & unit & source \\
    \midrule
    $E_{pump}$ & 2.42 & \SI{}{\micro\joule} & this work \\
    $E_g$ & 2.3 & \SI{}{\eV} & \cite{Lu2008} \\
    $t$ & 2.3 & \SI{}{\micro\meter} & \cite{Taylor2017:167104} \\
    $h\nu_p$ & 3.49 & \SI{}{\eV} & this work \\
    $r_0$ & 71 & \SI{}{\micro\meter} & this work \\
    $\lambda_p$ & 10000 & \SI{}{\per\centi\meter} & \cite{Newman:1959gr} \\
    $C_p$ & 594 & \SI{}{\joule\per\kilo\gram\per\kelvin} & \cite{Seltz1940} \\
    $\rho$ & 6.72 & \SI{}{\gram\per\cubic\centi\metre} & \cite{Lide2004} \\
    \bottomrule
    \end{tabular}
    \caption{Parameters used for the simulation of the pump-induced lattice heating.}
    \label{tabSI:parameters_calculation_heat_jump}
\end{table}

\subsection{From \emph{ab initio} calculations\label{subsecSI:ab_initio_heat_calculations}}

\begin{figure}
    \centering
    \includegraphics[width=\linewidth]{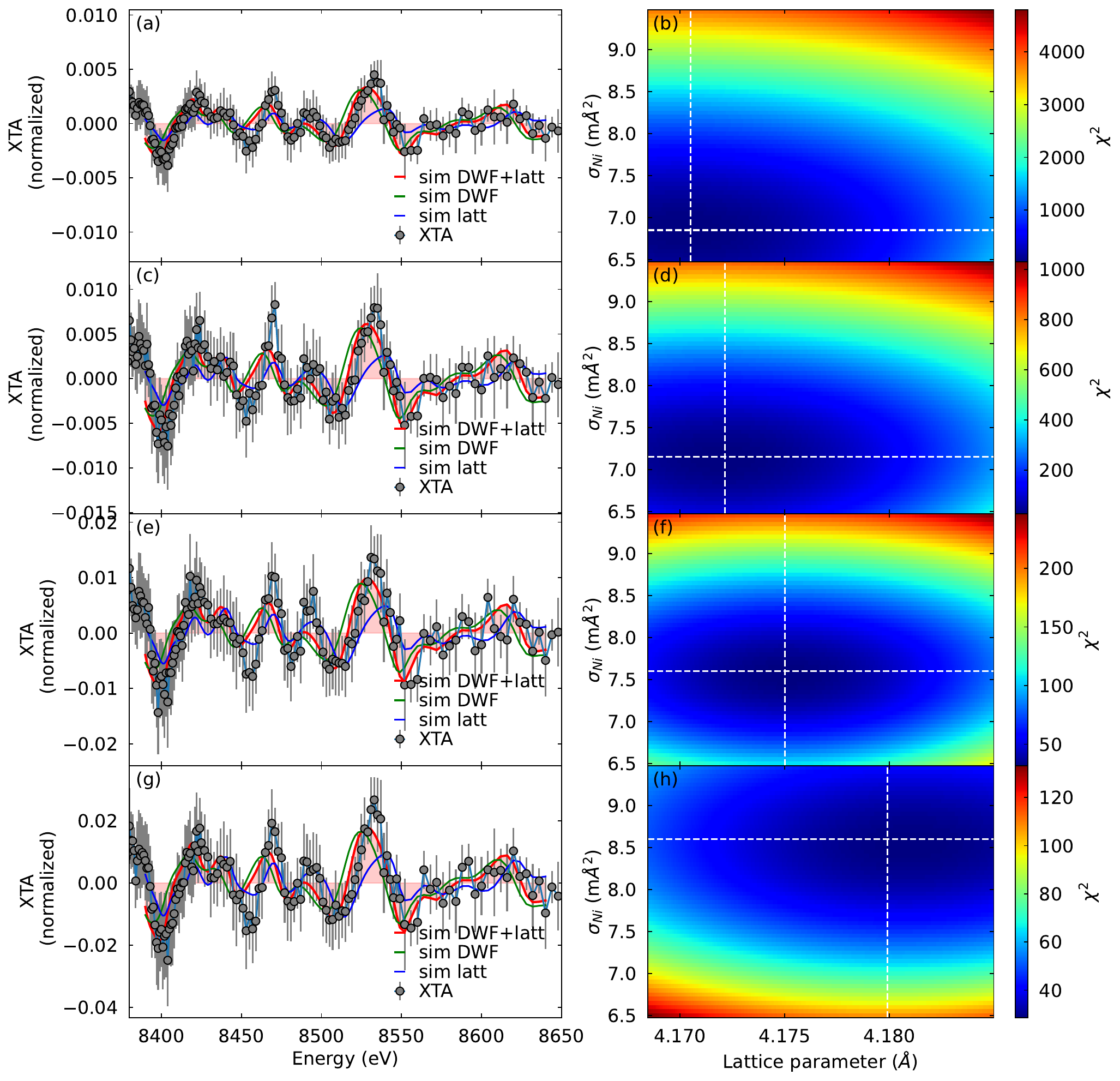}
    \caption{\textbf{Ab initio simulation of lattice heating contribution to the EXAFS.} (a, c, e, g) Best-matched calculated XTA spectra including an increased Debye-Waller factor (green curve), an increased lattice parameter (blue curve), and an increase of both Debye-Waller factor and lattice parameter (red shaded area) with respect to the XTA spectrum at \SI{100}{\pico\second} (grey circles with error bars). (b) Matrix of $\chi^2$ between calculated and experimental XTA spectra for different values of the Debye-Waller factor ($\sigma^2_{Ni}$) and lattice parameter in the excited state. The position of the minimum $\chi^2$ is given by the white dashed lines.}
    \label{figSI:ab_initio_lattice_heating_EXAFS}
\end{figure}

EXAFS spectra at different temperatures were simulated with the FDMNES code~\cite{Bunau:2009gp} (see \S\ref{subsecSI:FDMNES_calculations} for detail). The left column of Figure \ref{figSI:ab_initio_lattice_heating_EXAFS} displays the best-matched simulated spectra with the XTA in the EXAFS at increasing excitation fluences. Best agreement is always achieved by combining an increase of the Debye-Waller factor ($\Delta\sigma^2=\SI{0.42}{\milli\angstrom\squared}$ at excitation fluence \SI{24}{\milli\joule\per\centi\metre\squared}) and the lattice parameter ($\Delta a=\SI{0.00171}{\angstrom}$ at excitation fluence \SI{24}{\milli\joule\per\centi\metre\squared}). 

\cleardoublepage

%%%%%%%%%%%%%%%%%%%%%%%%%%%%%%%%%%%%%%%%%%%%%%%%%%%%%%%%%%%%%%%%%%%%%%%%%%%%
%%%%%%%%%%%%%%%%%%%%%%%%%%%%%%%%%%%%%%%%%%%%%%%%%%%%%%%%%%%%%%%%%%%%%%%%%%%%
%%%%%%%%%%%%%%%%%%%%%%%%%%%%%%%%%%%%%%%%%%%%%%%%%%%%%%%%%%%%%%%%%%%%%%%%%%%%
%%%%%%%%%%%%%%%%%%%%%%%%%%%%%%%%%%%%%%%%%%%%%%%%%%%%%%%%%%%%%%%%%%%%%%%%%%%%

\section{Simulation of chemical shift and/or spectral broadening\label{secSI:simulation_shift_broadening}}

\begin{figure}[!ht]
    \centering
    \includegraphics[width=\linewidth]{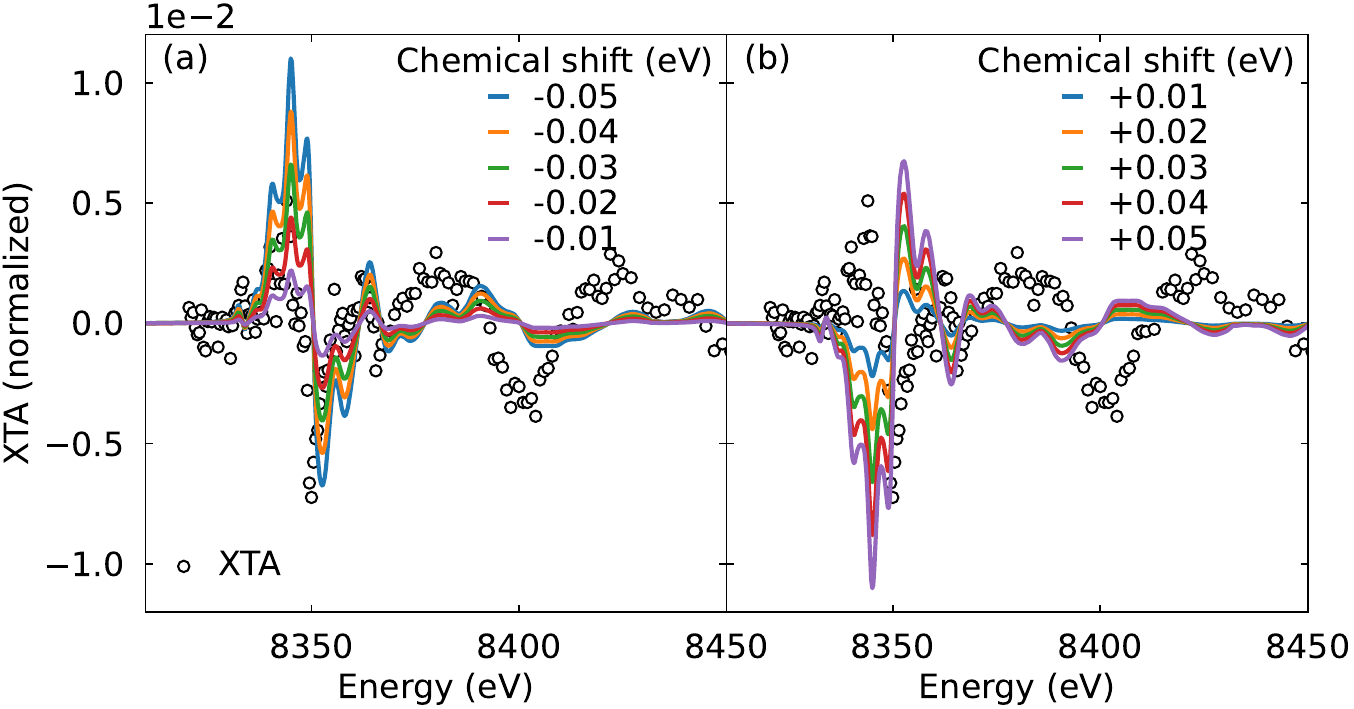}
    \caption{\textbf{Mismatch between simulated spectral shifts in the photoexcited state and experimental XTA spectra.} Simulated difference XAS spectra upon (a) negative, and (b) positive chemical shifts (colored curves). Experimental XTA spectra at \SI{100}{\pico\second} are shown for reference (black circles).}
    \label{figSI:chemical_shift_simulation}
\end{figure}

Simulations of chemical shifts in the photoexcited state were performed by applying an energy shift to the equilibrium XAS spectrum and computing the difference between the energy-shifted spectrum and the equilibrium one. Since the magnitude of the energy shift did not match the energy spacing between all data points, a cubic interpolation was performed on the equilibrium XAS spectrum to allow for the computation of differences between shifted and unshifted XAS spectra at the same energy points. Figure~\ref{figSI:chemical_shift_simulation} shows simulated XTA spectra with colored curves upon spectral red shifts (panel a) and spectral blue shifts (panel b).

\begin{figure}
    \centering
    \includegraphics[width=0.45\linewidth]{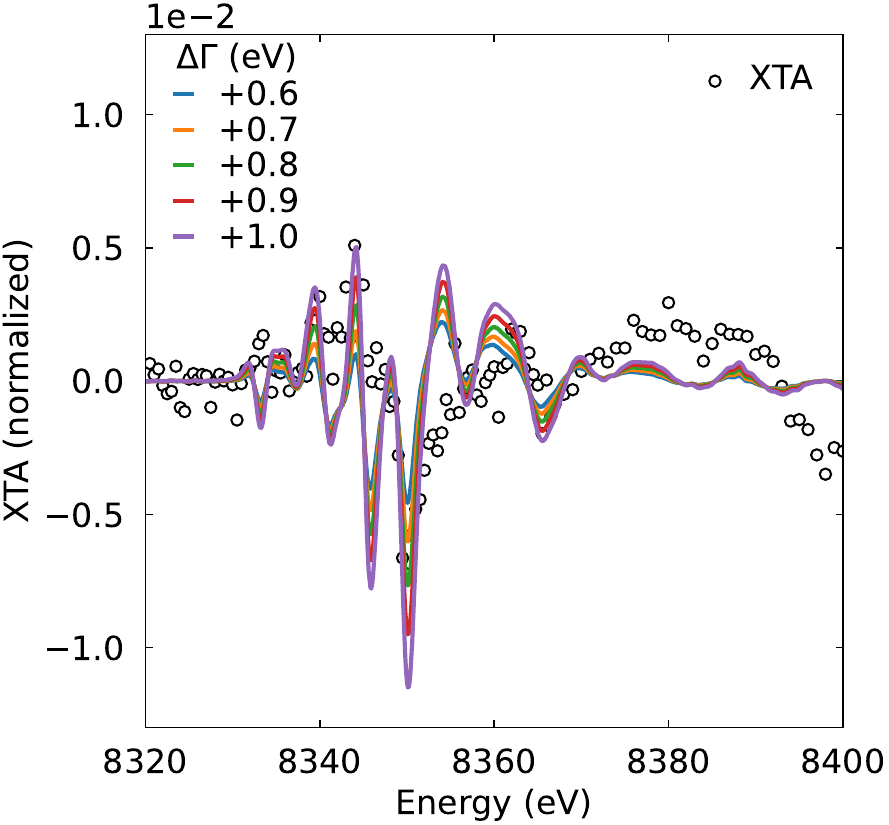}
    \caption{\textbf{Spectral broadening simulation.} Simulation of difference XAS spectra upon broadening of the equilibrium XAS spectrum (colored curves). Broadening is obtained by convolution with a Gaussian of specified FWHM. The XTA spectrum at \SI{100}{\pico\second} is shown for reference (black circles).}
    \label{figSI:broadening_simulation}
\end{figure}

Simulations of spectral broadening in the photoexcited state were performed by applying a convolution with a gaussian on the equilibrium XAS spectrum and then computing the difference between the resulting spectrum and the unbroadened equilibrium XAS spectrum. A normalization was applied on the broadened XAS spectrum to ensure spectral weight conservation. Energy points with equal spacing are required for the broadening convolution, a condition that was not fulfilled for the equilibrium XAS spectrum. A cubic interpolation was then performed first to allow for the convolution with a gaussian. The results of the convolution are displayed in Figure~\ref{figSI:broadening_simulation}. The broadening energy refers to the $\sigma$ value of a gaussian function with unity area $G(E, E_0, \sigma)$,

\begin{equation}
G(E, E_0, \sigma) = \frac{1}{\sqrt{2\pi}\sigma}e^{-\frac{(E-E_0)^2}{2\sigma^2}}
\end{equation}

\begin{figure}
	\centering
	\includegraphics[width=\linewidth]{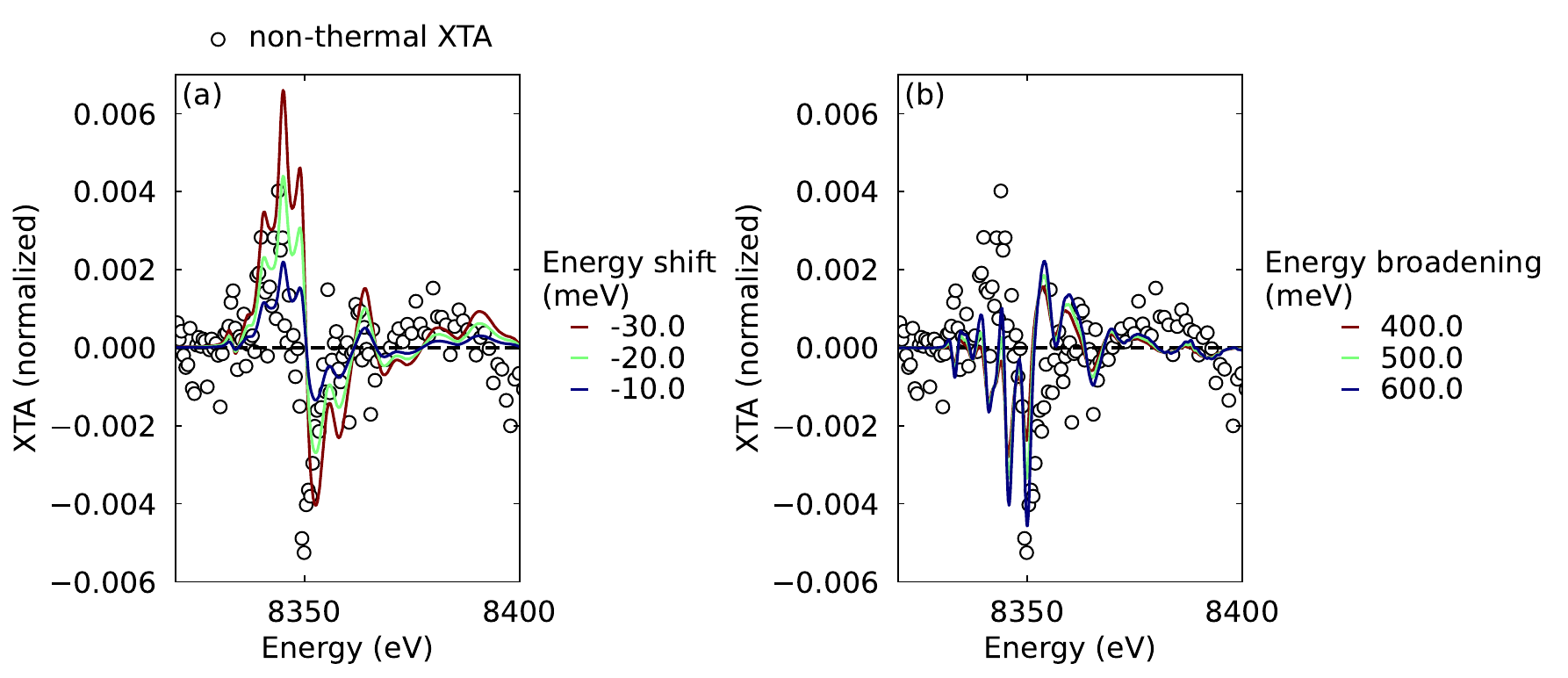}
	\caption{\textbf{Mismatch between experimental non-thermal XTA spectra and simulated XTA spectra upon energy shift or broadening.} Comparison between non-thermal XTA spectra and XTA spectra simulated upon (a) red shift of the XAS spectrum in the excited state by different magnitudes, and (b) spectral broadening of the XAS spectrum in the excited state by different magnitudes.}
	\label{figSI:chemical_shift_simulation_non_thermal_XTA}
\end{figure}

\begin{figure}
    \centering

    % Left column
    \begin{minipage}[t]{0.48\linewidth}
    	\vspace{0pt}
        \centering
        \includegraphics[width=\linewidth]{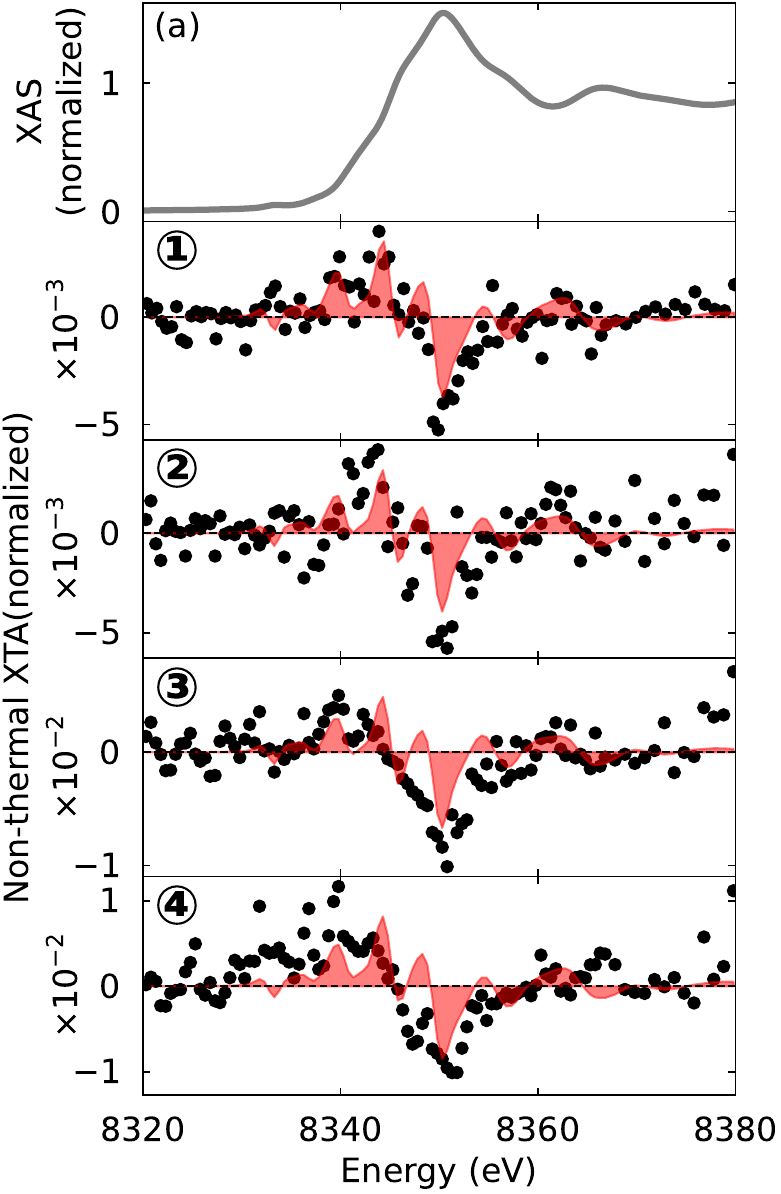}
    \end{minipage}\hfill
    % Right column
    \begin{minipage}[t]{0.45\linewidth}
        \vspace{0pt} % <-- forces top alignment

        % ---- Table ----
        \centering
        {\small
        \begin{tabular}{cccc}
            \toprule
            Label & Fluence (mJ/cm$^2$) & $\Delta$E (meV) & $\Delta\Gamma$ (meV) \\
            \midrule
            \ding{172} & 24 & -8 & 570 \\
            \ding{173} & 31 & -10 & 630 \\
            \ding{174} & 52 & -13 & 760 \\
            \ding{175} & 68 &  -17 & 870 \\
            \bottomrule
        \end{tabular}
        }

        \vspace{0.2cm}

        % ---- Second plot ----
        \includegraphics[width=1.13\linewidth]{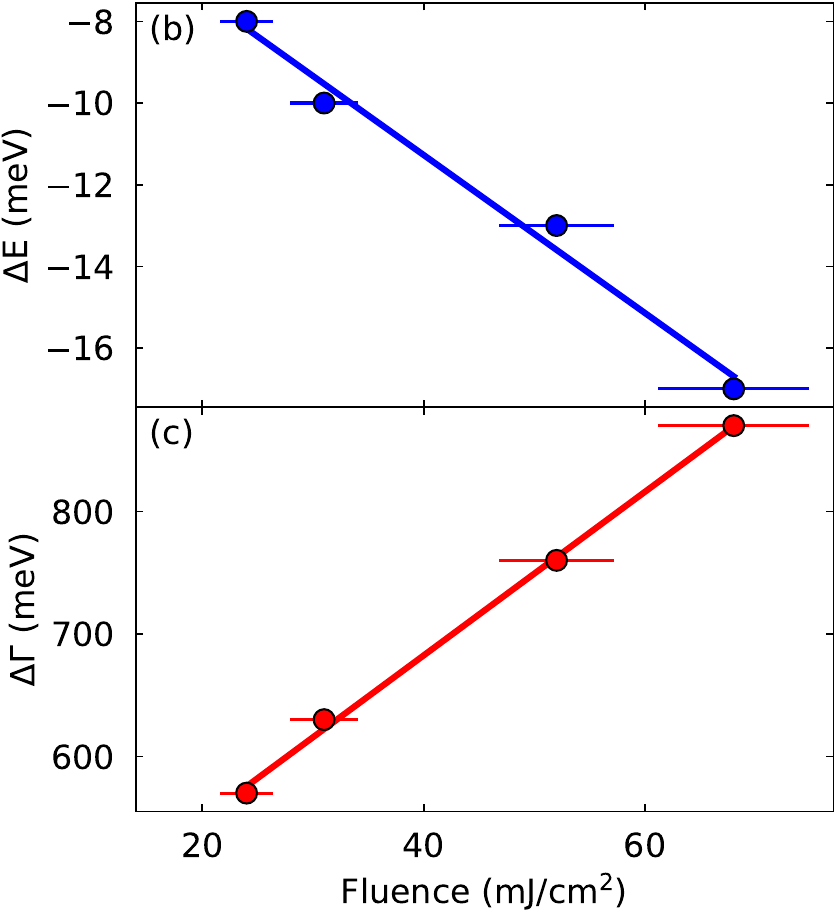}
    \end{minipage}

    \caption{\textbf{Non-thermal XTA spectrum qualitatively reproduced by a simultaneous shift and broadening of the XAS spectrum in the excited state.} (a) Non-thermal XTA spectra at various excitation fluences (black circles) and result of $\chi^2$ minimization combining an energy shift and a broadening in the excited state (red shaded area). The equilibrium XAS spectrum is shown for reference (top panel). Values of the energy shift and broadening are in the Table (top right). (b,c) Evolution of the broadening ($\Delta\Gamma$) and energy shift ($\Delta E$) of the excited-state XAS spectrum of NiO with excitation fluence (colored circles). Horizontal error bars indicate uncertainties in the excitation fluence. Linear fits are shown as continuous curves.}
    \label{figSI:shift_broadening_XAS}
\end{figure}

Simulations upon spectral shift and broadening were also compared with non-thermal XTA spectra (Figure~\ref{figSI:chemical_shift_simulation_non_thermal_XTA}). The simulations upon red shift of the XAS spectrum in the excited state are in qualitative agreement with the pump-probe data (panel a) while the simulations upon broadening do not reproduce the experimental trend (panel b). Simulations with a simultaneous spectral shift and broadening were performed using the same methods described above, including the cubic interpolation of the equilibrium spectrum to allow for the modeling of the broadening. Optimal values of the energy shift ($\Delta\text{E}$) and broadening ($\Delta\Gamma$) were determined via a $\chi^2$-minimization procedure that yielded the best agreement with the experimental non-thermal XTA spectra. Representative fits are shown in Figure~\ref{figSI:shift_broadening_XAS}a (red shaded areas), while the extracted values of $\Delta\text{E}$ and $\Delta\Gamma$ as a function of excitation fluence are summarized in Figure~\ref{figSI:shift_broadening_XAS}b,c. The extracted spectral red shifts range from~\SI{8}{} to \SI{17}{\milli\electronvolt}, while the corresponding broadenings span~\SI{570}{} to \SI{880}{\milli\electronvolt}. Both parameters exhibit an approximately linear dependence on the excitation fluence within the confidence intervals (Figure~\ref{figSI:shift_broadening_XAS}b,c). The magnitude of the energy shifts is comparable to the renormalization of the charge-transfer gap observed in previous ultrafast optical measurements at comparable excitation densities~\cite{Rossi2025:52718}, but is smaller than the shifts observed by femtosecond XAS at the Ni L$_{2,3}$ and O K-edges at a delay of \SI{500}{\femto\second}~\cite{Lojewski2023}. In contrast, the extracted spectral broadenings are several times larger than those reported in these earlier studies~\cite{Lojewski2023,Rossi2025:52718}. With increasing excitation fluences, the optimization gives a poorer agreement between the modeled and experimental XTA spectra (Figure~\ref{figSI:shift_broadening_XAS}a). This discrepancy suggests that a model based on homogeneous energy shifts and broadenings across the entire spectrum, although providing a reasonable agreement with experimental data, is insufficient.

\cleardoublepage

%%%%%%%%%%%%%%%%%%%%%%%%%%%%%%%%%%%%%%%%%%%%%%%%%%%%%%%%%%%%%%%%%
%%%%%%%%%%%%%%%%%%%%%%%%%%%%%%%%%%%%%%%%%%%%%%%%%%%%%%%%%%%%%%%%%
%%%%%%%%%%%%%%%%%%%%%%%%%%%%%%%%%%%%%%%%%%%%%%%%%%%%%%%%%%%%%%%%%
%%%%%%%%%%%%%%%%%%%%%%%%%%%%%%%%%%%%%%%%%%%%%%%%%%%%%%%%%%%%%%%%%

\section{Additional XTA spectra\label{secSI:additional_XTA_spectra}}

\begin{figure}[!ht]
	\centering
	\includegraphics[width=\linewidth]{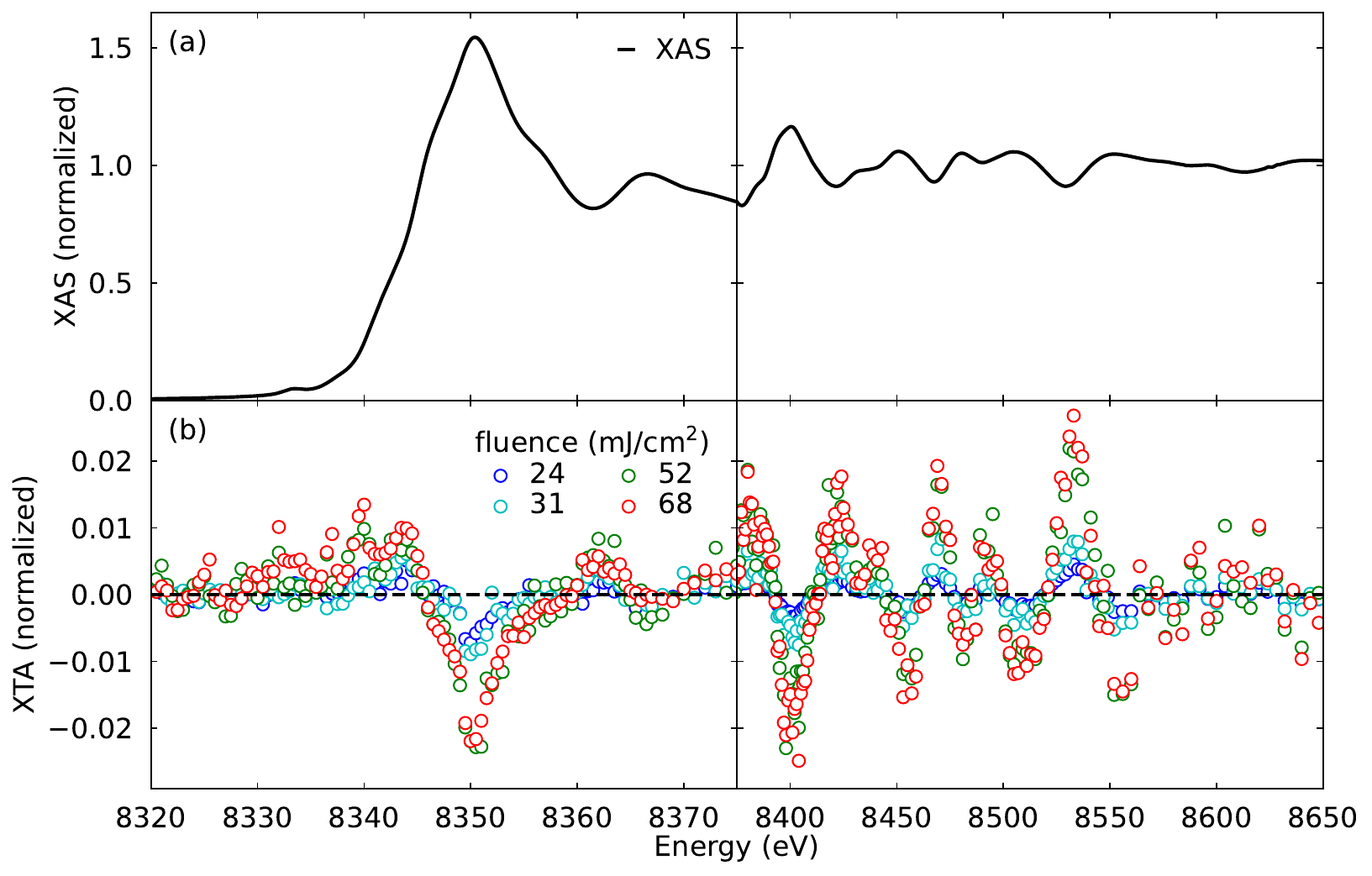}
	\caption{\textbf{Effect of pump excitation fluence on XTA spectrum.} (b) Evolution of XTA spectral traces at \SI{100}{\pico\second} at the Ni K-edge of NiO with excitation fluence. (a) The XAS spectrum is shown for reference. The pump photon energy is \SI{3.49}{\electronvolt}.}
	\label{figSI:XTA_with_fluence}
\end{figure}

\begin{figure}
	\centering
	\includegraphics[height=0.5\linewidth]{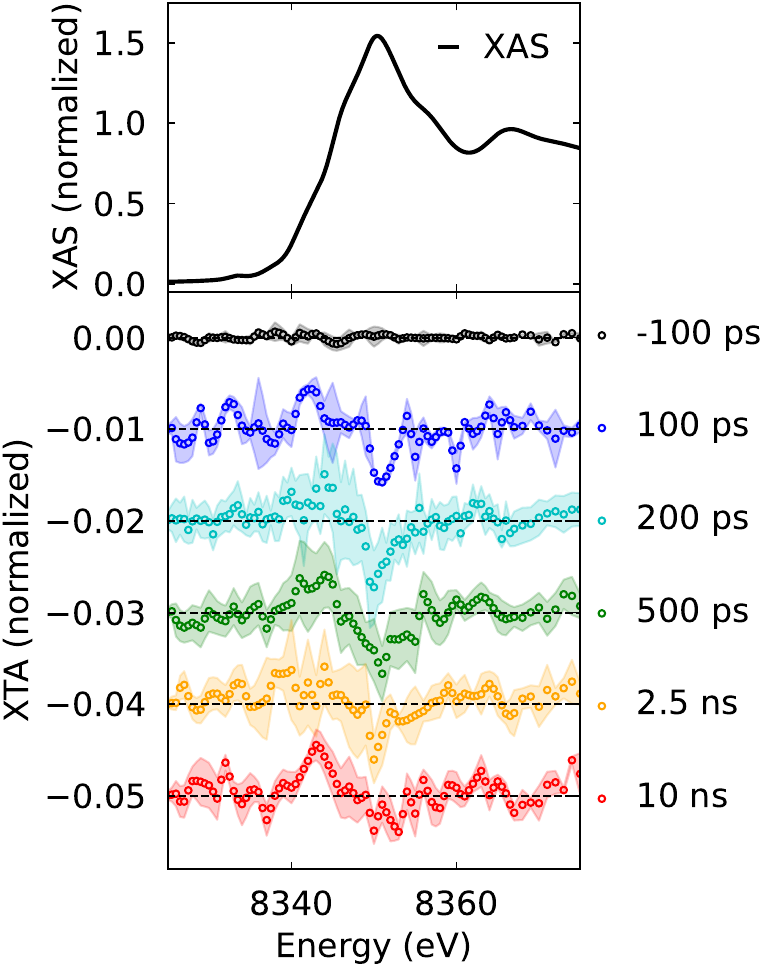}
	\caption{\textbf{Time evolution of XTA spectral traces.} Evolution of XTA spectra at different time delays (colored circles). Error bars correspond to standard deviations between individual measurements. The equilibrium XAS spectrum is shown in black for reference. The measurements at \SI{100}{\pico\second} and \SI{10}{\nano\second} are made of 3 consecutive scans, which means that the standard deviations are non-statistical.}
	\label{figSI:XTA_spectra_time_delays}
\end{figure}

This section provides additional XTA datasets. Figure~\ref{figSI:XTA_with_fluence} shows the evolution of XTA spectra at various pump excitation fluences. Figure~\ref{figSI:XTA_spectra_time_delays} shows the evolution of spectral traces with an excitation fluence of \SI{24}{\milli\joule\per\centi\metre\squared} at selected time delays between \SI{-100}{\pico\second} and \SI{10}{\nano\second}. 

\cleardoublepage

%%%%%%%%%%%%%%%%%%%%%%%%%%%%%%%%%%%%%%%%%%%%%%%%%%%%%%%%%%%%%%%%%
%%%%%%%%%%%%%%%%%%%%%%%%%%%%%%%%%%%%%%%%%%%%%%%%%%%%%%%%%%%%%%%%%
%%%%%%%%%%%%%%%%%%%%%%%%%%%%%%%%%%%%%%%%%%%%%%%%%%%%%%%%%%%%%%%%%
%%%%%%%%%%%%%%%%%%%%%%%%%%%%%%%%%%%%%%%%%%%%%%%%%%%%%%%%%%%%%%%%%

\section{XTA Kinetics\label{secSI:kinetics}}

\subsection{Time traces\label{subsecSI:kinetics}}

\begin{figure}[!ht]
    \includegraphics[width=\linewidth]{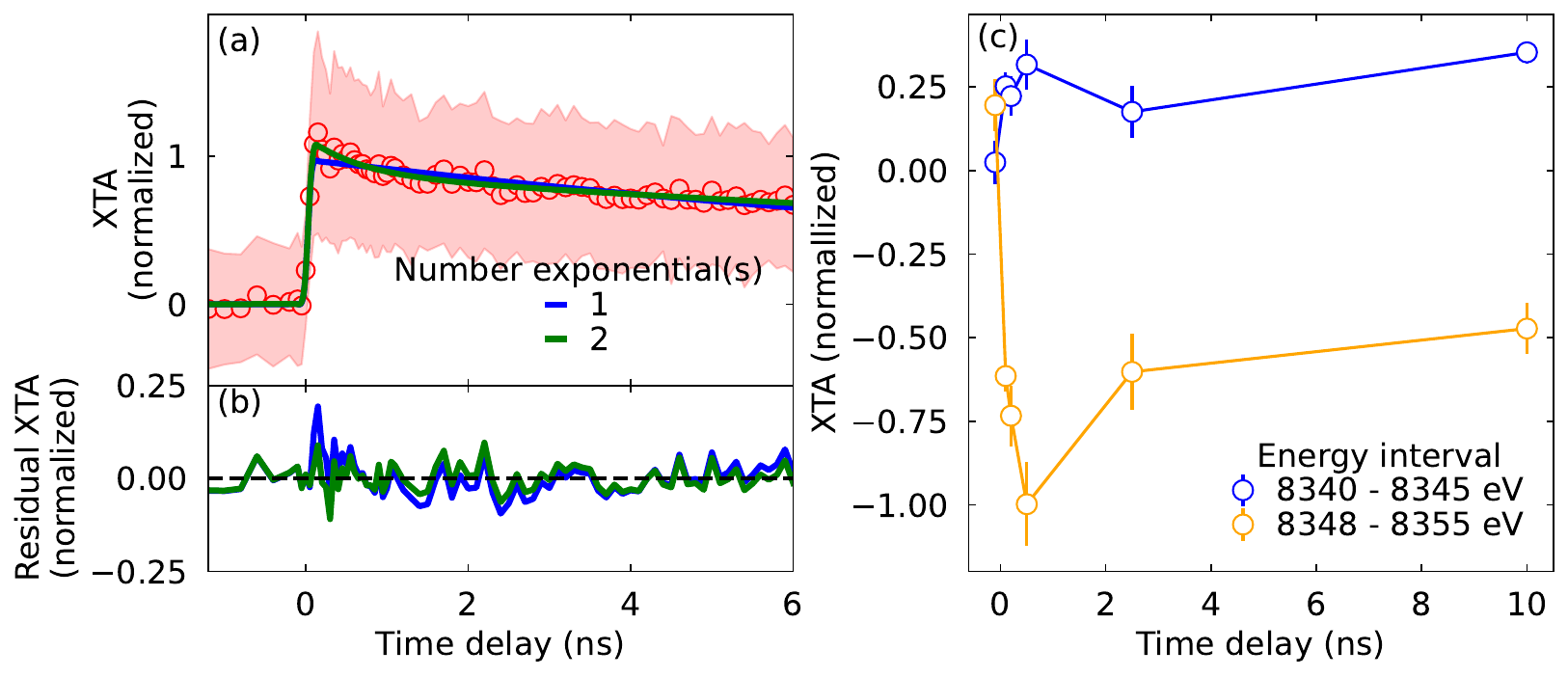}
    \vspace{-8mm}
    \caption{\textbf{Two decay components in the time traces.} (a) Decay of the XTA signal at \SI{8.35}{\kilo\electronvolt} (red circles). The pump photon energy is \SI{3.49}{\electronvolt} and the fluence is \SI{24}{\milli\joule\per\centi\metre\squared}. Shaded area represents the standard deviation between individual measurements. The data is best fitted with a parallel exponential decay mechanism with two components (green curve) with time constants $\tau_1=\SI{0.6\pm0.2}{\nano\second}$ (\SI{21}{\percent}) and $\tau_2=\SI{30\pm10}{\nano\second}$ (\SI{79}{\percent}). (b) Residuals of the fittings for the different models. (c) Time traces extracted from the evolution of the amplitude of spectral traces at selected time delays.}
    \label{figSI:time_traces}
\end{figure}

Figure~\ref{figSI:time_traces}a shows a time trace recorded at an energy of \SI{8.35}{\kilo\electronvolt}, which is best fitted with a biexponential decay model convoluted with the instrument response function (residuals in panel b). The time traces in Figure~\ref{figSI:time_traces}c are generated by extracting the amplitude of spectral traces recorded at different time delaysin Figure~\ref{figSI:XTA_spectra_time_delays}.

\subsection{Time constant of heat transfer from the NiO microcrystals to the solvent\label{subsecSI:heat_transfer_solvent}}

\begin{figure}[ht]
	\centering
	\includegraphics[width=0.5\linewidth]{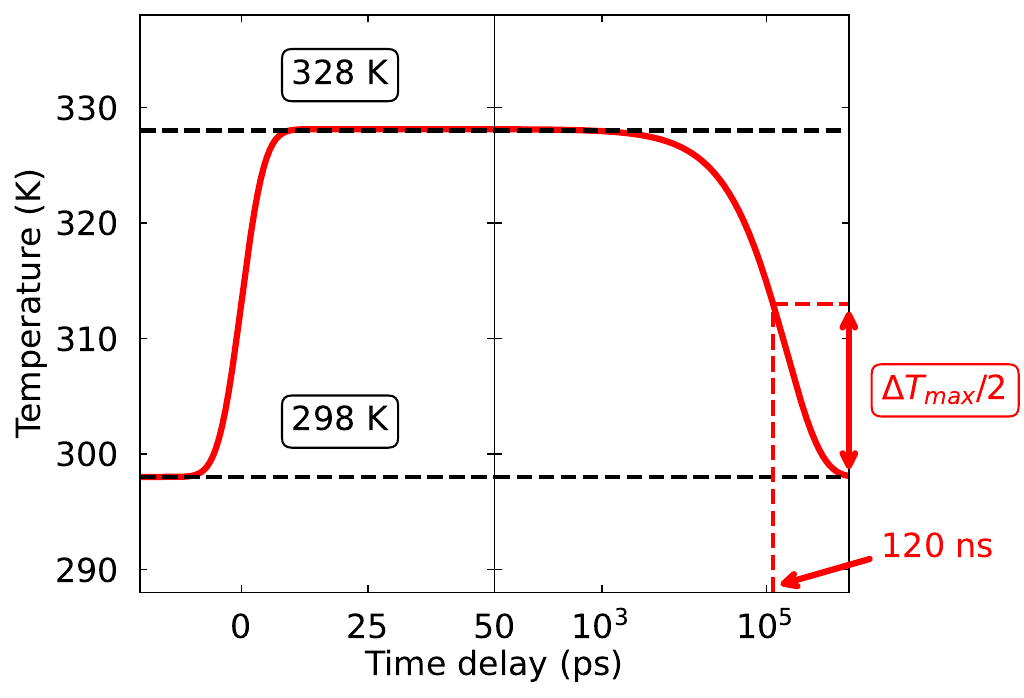}
	\caption{\textbf{The simulated time constant of heat transfer between the NiO lattice and the solvent is $\SI{120}{\nano\second}$ in XTA measurements.} Evolution of NiO lattice temperature following optical excitation with heat transfer to the solvent (water).}
	\label{figSI:heat_transfer_simulation}
\end{figure}

Interfacial heat transfer between photoexcited NiO particles and the solvent (water) is simulated for the XTA measurements. The evolution of the average particle temperature $T$ with a diameter $l$ (\SI{10}{\micro\meter}) in contact with water at room temperature ($T_R$) is modeled using the differential equation~\cite{Singer2015:86782},
\begin{equation}
l\cdot C(T)\cdot\frac{dT}{dt}=-G\cdot(T-T_R)+\dot{S}
\label{eqSI:interface_heat_transfer}
\end{equation}
with $C(T)$ the specific heat capacity of NiO, $G$ the thermal boundary (or Kapitza) conductance, and $\dot{S}$ the differential of the source term coming from the laser pump pulse excitation. The solvent temperature is kept constant at room temperature (\SI{298}{\kelvin}) due to its comparatively infinite heat capacity in a much larger volume than the NiO particles. The source term is made of a gaussian pulse with a FWHM of \SI{10}{\pico\second}, which corresponds to the duration of the pump pulse. The amplitude of the pump pulse (peak energy) is made such that the temperature of NiO reaches \SI{328}{\kelvin} at \SI{100}{\pico\second}, corresponding to the lattice temperature observed experimentally from the EXAFS analysis in Figure~\ref{fig:XTA} in the main text. There is no reported value for the thermal boundary conductance at the interface between NiO and water. We base our estimate from the existing literature, which has been recently reviewed in~\cite{Chen2022:260609}, showing that this quantity under ambient conditions depends on several parameters such as the particle surface roughness and patterning, the bonding strength and chemical functionalization. The reported conductance for the TiO$_2$/water interface is in the range of $\sim\SI{100}{\mega\watt\per\meter\squared\per\kelvin}$~\cite{Huang2025:225646}, which is the value considered for the simulations of the NiO/water interface. The heat capacity of NiO is modeled using a Shomate-type polynomial function $C(T)=a+bT+cT^2+dT^3$,  with $a=\SI{35}{\joule\per\mol\per\kelvin}$, $b=\SI{1.1e-2}{\joule\per\mole\per\kelvin\squared}$, $c=\SI{2e-6}{\joule\per\mole\per\kelvin\tothe{3}}$, and $d=\SI{5e-10}{\joule\per\mole\per\kelvin\tothe{4}}$~\cite{Seltz1940}. The differential equation \ref{eqSI:interface_heat_transfer} is solved with the finite difference method. The simulated evolution of the lattice temperature with the time delay is shown in Figure~\ref{figSI:heat_transfer_simulation}. It indicates that the characteristic time it takes for the NiO lattice temperature to reach half the maximum temperature rise generated by the laser excitation is \SI{120}{\nano\second}, which is in reasonable agreement with the longest time constant of the decay fitted in Figure~\ref{figSI:time_traces}.

%%%%%%%%%%%%%%%%%%%%%%%%%%%%%%%%%%%%%%%%%%%%%%%%%%%%%%%%%%%%%%%%%%%%%%%%%%%%
%%%%%%%%%%%%%%%%%%%%%%%%%%%%%%%%%%%%%%%%%%%%%%%%%%%%%%%%%%%%%%%%%%%%%%%%%%%%
%%%%%%%%%%%%%%%%%%%%%%%%%%%%%%%%%%%%%%%%%%%%%%%%%%%%%%%%%%%%%%%%%%%%%%%%%%%%
%%%%%%%%%%%%%%%%%%%%%%%%%%%%%%%%%%%%%%%%%%%%%%%%%%%%%%%%%%%%%%%%%%%%%%%%%%%%

\section{Discussion on the different contributions leading to spectral shifts of the XAS spectrum in photoexcited NiO\label{secSI:discussion_excited_XAS_contributions}}

% Possible interpretations of an energy shift of the spectrum
In photodoped materials, spectral shifts can arise from modifications of the Coulomb potential, which alters the energies of spectroscopic transitions through both mean-field and excitonic effects. Mean-field contributions originate from changes in the average Coulomb potential induced by photoexcited charge carriers~\cite{Golez2024:227781}. In charge-transfer insulators such as NiO, DMFT predicts that photoinduced charge redistribution between Ni $d$ and O $p$ orbitals generates an additional mean-field potential acting on the $d$ band. This so-called \emph{Hartree shift} results in a lowering of the energy of the UHB~\cite{Golez2024:227781}. In addition to this static effect, DMFT further predicts a dynamical screening of the on-site Coulomb interaction, which effectively reduces the Hubbard $U$. Both the Hartree shift and the renormalization of electronic correlations are expected to increase with photodoping density. However, simulations performed on cuprates indicate that correlation screening dominates over the Hartree contribution, with a magnitude approximately 4 times larger for photodoping levels between \SI{0.01}{} and \SI{0.1}{eh/uc}~\cite{Golez2024:227781}. Experimentally, such renormalization of electronic correlations has been observed in NiO as a reduction of the charge-transfer gap using time-resolved XAS~\cite{Lojewski2023,Cazali2025:166775} and optical spectroscopy~\cite{Rossi2025:52718}. Hence, Coulomb screening of energy states is expected to lead to spectral shifts if the corresponding states are involved in spectroscopic transitions. For the core-level transitions of XAS spectra sensitive to the unoccupied $d$ density of states, the screening of the unoccupied density of states is expected to lead to a red shift of the spectrum. We expect this could be observed with the pre-edge features of NiO due to transitions to Ni $3d$ orbitals~\cite{Gougoussis:2009kr}, if statistics was sufficient.

% Core-exciton interactions
In XAS spectra of metal oxides, strong excitonic interactions exist between the core hole and valence electrons localized on the same atomic site~\cite{Bassani1980:111585}. Enhanced screening of this interaction by photoexcited carriers increases the resonance energy of core excitons, resulting in a blue shift of the XAS spectrum. Such excitonic screening effects have recently been reported in photoexcited band insulators~\cite{Rossi2021,Rossi2025:87275}. Therefore, whereas Hartree shifts and the renormalization of electronic correlations tend to shift spectral features toward lower energies, excitonic screening drives spectral shifts in the opposite direction.

% Spectral broadening due to dynamic effects, challenge to find the electronic structure of excited because of competing effects
Beyond energy shifts, DMFT also predicts substantial spectral broadening following photoexcitation, driven by the coupling of electronic states to plasmon modes~\cite{Golez2024:227781}. Taken together, Hartree shifts, correlation renormalization, excitonic screening, and dynamical broadening give rise to competing and intertwined spectral signatures. This inherent complexity makes a direct and quantitative reconstruction of the excited-state electronic structure of NiO from experimental XTA spectra highly challenging.   

%%%%%%%%%%%%%%%%%%%%%%%%%%%%%%%%%%%%%%%%%%%%%%%%%%%%%%%%%%%%%%%%%%%%%%%%%%%%
%%%%%%%%%%%%%%%%%%%%%%%%%%%%%%%%%%%%%%%%%%%%%%%%%%%%%%%%%%%%%%%%%%%%%%%%%%%%
%%%%%%%%%%%%%%%%%%%%%%%%%%%%%%%%%%%%%%%%%%%%%%%%%%%%%%%%%%%%%%%%%%%%%%%%%%%%
%%%%%%%%%%%%%%%%%%%%%%%%%%%%%%%%%%%%%%%%%%%%%%%%%%%%%%%%%%%%%%%%%%%%%%%%%%%%

\section{\emph{Ab initio} calculations\label{secSI:ab_initio_calculations}}

\subsection{FDMNES simulation of the NiO hot lattice in the EXAFS}
\label{subsecSI:FDMNES_calculations}

% General information about the calculations
To simulate lattice heating at the Ni K-edge of NiO, X-ray absorption spectra at the Ni K-edge of NiO in the EXAFS region were calculated with the FDMNES code~\cite{Bunau:2009gp} and the MUMPS solver~\cite{Guda2015}. Calculations are performed with the full multiple scattering (FMS) on a NiO cluster of radius \SI{7.5}{\AA}, between \SI{-100}{} and \SI{350}{\eV} with respect to the Fermi level. The electronic structure is converged with a self-consistent field calculation including the core hole in the Ni $1s$ orbital. Absorption cross-sections are calculated in the dipole approximation, which is relevant for the EXAFS region of the spectrum. An electronic damping term of $S_0^2=\SI{0.87}{}$ is included in the calculation to reproduce the amplitude of the EXAFS oscillations in the experimental EXAFS spectrum. The NiO space group is set to $F_{m\bar{3}m}$, the equilibrium lattice parameter is set to \SI{4.1685}{\AA} and the equilibrium Debye-Waller factor of the Ni atoms to \SI{0.0065}{\AA\squared}, which correspond to the lattice parameter and the Debye-Waller factor retrieved from temperature-dependent measurements in the powder form (the $B_{iso}$ parameter of \SI{0.51}{\AA\squared} in Figure~\ref{figSI:XAS_XRD_fitting_parameters} corresponds to a Debye-Waller factor of \SI{0.0065}{\AA\squared}). The results of the simulations and the comparison with the experimental XTA spectra are detailed in \S\ref{subsecSI:ab_initio_heat_calculations}.

\subsection{Quantum Espresso}

\subsubsection{Self-consistent field calculations\label{subsubsecSI:QE_SCF}}

% General information about the calculations
The effect of changes in electronic correlations ($U$) and intersite electronic screening ($V$) is computed in Quantum-Espresso (PWscf)~\cite{Giannozzi:2009hx,Giannozzi:2017io}. The room temperature lattice parameter at \SI{25}{\degreeCelsius} is obtained by extrapolation of the temperature-dependent XRD measurement of the Ni-O bond distance (Figure \ref{figSI:XAS_XRD_fitting_parameters}) giving an expression $a(T)$ of the lattice parameter as a function of the lattice temperature $T$ given by $a(T)=4.1703-6.9864\times10^{-6}T+3.19\times10^{-9}T^2$ and thus $a=\SI{4.16850}{\angstrom}$ at room temperature (\SI{298}{\kelvin}) for the cubic unit cell (\SI{7.87732}{\bohr}). We neglected the rhombohedral distortion of the magnetic unit cell \cite{Eto:2000cw}.

\begin{figure}
    \centering
    \includegraphics[width=\linewidth]{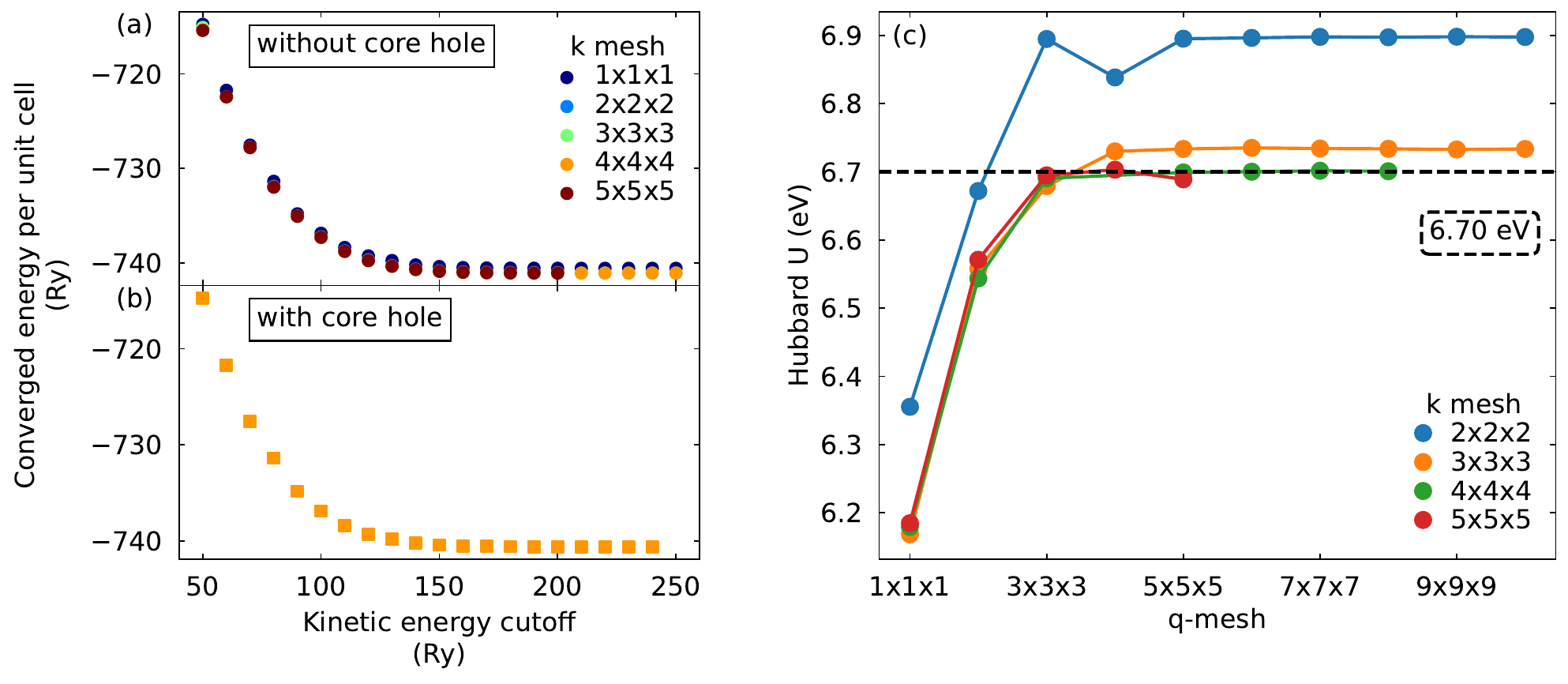}
    \caption{\textbf{Convergence calculations.} Evolution of the total energy per unit cell with the wavefunction energy cutoff and the density of k-points in the Monkhorst-Pack grid for a cell (a) without core hole, and (b) with core hole. (c) Evolution of the converged Hubbard U by DFPT.}
    \label{figSI:PWscf_convergence}
\end{figure}

\begin{figure}
    \centering
    \includegraphics[width=0.495\linewidth]{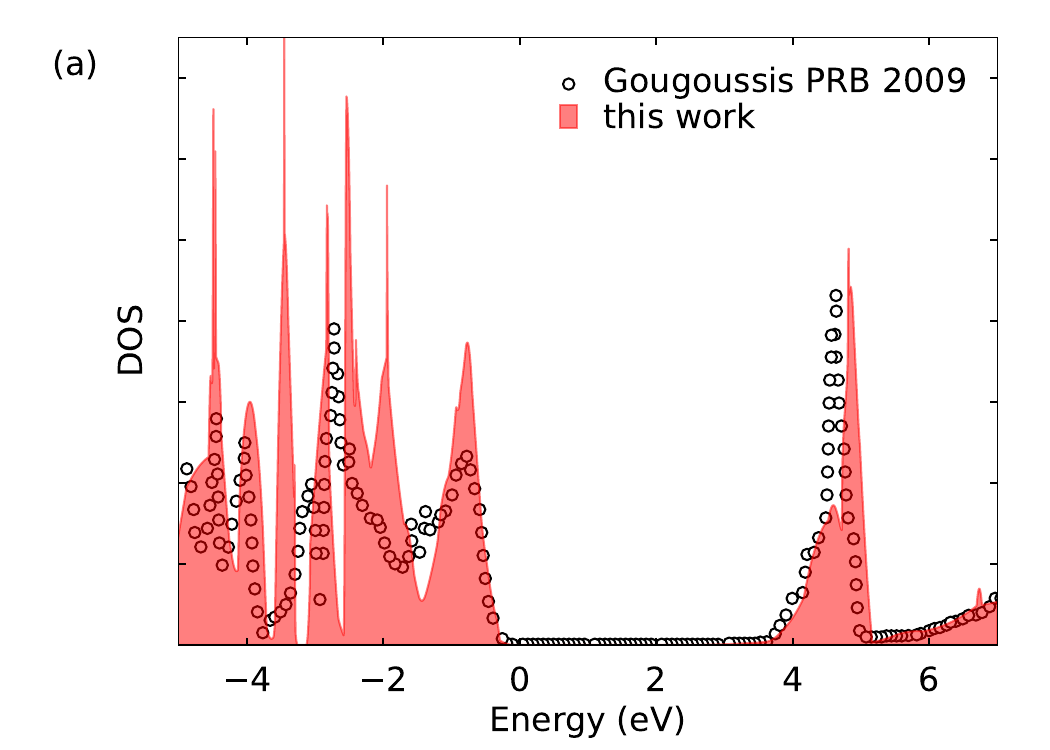}
    \includegraphics[width=0.495\linewidth]{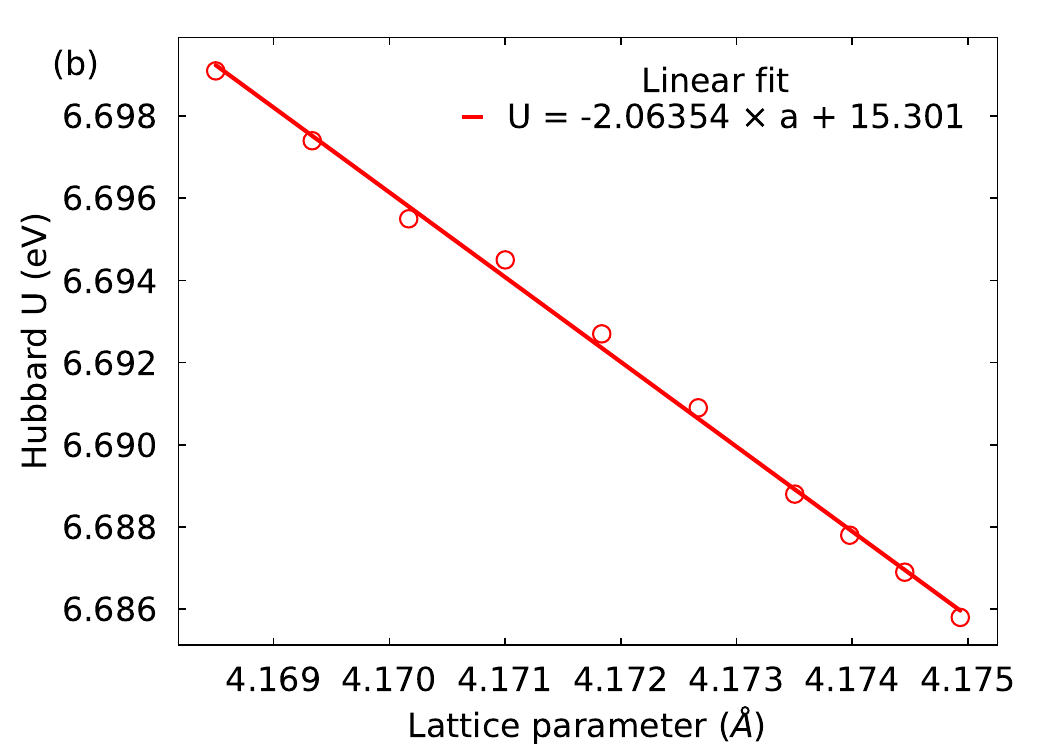}
    \caption{\textbf{DOS and Hubbard U with lattice parameter.} (a) Calculated density of states with $U=\SI{6.7}{\electronvolt}$ (without core hole, red shaded area). Comparison with previous calculations from reference~\cite{Gougoussis:2009kr} (black circles). (b) Calculated evolution of the Hubbard U with the lattice parameter of the cubic cell using DFPT (red circles). Linear fit to the data (continuous line).}
    \label{figSI:DOS}
\end{figure}

% Description of the PWscf calculations for convergence
Plane-wave self-consistent calculations were performed with scalar relativistic norm-conserving Troullier-Martins pseudopotentials for nickel and oxygen~\cite{Troullier:1991bm}, the spin-polarized generalized gradient approximation (GGA)~\cite{Perdew1996:145658}, and the DFT+U or DFT+U+V method~\cite{Gougoussis:2009dg,Timrov2020:73231}. The antiferromagnetic rhombohedral unit cell with parameter $a_R=\sqrt{6}a/2$ and angle $\cos\gamma=5/6$ was first considered to check the convergence of the total energy against the wavefunction energy cutoff and the Monkhorst-Pack k-mesh. In this unit cell, the atomic coordinates are provided in Table \ref{tabSI:QE_fractional_coordinates} with lattice parameter $a_R=\SI{5.10535}{\angstrom}$ (\SI{9.64771}{\bohr}). Input crystal structures were checked with the XcrySDen software~\cite{Kokalj1999}. The total magnetization was constrained to 0, occupations were fixed since NiO is insulating. The convergence threshold for the self-consistent calculation was set to \SI{1e-9}{} Ry and a mixing beta of \SI{0.3}{} in the "plain" mode. The evolution of the energy per unit cell with the energy and the k-mesh is displayed in Figure \ref{figSI:PWscf_convergence}a. Convergence is achieved for a kinetic energy cutoff of the wavefunction of \SI{210}{Ry} and a k-mesh of $4\times4\times4$. The calculated density of states is displayed in Figure~\ref{figSI:DOS}a and compared with a previous calculation~\cite{Gougoussis:2009dg}, showing large similarities in the calculated features despite a different value of the Hubbard $U$.

% Calculation of structural relaxation
Structural relaxation was checked self-consistently, with convergence achieved for a force convergence threshold of \SI{1e-5}{}, an energy convergence threshold of \SI{1e-10}{}, and a pressure convergence threshold of \SI{0.1}{\kilo\bar}. The converged cubic lattice parameter was \SI{7.96826}{\bohr} (\SI{4.21662}{\angstrom}), the converged rhombohedral lattice parameter wwas \SI{9.67155}{\bohr} (\SI{5.11796}{\angstrom}). The values were converted between cubic and rhombohedral lattice parameters and compared to experimental values from XRD and the analysis of the EXAFS spectrum in Table~\ref{tabSI:QE_structural_relaxation}. Converged lattice parameters are $\sim\SI{3}{\percent}$ higher than experimental values, which is a well-known effect from GGA functionals~\cite{Haas2009:23078}.

\begin{table}
    \centering
    \footnotesize
    \begin{tabular}{ccccc}
        \toprule
        & theory cubic cell ($a_0$/$\text{\AA}$) & theory rhombohedral cell ($a_0$/$\text{\AA}$) & exp. XRD ($a_0$/$\text{\AA}$) & exp. XAS ($\text{\AA}$) \\
        \midrule
        cubic $a$ & 8.168817/4.322752 & 8.128139/4.301226 & 7.877/4.169 & 7.88/4.17 \\
        rhombohedral $a$ & 10.004717/5.294269 & 9.954897/5.267905 & 9.624/5.093 & 9.65/5.11 \\ 
        \bottomrule
    \end{tabular}
    \caption{\textbf{Structural optimization.} Converged lattice parameters in the cubic and rhombohedral basis set compared with experimental values from XRD and XAS (EXAFS) in this work.}
    \label{tabSI:QE_structural_relaxation}
\end{table}

\begin{table}
    \centering
    \begin{tabular}{cccc}
        \toprule
        Atom & x & y & z \\
        \midrule
        Ni1 & 0.00000 & 0.00000 & 0.00000 \\
        Ni2 & -0.50000 & 1.50000 & -0.50000 \\
        O & 0.75000 & -0.25000 & -0.25000 \\
        O & -0.75000 & 0.25000 & 0.25000 \\
        \bottomrule
    \end{tabular}
    \caption{\textbf{Atomic positions} Fractional coordinates of the atoms in the rhombohedral magnetic NiO unit cell. Ni1/Ni2 is with majority spin up/down.}
    \label{tabSI:QE_fractional_coordinates}
\end{table}

\begin{figure}
    \centering
    \includegraphics[width=0.495\linewidth]{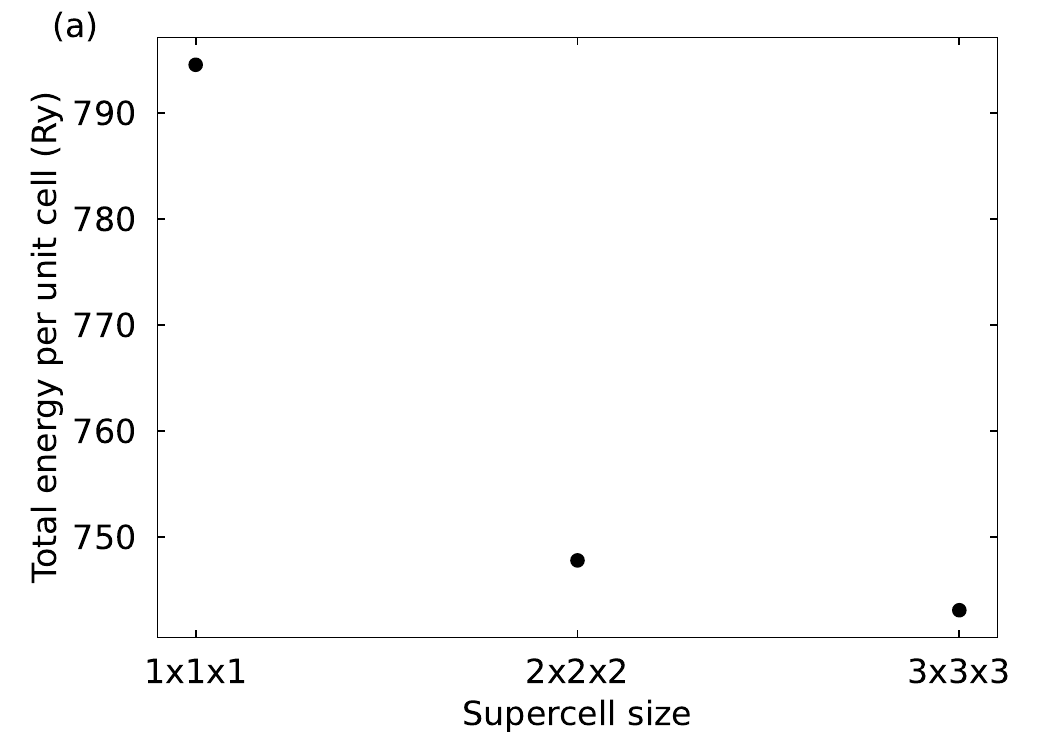}
    \includegraphics[width=0.495\linewidth]{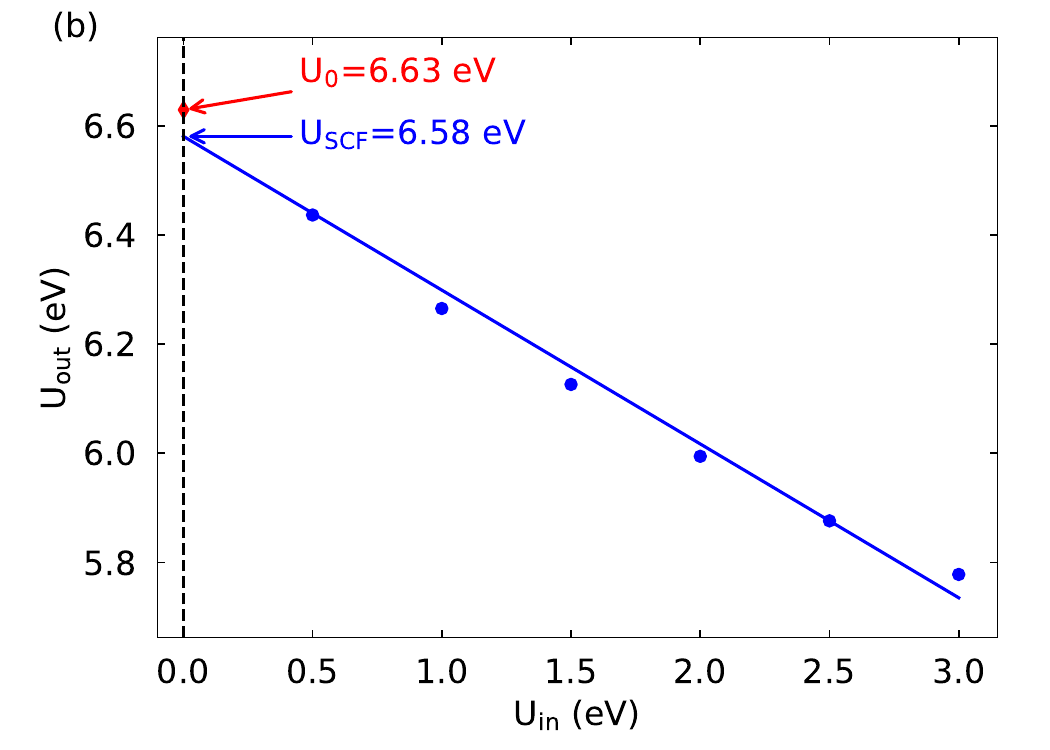}
    \caption{(a) Evolution of the total energy per unit cell with one core hole on one nickel site as a function of the supercell dimension (equivalent to increasing the distance between atoms with a core hole). (b) Calculated Hubbard U with the linear response method.}
    \label{figSI:PWscf_core_hole}
\end{figure}

% Description of the calculations to find the Hubbard U
The value of the Hubbard $U$ was determined by density functional perturbation theory (DFPT) following a method reported in reference~\cite{Timrov2018:85970} with the HP executable of Quantum-Espresso~\cite{Timrov2022:301065}. The Hubbard $U$ was applied to electrons in the $3d$ shell of the nickel atoms only. Figure~\ref{figSI:PWscf_convergence}b shows the evolution of the calculated Hubbard $U$ with the k- and q-mesh density. The converged value of the Hubbard $U$ is \SI{6.70}{\electronvolt}. We checked further the value of the Hubbard $U$ using the linear response method in a $2\times2\times2$ rhombohedral unit cell in order to avoid the interaction between perturbed atoms. The response function $\chi$ is the variation of the orbital occupation $n$ with $\alpha$ a perturbation over the Hubbard $U$ ($\chi=\partial n/\partial \alpha$). The value of the Hubbard $U$ is given by the difference between the inverse of the self-consistent response and the bare response ($U=\chi^{-1}-\chi_0^{-1}$). This equation solved for different values of a starting Hubbard $U$ ($U_{\text{in}}$) gave a value of a converged Hubbard $U$ upon perturbation ($U_{\text{out}}$), which is displayed in Figure~\ref{figSI:PWscf_core_hole}b. The value of the self-consistent Hubbard $U$ is $U_{\text{SCF}}=\SI{6.58}{\eV}$ whereas the value of the Hubbard $U$ calculated near $U_{\text{in}}=0$ is $U_0=\SI{6.63}{\eV}$, both values are in good agreement with the Hubbard $U$ value of \SI{6.70}{\eV} calculated with DFPT. We set $U=\SI{6.70}{\eV}$ in the following.

% Description of the calculations to find the Hubbard U and V self-consistently
The value of the equilibrium Hubbard V between Ni $3d$ and O $2p$ orbitals was computed using the workflow detailed in reference~\cite{Timrov2021:70354}, with the exception that the lattice structure of NiO is not computed at every SCF iteration. Only the values of the Hubbard U and the Hubbard V were updated after each calculation loop which consists of one PWscf calculation with input U and V values and one DFPT calculation providing new values of Hubbard U and V. The values of the Hubbard parameters are iterated self-consistently until convergence is achieved within \SI{1}{\milli\eV}. The results give $U=\SI{5.386}{\eV}$ and $V=\SI{2.10}{\eV}$.

% Effect of the Hubbard U on the electronic structure
\begin{figure}
    \centering
    \includegraphics[width=\linewidth]{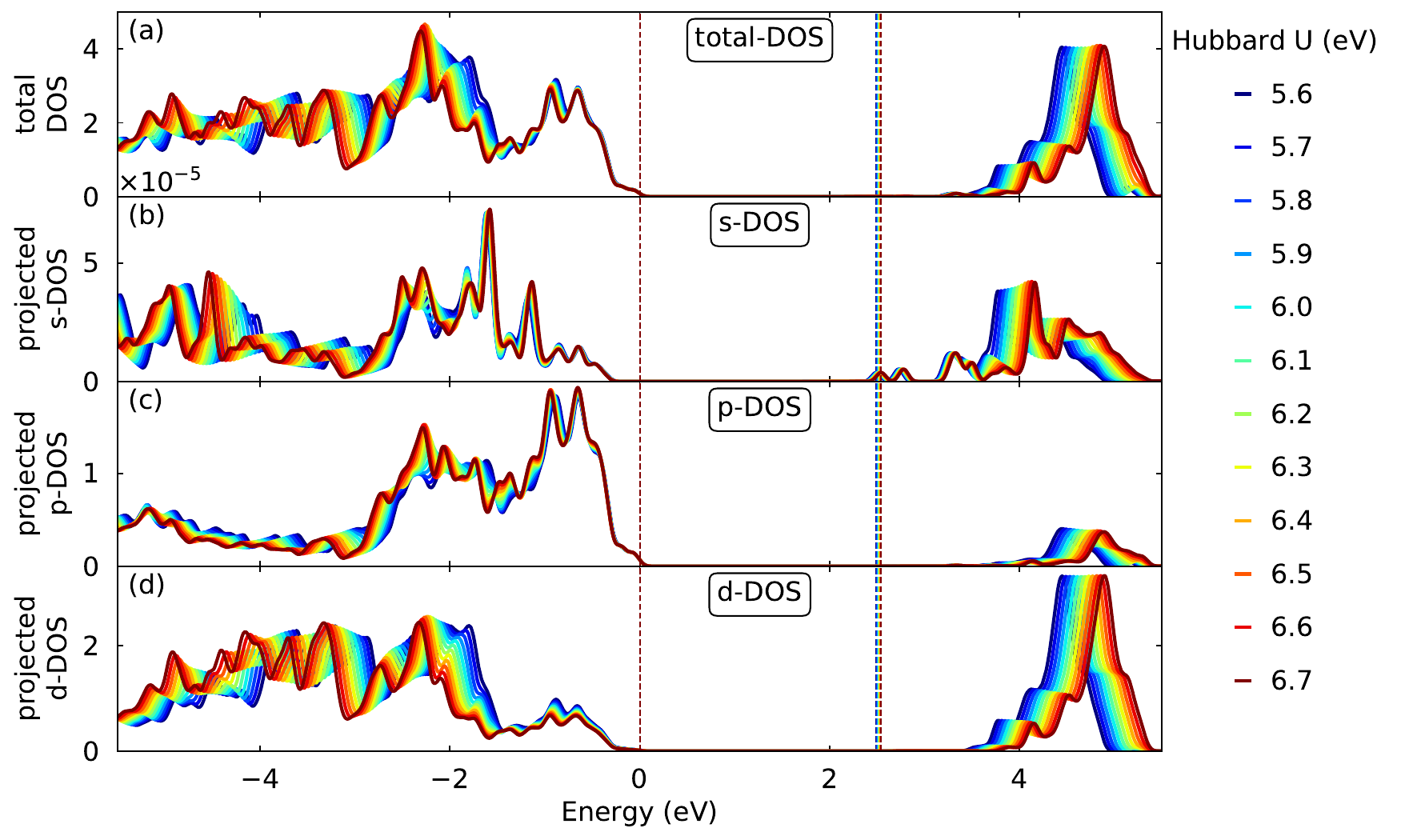}
    \caption{\textbf{Density of states with Hubbard U.} Effect of the Hubbard U on (a) the total density of states (DOS), and the projected (b) s-DOS, (c), p-DOS, and (d) d-DOS (continuous lines). The vertical dashed lines indicate the position of the highest occupied and lowest occupied energy levels for a given value of U. The zero energy is set to the highest occupied energy state before the broadening of the DOS.}
    \label{figSI:DOS_Hubbard_U}
\end{figure}

Figure~\ref{figSI:DOS_Hubbard_U}a shows the evolution of the density of states (DOS) as a function of the Hubbard $U$. It is clear that the effect of increasing the Hubbard $U$ is to push the density of states above the band gap to higher energies, while the effect on the density of states at the top of the valence band remains limited to much smaller value than the change of the Hubbard $U$ value. This is expected since the Hubbard $U$ is applied to the Ni $3d$ orbitals and the projected DOS in Figure \ref{figSI:DOS_Hubbard_U}b,c,d shows that the top of the valence DOS is made of a $p-d$ orbital mixture with a dominant contribution from $p$ DOS while the main changes in the conduction DOS originates from the $d$-DOS. 

% Effect of the Hubbard V between Ni 3d and O 2p on the electronic structure
\begin{figure}
    \centering
    \includegraphics[width=\linewidth]{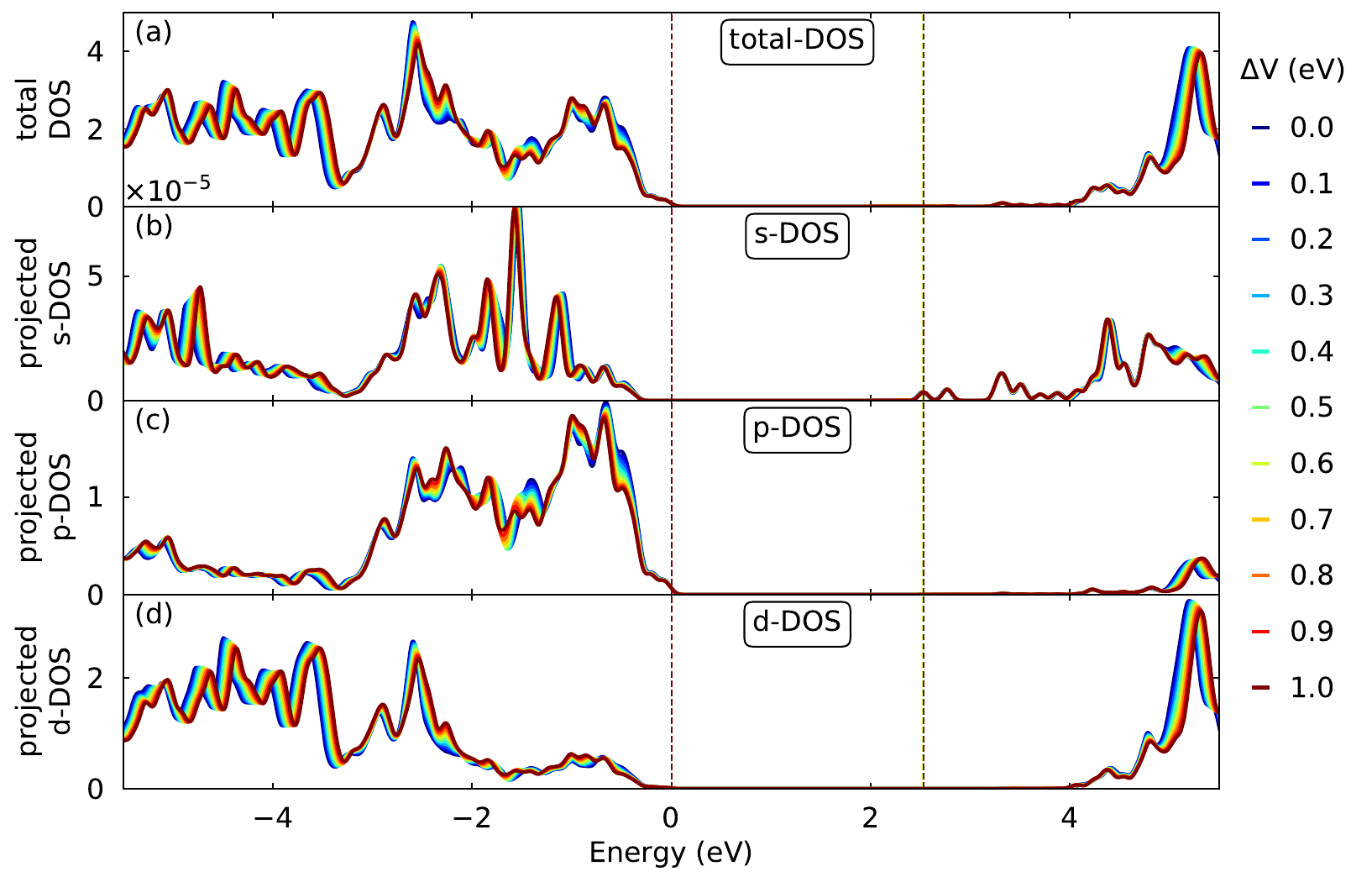}
    \caption{\textbf{Density of states with Hubbard V.} Effect of a change of the Hubbard V between Ni $3d$ and O $2p$ orbitals on (a) the total density of states (DOS), and the projected (b) s-DOS, (c), p-DOS, and (d) d-DOS (continuous lines). The vertical dashed lines indicate the position of the highest occupied and lowest occupied energy levels for a given value of V. The Hubbard U is set to a default value of $U=\SI{6.7}{\eV}$, the equilibrium $V$ values is \SI{2.1}{\eV}. The zero energy is set to the highest occupied energy state before the broadening of the DOS.}
    \label{figSI:DOS_Hubbard_V}
\end{figure}

Figure~\ref{figSI:DOS_Hubbard_V} shows the effect of the Hubbard V between Ni $3d$ and O $2p$ orbitals on the density of states. Contrary to the Hubbard $U$, the effect of this parameter on the energy of the states is rather limited near the band gap but more pronounced on energy states at higher energy in the unoccupied DOS.

\begin{figure}
    \centering
    \includegraphics[width=0.45\linewidth]{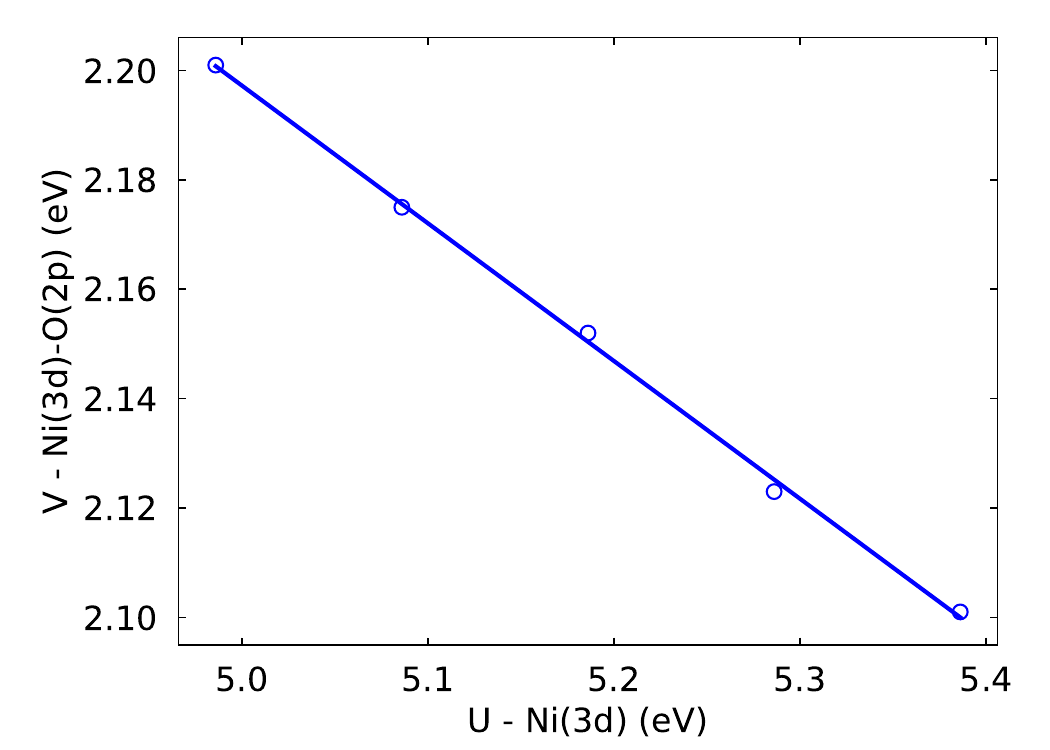}
    \caption{\textbf{Enhanced intersite screening with reduced on-site correlations.} Evolution of the converged value of the intersite screening between Ni $3d$ and O $2p$ orbitals for fixed values of the Hubbard U.}
    \label{figSI:intersite_screening}
\end{figure}

% Effect of the Hubbard U on the converged value of the Hubbard V
Figure~\ref{figSI:intersite_screening} shows the evolution of the intersite screening (V) converged by DFPT for fixed values of the Hubbard $U$. The results show an enhancement of the intersite screening for decreased on-site correlations. For the particular choice of $V$ parameter made in this work, this result implies that light excitation, while acting on the reduction of on-site interactions through Coulomb screening, leads to an enhancement of orbital hybridization.

%%%%%%%%%%%%%%%%%%%%%%%%%%%%%%%%%%%%%%%%%%%%%%%%%%%%%%%%%%%%%%%%%%%%%%%%%%%%%%%%%%%%%%%%%%%%%%%%%%%%%%%%%%%%%%%%%%%%%%%%%%%%%%

\subsubsection{XSpectra\label{subsubsecSI:XSpectra}}

% XSpectra calculations
Calculations of X-ray absorption spectra at the Ni K-edge of NiO were performed with the XSpectra package of Quantum-Espresso~\cite{Gougoussis:2009dg} using the norm-conserving pseudotentials described in \S\ref{subsubsecSI:QE_SCF}~\cite{Taillefumier:2002dea} and the DFT+U~\cite{Gougoussis:2009kr} or DFT+U+V approach~\cite{Timrov2020:73231}. The wavefunction of the core hole was calculated from the GIPAW information in the pseudopotential without the core hole. Convergence threshold for the Lanczos algorithm was set to \SI{1}{\milli\electronvolt}. A gaussian broadening $\Gamma$ of the spectra was performed with an arctan convolution to include the effect of the photoelectron mean-free path on the core-hole broadening following the equation, 
\begin{equation}
\Gamma=\Gamma_0+\Gamma_m\left(\frac{1}{2}+\frac{1}{\pi}\left(\frac{\pi}{3}\frac{\Gamma_m}{E_{\text{l}}}\left(e-\frac{1}{e^2}\right)\right)\right)
\end{equation}
with parameters $\Gamma_0=\SI{1.44}{\eV}$, $\Gamma_m=\SI{1.5}{\eV}$, $E_l=\SI{5}{\eV}$, and $E_c=\SI{2}{\eV}$. A PAW radius of \SI{1.5}{\angstrom} was applied to the $p$ component of the projector, which corresponds to 1.5 times the cutoff radius in the norm-conserving generated all-electron potential. Absorption cross-section from states below the Fermi level are set to 0. Convergence of the calculated XAS spectra was achieved for a k-grid of $5\times5\times5$. Calculations were performed in a supercell approach with a full core hole incorporated in the $1s$ orbital of the pseudopotential in one nickel atom (chosen arbitrarily as spin down). A calculation of the total energy in supercells with increasing dimensions was needed to ensure that core holes do not interact with each other in the periodic lattice, which is displayed in Figure~\ref{figSI:PWscf_core_hole}a. Convergence is assumed for a supercell of dimension $2\times2\times2$ since the energy per unit cell changes by a modest \SI{0.3}{\percent} when increasing the supercell dimension to $3\times3\times3$.

% Realignment of spectra for different values of U
XAS spectra calculated with different values of $U$ or $V$ required an energy realignment with respect to a reference spectrum to be comparable and to compute difference XAS spectra. The reference spectrum is taken for $U=\SI{6.700}{\electronvolt}$ in DFT+U and $U=\SI{5.386}{\eV}$/$V=\SI{2.10}{\eV}$ for DFT+U+V, which correspond to the converged DFPT values for the pseudopotentials used in this work. The value of V is in reasonable agreement with previous calculations at the same level on NiO~\cite{Campo2010:45010}. The SCF calculation with a full core hole in a nickel $1s$ orbital is referred to as FCH. The realignment procedure required two additional SCF calculations to calibrate the energy:
\begin{itemize}
\item \textbf{XCH calculation}: the total charge of the system was set to 0 with a $1s$ core hole in the pseudopotential of a nickel atom in the supercell. This charge constrains an additional charge in the lowest unoccupied energy level of the system, which required the use of the "smearing" keyword described by a Fermi-Dirac distribution with a degauss value of \SI{0.001}{Ry}.
\item \textbf{GS calculation}: Equilibrium calculation without core hole. 
\end{itemize}
From the GS, XCH, and FCH SCF calculations, the energy axis of a given spectrum $E^i$ is given with respect to the energy of a reference spectrum $E^{\text{ref}}$ as,
\begin{equation}
E^i=E^{\text{ref}}+\Delta^i_{\text{SCF}}-\Delta^i_{\text{gap}}
\end{equation}
with the following expression for $\Delta^i_{\text{SCF}}$ and $\Delta^i_{\text{gap}}$,
\begin{align}
\Delta^i_{\text{SCF}} &= \left(E^i_{\text{tot,XCH}}-E^i_{\text{tot,GS}}\right)-\left(E^{\text{ref}}_{\text{tot,XCH}}-E^{\text{ref}}_{\text{tot,GS}}\right) \\
\Delta^i_{\text{gap}} &= \left(E^i_{\text{LUMO}}-E^i_{\text{HOMO}}\right)-\left(E^{\text{ref}}_{\text{LUMO}}-E^{\text{ref}}_{\text{HOMO}}\right)
\end{align}
with LUMO/HOMO referring to the lowest/highest energy of an unoccupied/occupied state in $k$-space. With this procedure the energy required for spectral realignments was $\sim\SI{10}{\milli\electronvolt}$ for reduction of the Hubbard of $\sim\SI{100}{\milli\electronvolt}$.

Calculation of the partial density of states (pDOS) was performed to show the dominant component to the DOS contributing to the XAS spectrum near the absorption edge at the Ni K-edge. The workflow consisted of a SCF, a NSCF and a DOS calculation, all performed on an NiO magnetic unit cell without core hole in the Ni $1s$ orbital. The SCF calculation follows the procedure described in section \S\ref{subsubsecSI:QE_SCF}. The NSCF calculation was performed with a $12\times12\times12$ Monkhorst-Pack grid of k-points and 500 bands. The DOS calculation was performed in a final using a degauss parameter of \SI{0.3680}{Ry} and a tetrahedral integration over the Brillouin zone.

\begin{figure}
	\centering
    	\includegraphics[width=0.5\linewidth]{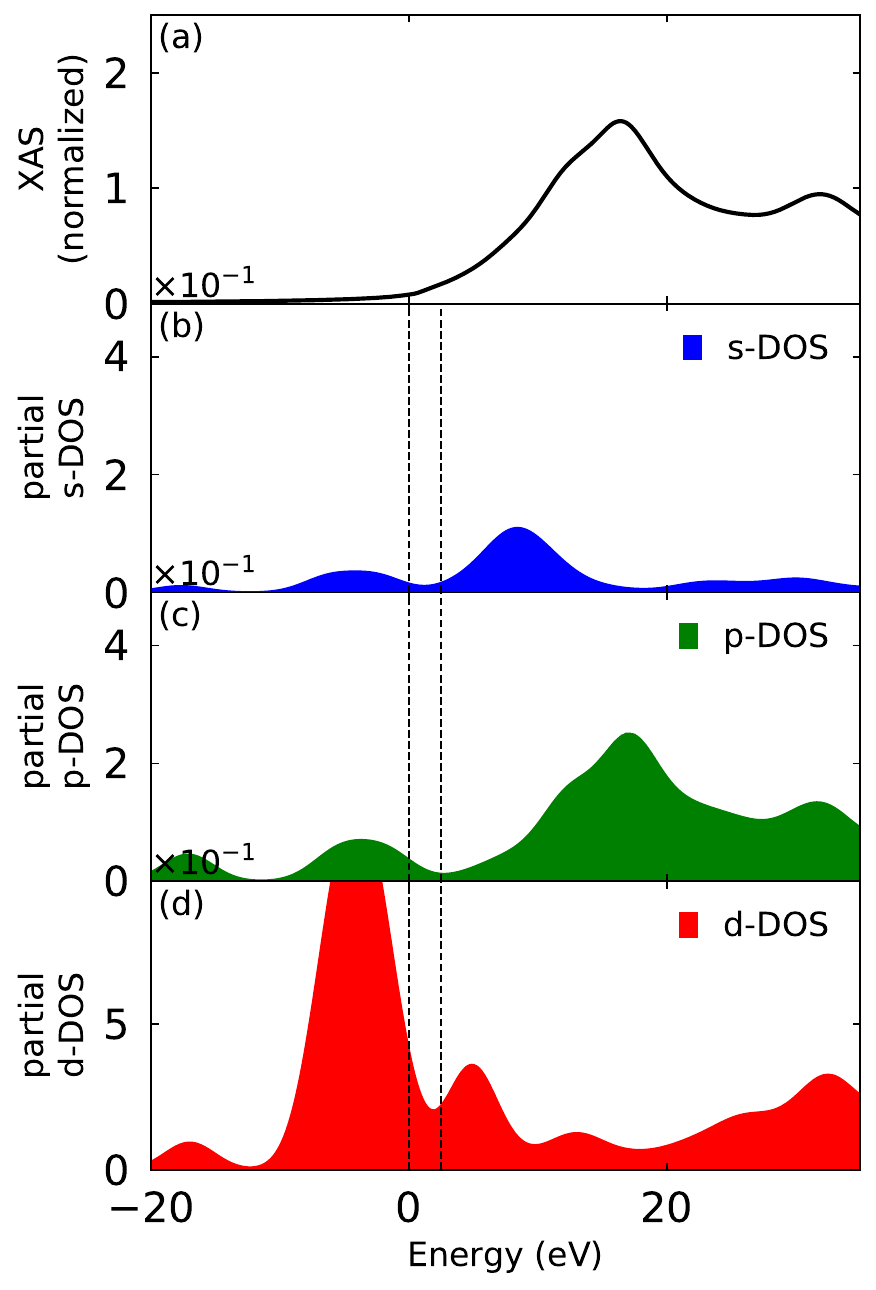}
	\caption{\textbf{Dominant contribution of the partial p-DOS in the near-edge region of the NiO XAS spectrum at the Ni K-edge.} Calculated XAS spectrum (black curve) and partial DOS of the $s$- (red shaded curve), $p$- (green shaded curve) and $d$-DOS (red shaded curve) at the Ni K-edge of NiO. Vertical dashed lines with the partial DOS indicate the position of the highest/lowest occupied/unoccupied states. The zero energy is set at the energy of the highest occupied state.}
    \label{figSI:pDOS_NiO}
\end{figure}

\subsubsection{Screening of the core-hole potential\label{subsecSI:screening_core_potential}}

\begin{figure}
	\centering
	\includegraphics[width=0.5\linewidth]{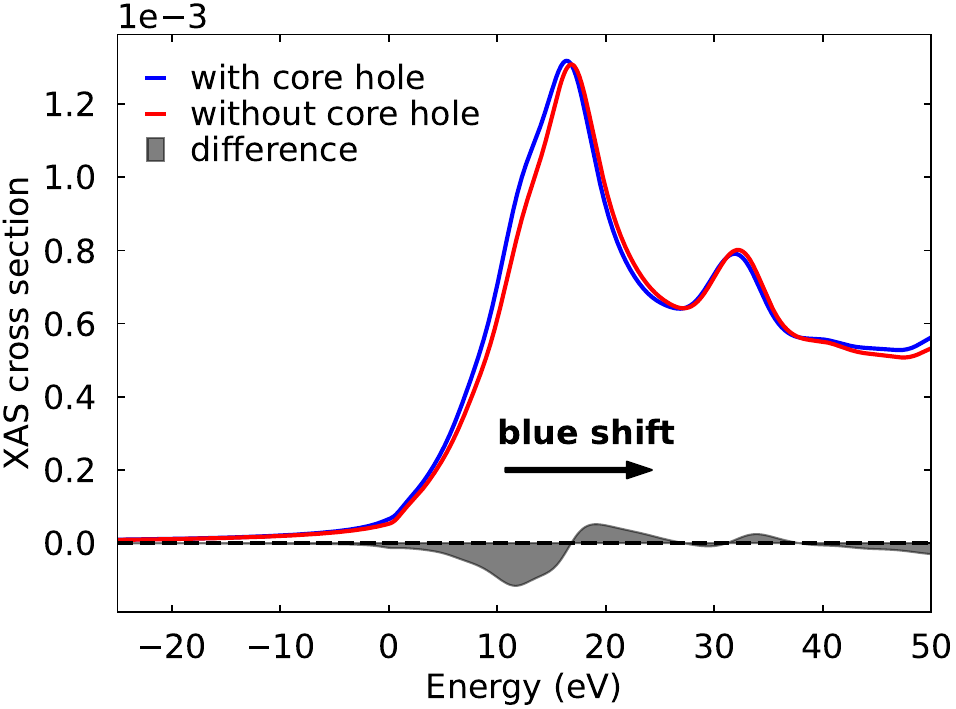}
	\caption{\textbf{Blue shift of the XAS spectrum upon core-hole screening.}
    Computed XAS spectra with a full core hole in the Ni $1s$ orbital (blue curve) and without a core hole (red curve). The difference between the spectra calculated without and with a core hole is shown as the shaded black curve and serves as a simulation of the XTA spectrum resulting from screening of the core-hole interaction. The energy axis is referenced to the Fermi level. The electronic structure was calculated at the DFT+$U$ level with $U=\SI{6.7}{\eV}$.}
    \label{figSI:XSpectra_core_hole_screening}
\end{figure}

To assess the impact of core-hole screening by photoexcited carriers, we simulated an XTA spectrum by taking the difference between computed XAS spectra with a full core hole (FCH) and without core hole, similar to a previous approach~\cite{Rossi2021} that yielded qualitative agreement with many-body perturbation theory calculations including an explicit treatment of the interactions between core and valence states~\cite{Rossi2025:87275}. The calculation, limited to the dipole matrix elements, reveals that screening by photoexcited carriers leads to a blue shift of the XAS spectrum in the excited state (black arrow in Figure \ref{figSI:XSpectra_core_hole_screening}). This behavior aligns with prior results in other metal oxides~\cite{Rossi2021,Rossi2025:87275}, where the screening reduces the binding energy between the core hole and the surrounding valence electrons~\cite{Rossi2021}. At the DFT level, this effect reflects a weakening of the core-hole potential due to changes in the mean-field in the excited state. This computed blue shift is in stark contrast to the experimental non-thermal XTA spectrum, which instead exhibits a signal indicating a red shift (Figure \ref{fig:non_thermal_XTA_evolution_fluence}b). This discrepancy shows that core-hole screening alone cannot account for the observed spectral changes, and suggests that other mechanisms must play a dominant role in the excited-state electronic response of photoexcited NiO.

\subsubsection{Screening on-site Coulomb interaction (Hubbard U)\label{subsecSI:screening_Hubbard_U}}

\begin{figure}
    \centering
    \includegraphics[width=0.48\linewidth]{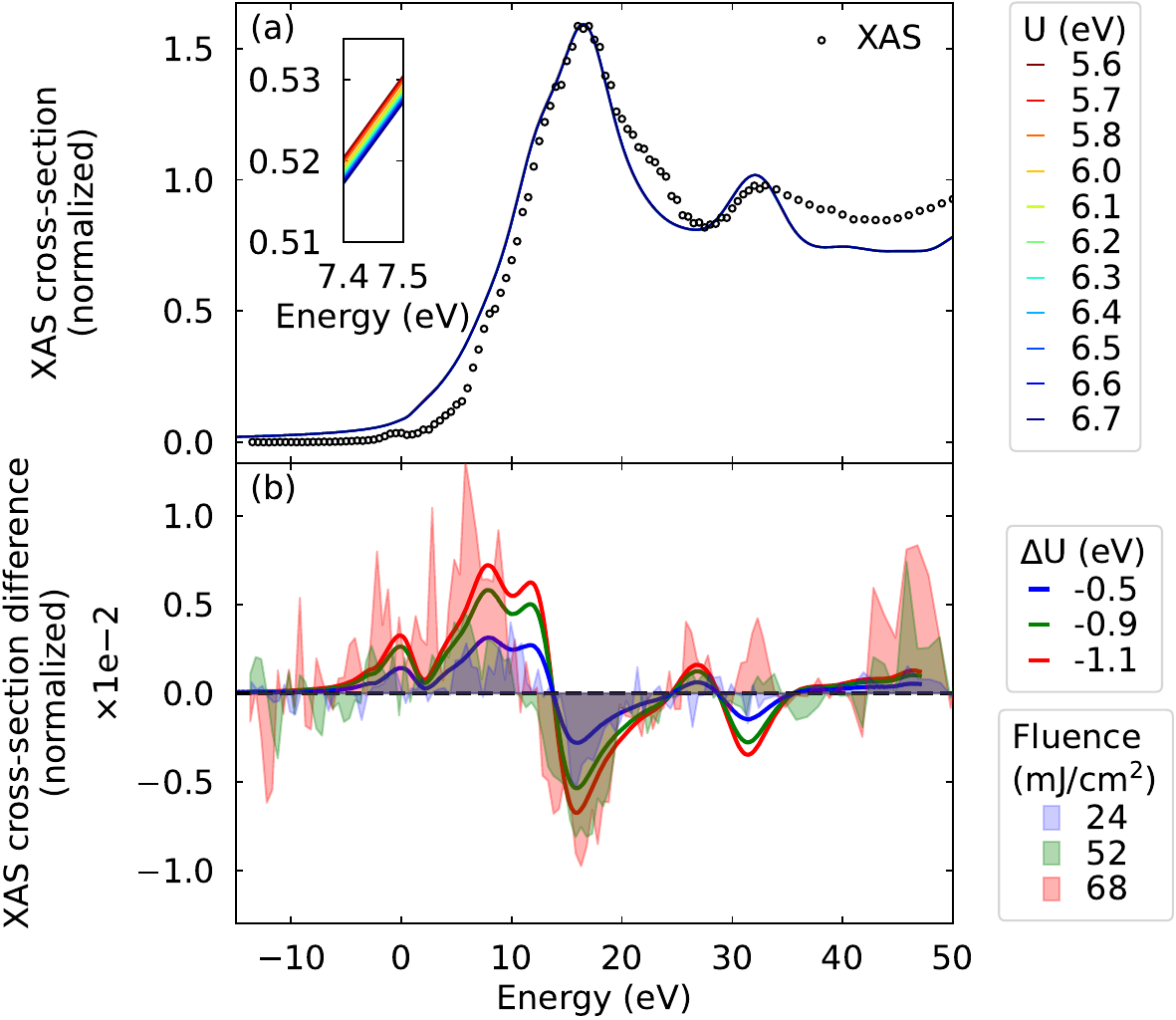}
    \caption{\textbf{Red shift of the XAS spectrum by reduction of the Hubbard U.} Computed (a) XAS and (b) difference XAS spectra upon reduction of the Hubbard $U$ from an equilibrium value (\SI{6.7}{\eV}) down to \SI{5.6}{\eV}. Difference XAS spectrum are shown in panel (b) that match best the experimental data for given reduction of the Hubbard $U$ ($\Delta U$).}
    \label{figSI:XSpectra_Hubbard_U}
\end{figure}

The effect of a screening of the on-site correlation in the Ni $3d$ orbitals was simulated by applying a reduction of the energy of the Hubbard $U$ with respect to the calculated equilibrium value ($U_{eq}=\SI{6.7}{\eV}$). Figure~\ref{figSI:XSpectra_Hubbard_U} shows the evolution of the XAS spectrum upon a reduction of the Hubbard $U$, which displays a small red shift of the absorption edge for decreasing values of $U$. Calculations were performed for reductions of the Hubbard interactions ranging from~\SI{100}{\milli\eV} to \SI{1.1}{\eV}, yielding red shifts between approximately a few meV to $\SI{40}{\milli\eV}$, i.e., more than two orders of magnitude smaller than the imposed reduction of $U$. We argue that the observed changes experimentally cannot be explained by a reduction of the Hubbard $U$ alone due to the limited effect it has on the position of the absorption edge.

\subsubsection{Screening and increase in orbital hybridization (Hubbard $V$)\label{subsecSI:screening_Hubbard_V}}

The effect of screening or the increase in hybridization between oxygen $2p$ and Ni $3d$ orbitals in the excited state is modeled by computing XAS spectra with reduced and increased values of the Hubbard parameter $V$ in a DFT+U+V approach~\cite{Timrov2020:73231}. The role of the Hubbard V is to model intersite electronic interactions, which is crucial in metal oxides with covalent interactions since it affects the hybridization between orbitals. 

\begin{figure}
	\centering
	\includegraphics[width=0.7\linewidth]{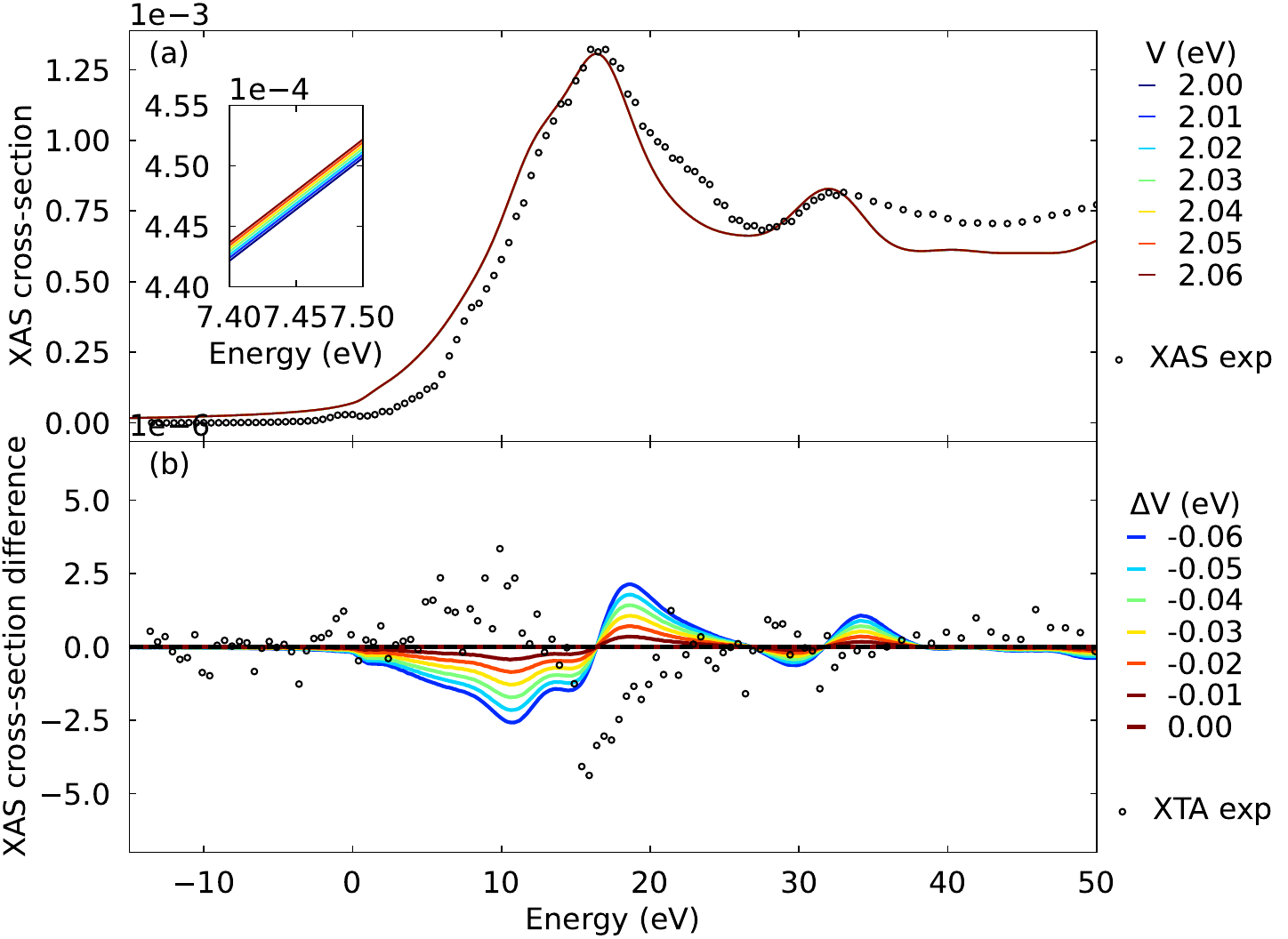}
	\caption{\textbf{Blue shift of XAS spectrum by reduction of the Hubbard V.} Computed (a) XAS and (b) difference XAS spectra upon reduction of the Hubbard V from a reference value of \SI{2.06}{\electronvolt} down to \SI{2.0}{\electronvolt}. The difference XAS spectrum simulates an XTA spectrum upon reduction of $V$. Experimental XAS and XTA spectra are shown for reference (black circles).}
	\label{figSI:XSpectra_Hubbard_V}
\end{figure}

Figure \ref{figSI:XSpectra_Hubbard_V}a shows that the effect of a reduction of $V$ is to blue shift the XAS spectrum, which is in contrast to experimental observations. We note that the effect of a reduction of $V$ leads to significantly larger differences than for a reduction of $U$ by the same magnitude (see main text). This difference originates from the contribution of the $p$ density of states at the absorption threshold, which is directly affected by a change in $V$.

\cleardoublepage

%%%%%%%%%%%%%%%%%%%%%%%%%%%%%%%%%%%%%%%%%%%%%%%%%%%%%%%%%%%%%%%%%
%%%%%%%%%%%%%%%%%%%%%%%%%%%%%%%%%%%%%%%%%%%%%%%%%%%%%%%%%%%%%%%%%
%%%%%%%%%%%%%%%%%%%%%%%%%%%%%%%%%%%%%%%%%%%%%%%%%%%%%%%%%%%%%%%%%
%%%%%%%%%%%%%%%%%%%%%%%%%%%%%%%%%%%%%%%%%%%%%%%%%%%%%%%%%%%%%%%%%

\cleardoublepage

\clearpage
\bibliography{bibliography}

\end{document}